\PassOptionsToPackage{dvipsnames}{xcolor}
\documentclass[11pt,a4paper]{article}
\usepackage[utf8]{inputenc}
\usepackage{jheppub}
\usepackage[english]{babel}
\usepackage{amsmath}
\usepackage{amsfonts}
\usepackage{amssymb}
\usepackage{graphicx}
\usepackage{lmodern}
\usepackage{enumitem}
\usepackage{braket}
\usepackage{subcaption}
\usepackage{float}
\usepackage{pdfpages}
\usepackage{slashed}
\usepackage{multirow}
\usepackage{array}
\usepackage{booktabs}

\usepackage{microtype}

\newcommand{\Fcal}{\mathcal{F}}

\newcommand{\Ccal}{\mathcal{C}}

\newcommand{\qb}{\bar{q}}
\newcommand{\qp}{{q^\prime}}
\newcommand{\qbp}{{\bar{q}^\prime}}

\newcommand{\dd}{\mathrm{d}}
\newcommand{\NC}{\mathrm{NC}}
\newcommand{\CC}{\mathrm{CC}}
\newcommand{\vh}{\hat{v}}
\newcommand{\ah}{\hat{a}}
\newcommand{\ih}{\hat{i}}
\newcommand{\xh}{\hat{x}}
\newcommand{\zh}{\hat{z}}
\newcommand\numberthis{\addtocounter{equation}{1}\tag{\theequation}}

\newcommand{\NoW}{\mathrm{NoW}}
\newcommand{\Fcon}{\mathrm{Fcon}}

\newcommand{\vSa}{\mathrm{A}}
\newcommand{\aSa}{\mathrm{A}}
\newcommand{\SaSa}{\mathrm{AA}}
\newcommand{\GeV}{\mathrm{GeV}}

\newcommand{\nf}{n_f}

\newcommand{\MSbar}{\overline{\textrm{MS}}}

\def\code#1{\texttt{#1}}

\allowdisplaybreaks

\title{Neutral and Charged Current Semi-Inclusive Deep-Inelastic Scattering at NNLO QCD}
\author[a]{Leonardo Bonino,}
\author[a,b]{Thomas Gehrmann,}
\author[a]{Markus Löchner,}
\author[a]{Kay Schönwald,}
\author[c]{and Giovanni Stagnitto}

\affiliation[a]{Universität Zürich, Physik-Institut, Winterthurerstrasse 190, CH-8057 Zürich, Switzerland}
\affiliation[b]{Institute of Nuclear Theory (INT), University of Washington, Seattle, WA 98195-1550, U.S.A.}
\affiliation[c]{Università degli Studi di Milano-Bicocca \& INFN, Piazza della Scienza 3, I-20126 Milano, Italy}

\emailAdd{leonardo.bonino@physik.uzh.ch}
\emailAdd{thomas.gehrmann@uzh.ch}
\emailAdd{markus.loechner@physik.uzh.ch}
\emailAdd{kay.schoenwald@physik.uzh.ch}
\emailAdd{giovanni.stagnitto@unimib.it}

\abstract{Semi-inclusive hadron production in deep inelastic lepton-nucleon scattering (SIDIS) provides important probes of
parton distributions and fragmentation functions.
We compute the next-to-next-to-leading order (NNLO) massless
QCD corrections to the full set of SIDIS coefficient functions in analytical form, accounting for
electroweak neutral current and charged current exchange.
Focusing on the kinematical setting of SIDIS measurements at the future Electron-Ion Collider (EIC),
we quantify the impact of these corrections on the phenomenological predictions and their associated uncertainties. We study the
impact of electroweak interference in the neutral current SIDIS process and investigate lepton polarisation asymmetries
designed to enhance the sensitivity on charged current SIDIS. }

\begin{document}

\begin{flushright}
 ZU-TH 45/25,   INT-PUB-25-018
\end{flushright}

\maketitle

\section{Introduction}

The BNL Electron-Ion Collider (EIC)~\cite{Accardi:2012qut,AbdulKhalek:2021gbh} will serve as a groundbreaking facility for probing the internal structure of hadrons and the dynamics of the strong interaction.
It will enable collisions between leptons and protons or ions at center-of-mass energies ranging from approximately $28\,\GeV$ to $140\,\mathrm{GeV}$, using both polarized and unpolarized beams.
These collisions will generate an unprecedented volume of deep-inelastic scattering (DIS) events, making the EIC a precision laboratory for QCD studies.
Historically, DIS experiments have played a central role in the development and testing of Quantum Chromodynamics (QCD), particularly during the emergence of the parton model~\cite{Bjorken:1968dy,Feynman:1969ej}.
They have been critical in mapping the partonic substructure of the proton.
In fact, the most stringent constraints on parton distribution functions (PDFs) come from DIS data, obtained in both fixed-target experiments such as COMPASS and HERMES and collider experiments like H1 and ZEUS at the DESY HERA collider.
Given this foundational role, it is unsurprising that theoretical predictions for DIS processes are known with exceptional precision.
The inclusive DIS cross section has been calculated up to next-to-next-to-next-to-leading order (N$^3$LO) in perturbative QCD~\cite{Moch:2004xu,Vermaseren:2005qc,Moch:2008fj,Blumlein:2022gpp}.
Also jet production in DIS has been computed at N$^3$LO~\cite{Gehrmann:2018odt}.
Recent developments have extended next-to-leading order (NLO) event generators, matched to parton showers, to the DIS framework~\cite{Carli:2010cg,Hoche:2018gti,Banfi:2023mhz,Borsa:2024rmh,Buonocore:2024pdv,FerrarioRavasio:2024kem}.

Among the EIC’s key physics targets is semi-inclusive DIS (SIDIS), where a specific hadron is identified in the final state.
The identification of a hadron in the final state offers valuable insight into the underlying scattering dynamics by helping to disentangle the contributions of quarks of different flavours originating from the incoming proton.
Moreover, SIDIS is particularly important for accessing fragmentation functions (FFs), which describe the non-perturbative transition of partons into hadrons.
When included in global fits, SIDIS data significantly enhance our ability to constrain the flavour decomposition of quark fragmentation functions.
Full results for the next-to-next-to-leading order (NNLO) SIDIS coefficient functions have recently been presented both for unpolarized beams~\cite{Goyal:2023zdi,Bonino:2024qbh,Goyal:2024emo} and for polarized beams~\cite{Bonino:2024wgg,Goyal:2024tmo}.
However, these results were obtained including only photon-mediated DIS processes.
At the higher end of the EIC’s energy range, up to $140\,\mathrm{GeV}$, neutral current effects involving Z-boson exchange and $\gamma$-Z interference become potentially relevant at the targeted precision.

In addition, thanks to the implementation of the Jacquet-Blondel reconstruction technique~\cite{Amaldi:1979yh}, the EIC may be capable of probing charged current SIDIS processes~\cite{AbdulKhalek:2021gbh}.
These processes were previously measured in fixed-target  neutrino-proton scattering (e.g.~\cite{Aachen-Bonn-CERN-Munich-Oxford:1982jrr})
and
provide complementary information on the flavour decomposition of quarks inside the proton~\cite{Helenius:2024fow,Paukkunen:2025kjb} and
on the quark-flavour dependence of fragmentation functions~\cite{Bonino:2025tnf}.
Future precision measurements at EIC will substantially extend the kinematical range of charged-current SIDIS
processes, and offer access to multiple hadron species.

For the interpretation of future EIC data for SIDIS processes, it
is  therefore  essential to extend the NNLO predictions
to neutral and charged electroweak current interactions.
The purpose of this paper is to present these NNLO corrections for the case of polarized lepton and
 unpolarised hadron beams. The resulting coefficient functions also apply to the case of  (anti-)neutrino scattering,
 for which they have been employed in our phenomenological results presented in~\cite{Bonino:2025tnf}.

 This paper is structured as follows. In Section~\ref{sec:kin}, we review the kinematics of SIDIS and provide the
 definitions of the basic cross sections for neutral and charged current exchange. The associated coefficient functions
 are defined in Sections~\ref{sec:NC} and~\ref{sec:CC} respectively, and the derivation of the NNLO corrections
 to these coefficient functions is described in Section~\ref{sec:calc}. We discuss the phenomenological implications of the
 newly derived results in Section~\ref{sec:num} and conclude with a brief outlook in Section~\ref{sec:conc}.

\section{Kinematics of SIDIS}
\label{sec:kin}
\begin{figure}[tb]
\ifx\nodiags\undefined
	\centering
	\includegraphics{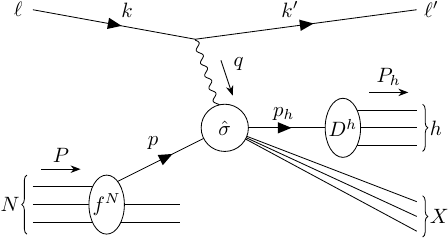}
\fi
\caption{Kinematics of semi-inclusive deep-inelastic scattering}
\label{fig:SIDIS_had}
\end{figure}

We consider the production of a final-state hadron $h$ from the scattering of a highly energetic lepton $\ell$ on a nucleon $N$, $\ell(k) N(P) \to \ell'(k') h(P_h) X$, with $X$ denoting any additional final-state radiation, Figure~\ref{fig:SIDIS_had},
and employ the commonly used kinematic variables
\begin{align*}
&q=k-k' &&\text{four-momentum of the exchanged virtual vector boson,} \\
&Q^2=-q^2 > 0 && \text{virtuality of the exchanged virtual vector boson,} \\
&x=\frac{Q^2}{2P\cdot q} &&\text{Bjorken variable,}  \\
&z=\frac{P\cdot P_h}{P\cdot q}  && \begin{tabular}{@{}p{0.7\textwidth}}\text{scaling variable related to the momentum fraction
picked up} \\ \text{by the identified hadron,} \end{tabular}  \\
&y=\frac{P\cdot q}{P\cdot k} &&\text{energy transfer to the virtual vector boson,}  \\
&W^2=(P+q)^2 &&\text{invariant mass of the hadronic final-state.} \numberthis
\label{eq:hadronic_kinematics}
\end{align*}

The above variables are related to the center-of-mass energy of the lepton-nucleon system, $s = (P+k)^2$, through $Q^2=x y s$. Semi-inclusive ($z$-differential) deep-inelastic scattering describes the kinematic regime $Q^2, W^2 \gg M_N^2$, with $M_N$ being the mass of the incoming nucleon.

\subsection{Hadronic tensor decomposition}
At leading order in the electroweak theory a single gauge boson is exchanged.
The leptonic and hadronic parts
consequently factorise, and the cross section is given by the general form
\begin{align}
\frac{\dd^3\sigma^h}{\dd x \, \dd y \, \dd z} = \frac{2\pi y \alpha^2}{Q^4} \sum_k \eta_k L^{k,\mu\nu} W_{\mu\nu}^{h,k},
\label{eq:lept_had}
\end{align}
with $\alpha$ being the QED coupling constant.
For neutral-current ($\NC$) interactions, the initial and final state leptons have equal flavor. The neutral current is mediated by $\gamma$ or $Z$ exchange at amplitude level, which can interfere at cross section level.
The index $k \in \{\gamma\gamma, ZZ, \gamma Z\}$ denotes the resulting different contributions.
Depending on the exchanged current, the cross section is modified by
\begin{align}
\eta_{\gamma\gamma} &= 1 \, , &
\eta_{\gamma Z} & = \left(\frac{G_FM_Z^2}{2\sqrt{2}\pi \alpha}\right) \left(\frac{Q^2}{Q^2+M_Z^2}\right)\, ,&
\eta_{ZZ} = \eta_{\gamma Z}^2 \,  ,
\end{align}
with the Fermi coupling $G_F$ and the $Z$ boson mass $M_Z$.
In charged-current ($\CC$) interactions the flavours of the initial and final state leptons are different.
There are four different leptonic transitions, $\ell^-\to \nu_\ell$ and $\bar{\nu}_\ell\to \ell^+$ with a mediating $W^-$, and $\ell^+\to \bar{\nu}_\ell$ and $\nu_\ell \to \ell^-$ with a mediating $W^+$.
Our convention for the charge of the mediating $W$ boson is given by the electric charge flow from the lepton to the quark line.
Again, the total cross section takes the form of eq.~\eqref{eq:lept_had} with
\begin{align}
\eta_W = \frac{1}{2}\left( \frac{G_F M_W^2}{4\pi \alpha}\frac{Q^2}{Q^2+M_W^2}\right)^2 \, ,
\end{align}
for the $W$ boson mass $M_W$.

The leptonic tensor $L_{\mu\nu}^{k}$ for an initial state lepton $\ell$ of lepton number $l_{\ell}=\pm 1$ and helicity $\lambda_{\ell}=\pm 1$ is given by
\begin{align}
& L_{\mu\nu}^{\gamma\gamma} = 2\left( k_\mu^{\phantom{\prime}} k_\nu^\prime + k_\mu^\prime k_\nu^{\phantom{\prime}} - (k\cdot k^\prime - m_\ell^2) g_{\mu\nu}^{\phantom{\prime}} - i \lambda_{\ell} \varepsilon_{\mu\nu\rho\sigma}^{\phantom{\prime}}k^\rho k^{\prime \sigma}\right)\, , \nonumber \\
& L_{\mu\nu}^{\gamma Z} = \left( g_V^\ell - l_{\ell} \lambda_{\ell} g_A^\ell \right) L_{\mu\nu}^{\gamma\gamma} \,
 , \quad L_{\mu\nu}^{ZZ} = (g_V^\ell - l_{\ell} \lambda_{\ell} g_A^\ell)^2 L_{\mu\nu}^{\gamma\gamma} \, , \quad
L_{\mu\nu}^{W} =& (1 - l_{\ell}\lambda_{\ell})^2 L_{\mu\nu}^{\gamma\gamma} \, .
\end{align}
with
\begin{align}
g_V^{e,\mu,\tau} =& -\frac{1}{2} +2\sin^2\theta_\mathrm{W}\, , &
g_A^{e,\mu,\tau} =& -\frac{1}{2}\, , &
g_V^{\nu,\bar{\nu}} &= g_A^{\nu,\bar{\nu}} = \frac{1}{2} \, .
\end{align}

The hadronic tensor $W_{\mu\nu}^{h,k}$ for a resolved final state hadron $h$ is defined by the current insertion of the electromagnetic or electroweak current belonging to $k$,
\begin{align}
W^{h,k}_{\mu\nu} 
= \frac{1}{4\pi} \int \dd^4r \, e^{iq\cdot r} \sum\limits_{X} \Braket{P | J_{\mu,k}^\dagger (r) | h,X } \Braket{ h,X | J_{\nu,k}^{\phantom{\dagger}}(0) | P} \,,
\end{align}
for a nucleon with momentum $P$ where $P^2=M_{N}^{2}$.

The hadronic tensor for an unpolarized (spin-averaged) hadron
is decomposed into structure functions by~\cite{ParticleDataGroup:2024cfk,deFlorian:2012wk}
\begin{align}
W_{\mu \nu}^{h,k} =& \bigg(-g_{\mu \nu} + \frac{q_\mu q_\nu}{q^2}\bigg) F_1^{h,k}(x,z,Q^2) + \frac{\hat{P}_\mu \hat{P}_\nu}{P\cdot q} F_2^{h,k}(x,z,Q^2) - i \varepsilon_{\mu\nu\rho\sigma} \frac{q^{\rho} P^\sigma}{2P\cdot q} F_3^{h,k}(x,z,Q^2)
\label{eq:had_tens}
\end{align}
with
\begin{align}
\hat{P}_\mu=& P_\mu - \frac{P\cdot q}{q^2}q_\mu \, .
\end{align}

Instead of $F_1^{h,k}$, $F_2^{h,k}$, $F_3^{h,k}$ our results are conventionally re-expressed in terms of the structure functions
\begin{align}
\Fcal_T^{h,k} &= 2F_1^{h,k} \, , &
\Fcal_L^{h,k} &= F_2^{h,k}/x - 2 \, F_1^{h,k}, \, &
\Fcal_3^{h,k} &= F_3^{h,k} \, .
\end{align}
The structure functions associated with an \mbox{(anti-)symmetric} tensor structure will be referred to as \mbox{(anti-)symmetric} in the following.

The structure functions depend  on $x$, $z$, and $Q^2$.
In the NC case they apply to scattering processes with conserved lepton flavour.
For a charged lepton, they are obtained by summing $\gamma$, $Z$ and interference contributions:
\begin{align} \label{eq:SFNC}
\mathcal{F}_{T,L}^{h,\NC}&= \mathcal{F}_{T,L}^{h,\gamma\gamma}-\left(g^{\ell}_V - l_{\ell} \lambda_{\ell} g^{\ell}_A\right)\eta_{\gamma Z}\,\mathcal{F}_{T,L}^{h,\gamma Z}+\left(g^{\ell \, 2}_V+g^{\ell \, 2}_A - 2 l_{\ell} \lambda_{\ell} g^{\ell}_Vg^{\ell}_A\right)\eta_{ZZ}\,\mathcal{F}_{T,L}^{h,ZZ} \, , \nonumber \\
\mathcal{F}_{3}^{h,\NC}&= -\left(g^{\ell}_A - l_{\ell} \lambda_{\ell} g^{\ell}_V\right)\eta_{\gamma Z}\,\mathcal{F}_{3}^{h,\gamma Z} + \left[2 g^{\ell}_V g^{\ell}_A - l_{\ell} \lambda_{\ell} \left(g^{\ell \, 2}_V+g^{\ell \, 2}_A \right)\right]\eta_{ZZ}\, \mathcal{F}_{3}^{h,ZZ}\,  .
\end{align}
The lepton-flavour changing $\CC$ structure functions are given by
\begin{align}
\mathcal{F}_{T,L}^{h,\CC}&= \mathcal{F}_{T,L}^{h,W^{\pm}} \, , &
\mathcal{F}_{3}^{h,\CC}&= \mathcal{F}_{3}^{h,W^{\pm}} \,  .
\end{align}
The neutral current/charged current SIDIS cross section in terms of these structure functions is then given by
\begin{align}\label{eq:xsNCCC}
\frac{\dd^3 \sigma^{h,\NC / \CC}}{\dd x\,\dd y\,\dd z} = \frac{4\pi \alpha^2}{Q^2} \eta^{\NC/\CC} \Bigg[  \frac{1+(1-y)^2}{2y}\Fcal_T^{h,\NC / \CC}+\frac{1-y}{y}\Fcal_L^{h,\NC / \CC} & \nonumber \\
+ l_{\ell} \frac{1-(1-y)^2}{2y}\Fcal_3^{h,\NC / \CC}&  \Bigg] \,  .
\end{align}
Furthermore $\eta^{\NC}=1$, whereas $\eta^{\CC}=(1 - l_{\ell}\lambda_{\ell})^2\eta_W$, with the helicity of the (anti-)lepton $\lambda_{\ell}=\pm 1$.

\subsection{Parton-model expressions}

Factorization in SIDIS separates the cross section into three pieces:
\begin{enumerate}[label=(\roman*)]
\item Long-range initial-state interactions, with a parton $p$ extracted from the incoming nucleon $N$ according to a collinear parton distribution function (PDF) $f_p^N$ evaluated at the initial state factorization scale $\mu_F$.
\item Long-range final-state interactions, with a parton $p'$ fragmenting into the detected hadron $h$ through the fragmentation function (FF) $D_{p'}^h$ evaluated at the final state factorization scale $\mu_A$.
\item The short-distance cross section $\hat{\sigma}$ at the hard scale $Q$, which is calculated at the partonic level in perturbative QCD. The renormalization of $\hat{\sigma}$ introduces a third scale, the renormalization scale $\mu_R$.
\end{enumerate}
This factorization is expressed at the level of the structure functions:
\begin{align}\label{eq:Fcal_Gcal}
\Fcal_i^{h,k}(x,z,Q^2) =& \sum_{p,p'}\int_x^1 \frac{\dd \xh}{\xh} \int_z^1 \frac{\dd \zh}{\zh} \, f_p^N\bigg(\frac{x}{\xh},\mu_F^2\bigg)\, D^h_{p'}\bigg(\frac{z}{\zh},\mu_A^2\bigg) \, \Ccal^{i,k}_{p'p}(\xh,\zh,Q^2,\mu_R^2,\mu_F^2,\mu_A^2) \, ,
\end{align}
where $\Ccal^{i,k}_{p'p}$ denotes the parton-level coefficient function.
The indices $i\in\{T,L,3\}$ refer to the respective structure function, and $k\in\{\gamma \gamma, \gamma Z, ZZ, W^\pm \}$ refers to the electroweak gauge bosons exchanged in the process.
For this purpose, we introduce the partonic kinematics in analogy with eq.~\eqref{eq:hadronic_kinematics} as
\begin{align*}
&\xh=\frac{Q^2}{2p\cdot q} &&\text{momentum fraction of the initial state parton,}  \\
&\zh=\frac{p\cdot p_h}{p\cdot q}  &&\text{momentum fraction
picked up by the identified parton.} \numberthis
\label{eq:partonic_kinematics}
\end{align*}
The coefficient functions describing the hard scattering process can be calculated in perturbation theory by expanding in the strong coupling constant $\alpha_s$,
\begin{align}
\Ccal^{i,k}_{p'p}=\Ccal^{i,k, (0)}_{p'p}+\frac{\alpha_s(\mu_R^2)}{2\pi} \Ccal^{i,k, (1)}_{p'p}+\left(\frac{\alpha_s(\mu_R^2)}{2\pi}\right)^2 \Ccal^{i,k, (2)}_{p'p} +\mathcal{O}(\alpha_s^3)\,  .
\end{align}

In the following sections we describe the individual contributions to the SIDIS coefficient functions up to NNLO.

\section{SIDIS Coefficient Functions up to NNLO}

\subsection{Electroweak Channel Decomposition}
In previous calculations of SIDIS at NNLO~\cite{Goyal:2023zdi,Bonino:2024qbh,Bonino:2024wgg,Goyal:2024emo,Goyal:2024tmo} the channel decomposition of~\cite{Anderle:2016kwa} was employed.
In this section we will discuss why this decomposition is unsuited for describing the electroweak case.
We will instead present our results in a new decomposition of SIDIS coefficient functions which simultaneously applies to neutral and charged current SIDIS.

One problem at NNLO in charged current SIDIS is the flavour changing nature of the $W$ boson vertex, which introduces further structure in all channels containing two quark lines at amplitude level:
forward scattering (squared amplitude) diagrams appearing in neutral current SIDIS are diagrammatically forbidden in charged current SIDIS if the two electroweak vector bosons couple to different quark lines.
Such contributions appearing only in $\NC$ SIDIS but not in $\CC$ SIDIS will be denoted as ``$\NoW$''.

\begin{figure}[tb]
\ifx\nodiags\undefined
	\centering
	\subcaptionbox{\label{fig:SIDIS_qC40q_Fcha}
}[.33\textwidth]{
		\centering
		\includegraphics[scale=.8]{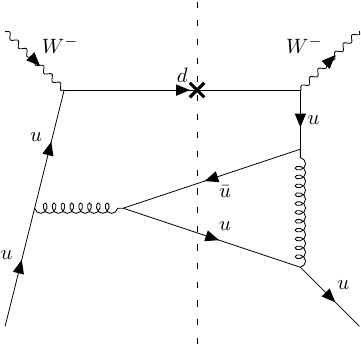}
	}
	\subcaptionbox{\label{fig:SIDIS_qC40q_Fcon}
}[.33\textwidth]{
		\centering
		\includegraphics[scale=.8]{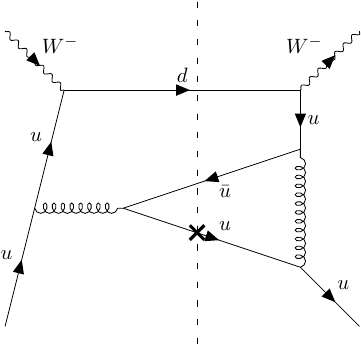}
	}
	\subcaptionbox{\label{fig:SIDIS_qB40q_Fcon}
}[.32\textwidth]{
		\centering
		\includegraphics[scale=.8]{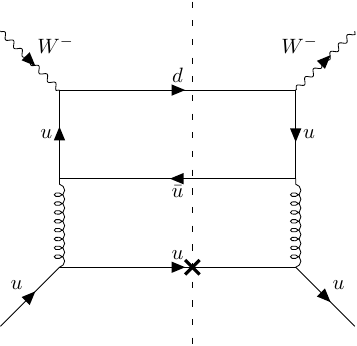}
	}
\fi
\caption{Representative diagrams for contributions to the flavour changing (a) and flavour-conserving $C_{qq,\Fcon}^{(2)}$ coefficient functions (b),(c).
The identified final state parton is marked with an `x'.
}
\end{figure}

In the following we discuss the effects of flavour change at the $W$ boson vertex to the $q\to q$ channel.
We consider only contributions to the $q\to q$ channel where the identified final state quark is connected to the initial state quark through a fermion line.
As the final state quark is identified in SIDIS we become sensitive to its flavour.
Up to NLO the transition is always flavour conserving in $\NC$ SIDIS and flavour changing in $\CC$ SIDIS, e.g.\ $u\to d$ or $d \to u$, due the presence of only a single quark line in the forward scattering amplitude.
At NNLO however, it is possible to have a second quark line.
Therefore both flavour conserving and flavour changing transitions in $\CC$ SIDIS are allowed in this channel, while the corresponding $\NC$ forward scattering diagrams remain flavour conserving.
The flavour-conserving (``Fcon'') contributions in CC arise from two distinct cases at NNLO:
\begin{enumerate}
\item Forward scattering amplitudes generated by the interference of a non-singlet-type amplitude, for which the vector boson couples to the fermion line of the initial quark, with a singlet-type amplitude, where the boson couples to a secondary quark line.
Examples are depicted in Fig.~\ref{fig:SIDIS_qC40q_Fcha} and \ref{fig:SIDIS_qC40q_Fcon}.
Depending on the identified quark on the cut, these diagrams give rise to the flavour changing ($u\to d$) or to the flavour conserving ($u\to u$) channel.
\item Pure-singlet-type forward scattering amplitudes arising from the modulus square of singlet amplitudes, cf.\ Fig.~\ref{fig:SIDIS_qB40q_Fcon}.
In such contributions the flavour change takes place on an unidentified quark line.
\end{enumerate}

\begin{figure}[tb]
\ifx\nodiags\undefined
	\centering
	\subcaptionbox{\label{fig:SIDIS_qC40qb_Fcha}
}[.4\textwidth]{
		\centering
		\includegraphics[scale=.8]{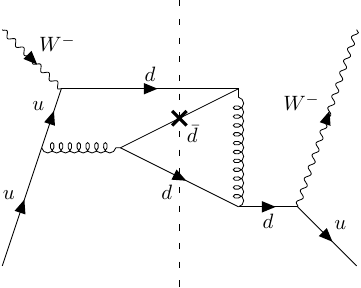}
		\vspace{1em}
	}
	\subcaptionbox{\label{fig:SIDIS_qC40qb_Fcon}
}[.4\textwidth]{
		\centering
		\includegraphics[scale=.8]{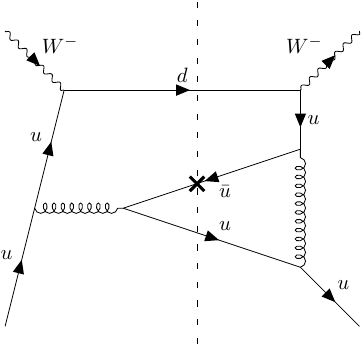}
	}
\fi
\caption{Representative diagrams for the flavour-changing and flavour-conserving (non-singlet) contribution to the $C_{\qb q}^{(2)}$ coefficient function.
The identified final state parton is marked with an `x'.
}
\end{figure}

A similar distinction in charged-current SIDIS arises in the $q\to \qb$ channel.
Depending on the position of the electroweak coupling along the quark line in the forward-scattering diagram, one has flavour-changing (Fig.~\ref{fig:SIDIS_qC40qb_Fcha}) or flavour-conserving contributions (Fig.~\ref{fig:SIDIS_qC40qb_Fcon}) in CC SIDIS.

\begin{figure}[tb]
\ifx\nodiags\undefined
	\centering
	\subcaptionbox{\label{fig:SIDIS_qA31q_vSa}
}[.4\textwidth]{
		\includegraphics[scale=.8]{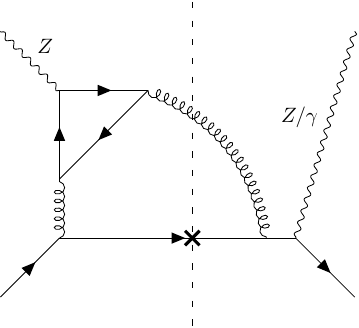}
	}
	\subcaptionbox{\label{fig:SIDIS_gA21axx21g_SaSa}
}[.4\textwidth]{
		\includegraphics[scale=.8]{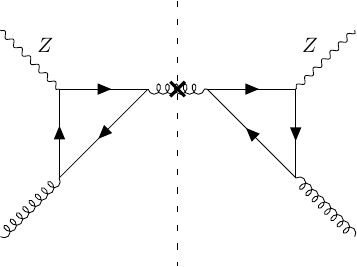}
		\vspace{1em}
	}
\fi
\caption{Example diagrams for anomaly contributions to the neutral current case, vanishing only for even $\nf$.
The identified final state parton is marked with an `x'.
}
\label{fig:anomaly_diagrams}
\end{figure}

Finally, due to the presence of axial couplings, also anomaly contributions arise at NNLO in $\NC$ SIDIS with a new coefficient structure proportional to the sum of axial vector couplings.
Representative diagrams are shown in Fig.~\ref{fig:anomaly_diagrams}.
Owing to the flavour changing nature of the $W$ vertex such contributions are forbidden in $\CC$ SIDIS.
In the following contributions with a single anomaly insertion are denoted as ``$\mathrm{A}$'' and contributions with a double anomaly insertion as ``$\SaSa$''.
Apart from new structures in the channels $q\to q$, $q\to g$, and $g\to q$, also a finite contribution in the $g\to g$ channel with a double anomaly arises. 
In Table \ref{tab:chanlab} we summarize the labels of the above mentioned channel contributions
\begin{center}
\begin{table}[h]
\centering
\begin{tabular}{ |m{2.5em}|m{25em}| } 
 \hline
 \multicolumn{2}{|c|}{Channel labels} \\ 
 \hline
 \hline
 NoW & NC contribution only: the electroweak vector bosons couple to two different quark lines  \\ 
 \hline
 Fcon & Flavour conserving contribution \\ 
 \hline
 A & Single anomaly insertion contribution\\
 \hline
 AA & Double anomaly insertion contribution \\
 \hline
\end{tabular}
\caption{List and description of channel labels.}\label{tab:chanlab}
\end{table}
\end{center}

Following the discussion, we are ready to formulate the electroweak channel decomposition
\begin{alignat*}{3}
& \Ccal^{i,k,(2)}_{qq} &&= C_{qq}^{i,k,(2)} + C^{i,k,(2)}_{qq, \Fcon} + C^{i,k,(2)}_{qq,\NoW} + \left. C^{i,k,(2)}_{\qp q} \right|_{\qp=q} + \left. C^{i,k,(2)}_{\qp q, \NoW} \right|_{\qp=q} \,  , \\
& \Ccal^{i,k,(2)}_{\qb q} &&= C^{i,k,(2)}_{\qb q} + C^{i,k,(2)}_{\qb q, \Fcon} + \left. C^{i,k,(2)}_{\qbp q} \right|_{\qbp = \qb} + \left. C^{i,k,(2)}_{\qbp q, \NoW} \right|_{\qbp = \qb} \,  , \\
& \Ccal^{i,k,(2)}_{\qp q} &&= C^{i,k,(2)}_{\qp q} + C^{i,k,(2)}_{\qp q, \NoW}   && \hspace{-6.5cm}(\text{for } \qp \neq q) \,  , \\
& \Ccal^{i,k,(2)}_{\qbp q} &&= C^{i,k,(2)}_{\qbp q} + C^{i,k,(2)}_{\qbp q, \NoW}  && \hspace{-6.5cm} (\text{for } \qbp \neq \qb) \, , \\
& \Ccal^{i,k,(2)}_{gq} &&= C^{i,k,(2)}_{gq} + C^{i,k,(2)}_{gq,\NoW} \, , \\
& \Ccal^{i,k,(2)}_{qg} &&= C^{i,k,(2)}_{qg} + C^{i,k,(2)}_{qg,\NoW} \, , \\
& \Ccal^{i,k,(2)}_{gg} &&= C^{i,k,(2)}_{gg} + C^{i,k,(2)}_{gg,\NoW} \numberthis \, ,
 \label{eq:channel_decomp}
\end{alignat*}
where $q,\qp$ denote quarks and $\qb,\qbp$ denote anti-quarks.

In Section~\ref{sec:NC} the precise composition of the $\Ccal^{i,k,(2)}_{p p'}$ for $\NC$ exchange in terms of unique building blocks is discussed.
In Section~\ref{sec:CC} we use these building blocks to assemble the $\CC$ coefficient functions.

\subsection{Neutral Current SIDIS}
\label{sec:NC}

In the Standard Model the $\NC$ SIDIS coefficient functions describe the partonic hard scattering in the processes
\begin{align}
& \ell^- N \to \ell^- h X\, , & & \ell^+ N \to \ell^+ h X \, , & &  \nu_\ell N \to \nu_\ell h X \, , & & \bar{\nu}_\ell N \to \bar{\nu}_\ell h X\,.
\end{align}
In the following we establish the notation and present the coefficient functions for NC SIDIS up to NNLO in QCD.

The $\NC$ structure functions are present in photon and $Z$ boson exchange, and in their interference ($k\in\{\gamma \gamma , ZZ ,\gamma Z\}$).
We introduce generic vector or axial vector couplings
\begin{align}
v^{\gamma}_q &=e_q^{\phantom{\gamma}} \,, & v^{Z}_q &= g^q_V \,, &
a^{\gamma}_q &=0\,, & a^{Z}_q &= g^q_A\,,
\end{align}
with
\begin{align*}
e_u &= \phantom{+}\frac{2}{3}\, , &
g^u_V &= \phantom{+}\frac{1}{2} - \frac{4}{3}\sin^2\theta_\mathrm{W} \, , &
g^u_A &= \phantom{+}\frac{1}{2} \, , \\
e_d &= -\frac{1}{3} \, , &
g^d_V &= -\frac{1}{2} + \frac{2}{3}\sin^2\theta_\mathrm{W} \, ,&
g^d_A &= -\frac{1}{2} \numberthis
\end{align*}
for up- and down-like quarks with the weak mixing angle $\sin^2 \theta_\mathrm{W}$.
To write the coefficient functions for $\NC$ SIDIS in uniform notation, we define
\begin{align*}
\vh^{\gamma\gamma}_{q\qp} &=e_q e_\qp \,, & \vh^{\gamma Z}_{q\qp} &= e_q g_V^{\qp} + e_\qp g_V^q \, , & \vh^{ZZ}_{q\qp} &= g^q_V\, g^{\qp}_V\,, \\
\ah^{\gamma\gamma}_{q\qp} &=0\,, & \ah_{q\qp}^{\gamma Z} &= 0 \, , & \ah^{ZZ}_{q\qp} &= g^q_A \, g^{\qp}_A \,, \\
\ih^{\gamma\gamma}_{q\qp} &=0\,, & \ih^{\gamma Z}_{q\qp} &= 2 e_q^{\phantom{q}} g^{\qp}_A\,,  & \ih^{ZZ}_{q\qp} &= g^q_V g^{\qp}_A + g^{\qp}_V g^q_A  \numberthis
\label{eq:couplings}
\end{align*}
for two arbitrary quarks $q,\qp$.
Here, $\vh$ denotes combinations of vector couplings, $\ah$ denotes combinations of axial vector couplings, and $\ih$ denotes interferences between vector and axial vector couplings.
Note that $\ih_{q\qp}^{\gamma Z}$ is not symmetric in the quark indices: $\ih_{q\qp}^{\gamma Z}\neq \ih_{\qp q}^{\gamma Z}$.

\subsubsection{Symmetric Coefficient Functions}\label{sec:NCSymmCF}
In this section we describe the neutral current (NC) coefficient functions for the structure functions $\mathcal{F}^h_T$ and $\mathcal{F}^h_L$ up to NNLO.

At LO only the $q\to q$ channel contributes,
\begin{align*}
\Ccal^{\{T,L\},k,(0)}_{qq} = \left[ \vh_{qq}^k + \ah_{qq}^k \right] C^{\{T,L\},(0)}_{qq} \,  , \numberthis
\end{align*}
with $C^{T,(0)}_{qq} = \delta(1-\zh)\delta(1-\xh)$ and $C^{L,(0)}_{qq} = 0$.

At LO the symmetric $\NC$ SIDIS structure functions are therefore given by
\begin{alignat*}{2}
& \Big[ \Fcal_{T}^{h,\gamma\gamma},~ \Fcal_{T}^{h,\gamma Z},~ \Fcal_{T}^{h,ZZ} \Big] &&= \sum_{q} \Big[ e_q^2,~ 2 \, e_q \, g_V^q ,~ (g_V^q)^2 + (g_A^q)^2 \Big] \big( f_q^N \, D_q^h + f_{\qb}^N \, D_{\qb}^h \big) \, , \\
& \Big[ \Fcal_{L}^{h,\gamma\gamma},~ \Fcal_{L}^{h,\gamma Z},~ \Fcal_{L}^{h,ZZ} \Big] &&= 0 \, . \numberthis
\end{alignat*}

The NLO SIDIS coefficient functions $\Ccal^{\{T,L\},(1)}_{p'p}$ were first computed in \cite{Altarelli:1979kv,Baier:1979sp}.
At NLO, the $q\to g$ and $g\to q$ channels open up, and the longitudinal components no longer vanish:
\begin{align*}
\Ccal^{\{T,L\},k,(1)}_{qq} &= \left[ \vh_{qq}^k + \ah_{qq}^k \right] C^{\{T,L\},(1)}_{qq} \,  , \\
\Ccal^{\{T,L\},k,(1)}_{gq} &= \left[ \vh_{qq}^k + \ah_{qq}^k \right] C^{\{T,L\},(1)}_{gq} \,  , \\
\Ccal^{\{T,L\},k,(1)}_{qg} &= \left[ \vh_{qq}^k + \ah_{qq}^k \right] C^{\{T,L\},(1)}_{qg} \,  . \numberthis
\end{align*}

At NNLO, each of the contributions in eq.~\eqref{eq:channel_decomp} is present in the transverse and longitudinal $\NC$ structure functions.
The missing coefficient functions with anti-quarks in the initial state are trivially obtained by charge conjugation through
\begin{equation}
\Ccal^{\{T,L\},k,(j)}_{p'p} = \Ccal^{\{T,L\},k,(j)}_{\bar{p}'\bar{p}}\quad \mathrm{for} \quad j \in \{0,1,2\} \, ,
\end{equation}
where $p$ and $p'$ are understood as generic partons (quark, antiquarks or gluons).

The contributions entering the right-hand-side of~\eqref{eq:channel_decomp} are given in terms of the electroweak couplings from eq.~\eqref{eq:couplings},
multiplying building blocks of different diagrammatic origin:
\begin{align*}
C^{\{T,L\},k,(2)}_{qq} & = \left[ \vh_{qq}^k + \ah_{qq}^k \right]C^{\{T,L\},(2)}_{qq}  \,  , \\
C^{\{T,L\},k,(2)}_{qq, \Fcon} &= \left[ \vh_{qq}^k + \ah_{qq}^k \right] C^{\{T,L\},(2)}_{qq,\Fcon,1} + \sum_{\qp} \left[ \vh_{\qp\qp}^k + \ah_{\qp\qp}^k \right] C^{\{T,L\},(2)}_{qq,{\Fcon},2}  \,  , \\
C^{\{T,L\},k,(2)}_{qq,\NoW} &= a^k_q \sum_{\qp} \left( a^k_{\qp} \right) C^{\{T,L\},(2)}_{qq, \aSa,\NoW}  \,  , \\
C^{\{T,L\},k,(2)}_{\qb q} & = \left[ \vh_{qq}^k + \ah_{qq}^k \right]C^{\{T,L\},(2)}_{\qb q}  \,  , \\
C^{\{T,L\},k,(2)}_{\qb q, \Fcon} &= \left[ \vh_{qq}^k + \ah_{qq}^k \right] C^{\{T,L\},(2)}_{\qb q, \Fcon}  \,  , \\
C^{\{T,L\},k,(2)}_{\qp q} & =  \left[ \vh_{qq}^k + \ah_{qq}^k \right]C^{\{T,L\},(2)}_{\qp q, 1} + \left[ \vh_{\qp\qp}^k + \ah_{\qp\qp}^k \right]C^{\{T,L\},(2)}_{\qp q, 2} \,  , \\
C^{\{T,L\},k,(2)}_{\qp q, \NoW} &= \vh_{q\qp}^k \,  C^{\{T,L\},(2)}_{\qp q, \NoW, 3} + \ah_{q\qp}^k \, C^{\{T,L\},(2)}_{\qp q, \NoW, 4}  \,  , \\
C^{\{T,L\},k,(2)}_{\qbp q} & = \left[ \vh_{qq}^k + \ah_{qq}^k \right]C^{\{T,L\},(2)}_{\qp q, 1} + \left[ \vh_{\qp\qp}^k + \ah_{\qp\qp}^k \right]C^{\{T,L\},(2)}_{\qp q, 2} \,  , \\
C^{\{T,L\},k,(2)}_{\qbp q, \NoW} &= - \vh_{q\qp}^k \, C^{\{T,L\},(2)}_{\qp q, \NoW, 3}+ \ah_{q\qp}^k \, C^{\{T,L\},(2)}_{\qp q, \NoW, 4}  \,  , \\
C^{\{T,L\},k,(2)}_{gq} &= \left[ \vh_{qq}^k + \ah_{qq}^k \right] C^{\{T,L\},(2)}_{gq} \, ,  \\
C^{\{T,L\},k,(2)}_{gq,\NoW} &= a^k_q \sum_{q'} \left( a^k_{q'} \right) C^{\{T,L\},(2)}_{gq, \aSa,\NoW}  \,  , \\
C^{\{T,L\},k,(2)}_{qg} &= \left[ \vh_{qq}^k + \ah_{qq}^k \right] C^{\{T,L\},(2)}_{qg} \, ,  \\
C^{\{T,L\},k,(2)}_{qg,\NoW} &= a^k_q \sum_{q'} \left( a^k_{q'} \right) C^{\{T,L\},(2)}_{q g, \aSa,\NoW}  \,  , \\
C^{\{T,L\},k,(2)}_{gg} &=\sum_{q'} \left[ \vh_{\qp\qp}^k + \ah_{\qp\qp}^k \right] C^{\{T,L\},(2)}_{gg} \, ,  \\
C^{\{T,L\},k,(2)}_{gg,\NoW} &= \left[ \sum_{q'} a^k_{\qp} \right]^2  C^{\{T,L\},(2)}_{g g, \SaSa,\NoW}  \,  . \numberthis
\label{eq:NC_CoeffF_Full}
\end{align*}
Note that each of the coefficient function building blocks on the right hand side of eq.~\eqref{eq:NC_CoeffF_Full} is finite and mass factorizes on its own.

A few comments on some building blocks appearing in~\eqref{eq:NC_CoeffF_Full} are in place.
While neutral current interactions do not change flavour at the vertices, it is nevertheless sensible to separate contributions according to their flavour structure in anticipation of the charged current scenario in Section~\ref{sec:CC}.
This way we can express all structure functions in terms of a minimal set of non-redundant building blocks.
As mentioned previously, the contributions labelled as ``$\NoW$'' are not present if the quark flavour is not conserved at the vector or axial vector vertex.
For this reason the $\NoW$ contributions mass factorize independently.

The contribution $C^{T,ZZ,(2)}_{qq,\NoW}$, depicted with a representative diagram in Fig.~\ref{fig:SIDIS_qA31q_vSa}, depends on the renormalization scale $\mu_R$, which is not predicted by the renormalization group equation.
This dependency originates because our theory is not anomaly free for an arbitrary number of quarks.
If an even number of active quark flavours is considered, the sum of axial charges $\sum_{q'} \left(a^k_{q'} \right) = 0$ cancels the anomaly and removes the renormalization scale dependence from the cross section.
We have nevertheless decided to quote the contributions proportional to $\sum_{q^\prime} a^k_{q'}$ since they are required in the context of variable flavour number schemes.
In this case the renormalization scale  dependence from the massless quark flavours has to be cancelled by the corresponding anomaly contribution from the massive quark triangle diagrams.

The contribution $C_{gg,\NoW}^{\{T,L\},k,2}$ shown in Fig.~\ref{fig:SIDIS_gA21axx21g_SaSa} only arises in the case of two axial vector couplings in $Z$ boson exchange, i.e.\ $k=ZZ$.
It contains the anomaly diagram twice and is finite even before renormalization and mass factorization.
Notably, it is the only non-vanishing two-loop (virtual-virtual) contribution in $\Fcal^h_L$.

The $q\to q'$ channel with the electroweak vector bosons coupling to different quark lines is decomposed non-trivially into two distinctly separate contributions $\ih_{q\qp}^k$, $\ih_{\qp q}^k$.
This is due to the different electroweak boson couplings with quarks in vector and axial vector interference contributions, and in contrast to photon exchange, where both contributions reduce to the common factor $e_q e_\qp$.

The results for photon exchange, i.e.\ $k=\gamma\gamma$, have been determined previously in~\cite{Goyal:2023zdi,Bonino:2024qbh,Goyal:2024emo}, for which we find full agreement.

For parts of the coefficient functions, where the external current couples to the
same quark line, we expect the coefficient functions to be the same between
purely vector and purely axial vector couplings, since one can na\"ively anticommute
the $\gamma_5$ matrices on the respective spin line.
This is explicitly verified in the calculation and reflected in eq.~\eqref{eq:NC_CoeffF_Full} by the
factorization of the factor $\left[ \vh_{qq}^k + \ah_{qq}^k \right]$.
For the four quantities $C^{\{T,L\},(2)}_{\qp q, 1}$, $C^{\{T,L\},(2)}_{\qp q, 2}$, $C^{\{T,L\},(2)}_{\qp q, \NoW, 3}\,$, and $C^{\{T,L\},(2)}_{\qp q, \NoW, 4}\,$
contributing to the $q\to \qp$ transition there also exist relations from swapping the identified final state quark $\qp$ with an anti-quark $\qbp$,
which are taken into account in eq.~\eqref{eq:NC_CoeffF_Full}.

\subsubsection{Antisymmetric Coefficient Functions}\label{sec:NCAntisymmCF}

In this section we present the first NNLO calculation of the $\NC$ SIDIS coefficient functions for the structure function $\Fcal_3^h$.
The $\Fcal_3^h$ structure function corresponds to the contributions arising in the interference between vector and axial vector coupling.
In $\NC$ exchange it therefore only contributes for $k=\{\gamma Z, ZZ\}$.

The LO contributions to $\Fcal_3^h$ in electroweak SIDIS are given by
\begin{align*}
\Ccal^{3,k,(0)}_{qq} &= \ih_{qq}^k C^{3,(0)}_{qq} \,  ,  \numberthis
\end{align*}
with $C^{3,(0)}_{qq} = \delta(1-\zh)\delta(1-\xh)$.
At LO we therefore have
\begin{alignat*}{2}
& \Big[ \Fcal_{3}^{h,\gamma\gamma},~ \Fcal_{3}^{h,\gamma Z},~ \Fcal_{3}^{h,ZZ} \Big] &&= \sum_{q} \Big[ 0,~ 2 \, e_q \, g_A^q ,~ 2 \, g_V^q\, g_A^q \Big] \big( q \, D_q^h - \qb \, D_{\qb}^h \big) \, . \numberthis
\end{alignat*}

At NLO we have the same channels as for the symmetric NC coefficient functions,
\begin{align*}
\Ccal^{3,k,(1)}_{qq} &=  \ih_{qq}^k C^{3,(1)}_{qq} \,  , \\
\Ccal^{3,k,(1)}_{gq} &=  \ih_{qq}^k C^{3,(1)}_{gq} \,  , \\
\Ccal^{3,k,(1)}_{qg} &=  \ih_{qq}^k C^{3,(1)}_{qg} \,  .  \numberthis
\end{align*}
The NLO coefficient functions $\Ccal^{3,(1)}_{p'p}$ above were previously computed in \cite{Furmanski:1981cw}.
Note that the gluon-induced coefficient functions $\Ccal^{3,(1)}_{qg} = -\Ccal^{3,(1)}_{\bar{q}g}$ are non-vanishing in SIDIS, but in inclusive DIS they cancel identically due to symmetry arguments.

The missing coefficient functions with initial-state anti-quarks are obtained by charge conjugation, which now introduces a sign flip,
\begin{equation}
\Ccal^{3,k,(j)}_{p'p} = - \Ccal^{3,k,(j)}_{\bar{p}'\bar{p}}\quad \mathrm{for} \quad  j \in \{0,1,2\} \,  .
\end{equation}

At NNLO, the contributions appearing in~\eqref{eq:channel_decomp} in terms of elementary building blocks read as follow:
\begin{alignat*}{2}
& C^{3,k,(2)}_{qq} &&= \ih_{qq}^k C^{3,(2)}_{qq}  \,  ,  \\
& C^{3,k,(2)}_{qq,\Fcon} && = \ih_{qq}^k C^{3,(2)}_{qq,\Fcon, 1} \,  , \\
& C^{3,k,(2)}_{qq,\NoW} &&= 2 v_q^k \sum_{q'} \left( a_{q'}^k \right) C^{3,(2)}_{qq,\vSa,\NoW} \,  , \\
& C^{3,k,(2)}_{\qb q} && = \ih_{qq}^k C^{3,(2)}_{\qb q}  \,  ,  \\
& C^{3,k,(2)}_{\qb q,\Fcon} && = \ih_{qq}^k C^{3,(2)}_{\qb q,\Fcon} \,  , \\
& C^{3,k,(2)}_{\qp q} && = \ih_{qq}^k C_{\qp q, 1}^{3,(2)} + \ih_{\qp\qp}^k \, C_{\qp q, 2}^{3,(2)}  \,  ,  \\
& C^{3,k,(2)}_{\qp q,\NoW} && = \ih_{q\qp}^k \, C_{\qp q, \NoW, 3}^{3,(2)} + \ih_{\qp q}^k C_{\qp q, \NoW, 4}^{3,(2)}  \,  ,  \\
& C^{3,k,(2)}_{\qbp q} && = \ih_{qq}^k C_{\qp q, 1}^{3,(2)} - \ih_{\qp\qp}^k \, C_{\qp q, 2}^{3,(2)}  \,  ,  \\
& C^{3,k,(2)}_{\qbp q,\NoW} && = \ih_{q\qp}^k \, C_{\qp q, \NoW, 3}^{3,(2)} - \ih_{\qp q}^k C_{\qp q, \NoW, 4}^{3,(2)}  \,  ,  \\
& C^{3,k,(2)}_{g q} && = \ih_{qq}^k C^{3,(2)}_{g q}  \,  , \\
& C^{3,k,(2)}_{g q,\NoW} && = 2 v_q^k \sum_{q'} \left( a_{q'}^k \right) C^{3,(2)}_{g q,\NoW}  \,  , \\
& C^{3,k,(2)}_{qg} && = \ih_{qq}^k C^{3,(2)}_{qg}  \,  , \\
& C^{3,k,(2)}_{qg,\NoW} && = 2 v_q^k \sum_{q'} \left( a_{q'}^k \right) C^{3,(2)}_{qg,\NoW}  \, . \numberthis \label{eq:NCF3_blocks}
\end{alignat*}
With respect to the symmetric case, eq.~\eqref{eq:NC_CoeffF_Full}, the singlet contributions $C^{3,k,(2)}_{gg}$, $C^{3,k,(2)}_{g g,\NoW}$, $C^{3,k,(2)}_{qq,\Fcon,2}$ are found to vanish like in inclusive DIS by a line reversal symmetry argument.
For the remaining singlet SIDIS coefficient functions, the same line reversal argument may not be applied as the symmetry along the singlet line is broken by the now-identified parton.

The building blocks $C_{\qp q, 1}^{3,(2)}$, $C_{\qp q, 2}^{3,(2)}$, $C_{\qp q, \NoW, 3}^{3,(2)}$, and $C_{\qp q, \NoW, 4}^{3,(2)}$ in eq.~\eqref{eq:NCF3_blocks} reappear in different coefficient functions, which are related by charge conjugation from $\qp$ to $\qbp$.

\subsection{Charged Current SIDIS}
\label{sec:CC}

We now turn to identified hadron production in $\CC$ SIDIS, describing
 the processes
\begin{align}
& \ell^- N \to \nu_\ell h X \, , &
& \ell^+ N \to \bar{\nu}_\ell h X \, ,&
& \nu_{\ell}  N \to \ell^{-} h X \,  &
& \bar{\nu}_{\ell}  N\to \ell^{+} h X\, .
\end{align}
In the following we present the coefficient functions in $\CC$ SIDIS at NNLO in QCD corresponding to the symmetric and antisymmetric structure functions.

\subsubsection{Symmetric Coefficient Functions}\label{sec:CCSymmCF}

In this section we describe the CC coefficient functions for the symmetric structure functions $\Fcal^h_T$ and $\Fcal^h_L$, up to NNLO.
In $\CC$ SIDIS the allowed quark transitions are restricted by the charge of the current entering the hadronic line.
In the following we explicitly give the contributions for $W^-$ exchange.
The contributions for $W^+$ exchange can be inferred.

We denote the set of up-type quarks with $\mathfrak{u}=\{u,c,(t)\}$ and the set of down-type quarks with $\mathfrak{d}=\{d,s,(b)\}$. At LO we have
\begin{align*}
\Ccal^{\{T,L\},W^{-},(0)}_{\beta\alpha} = 2\left|V_{\alpha\beta}\right|^2 C^{\{T,L\},(0)}_{qq} \,  ,  \\
\Ccal^{\{T,L\},W^{-},(0)}_{\bar{\alpha}\bar{\beta}} = 2\left|V_{\alpha\beta}\right|^2 C^{\{T,L\},(0)}_{qq} \,  , \numberthis
\end{align*}
where $\alpha \in \mathfrak{u}$ and  $\beta \in \mathfrak{d}$.
The factors of $2$ arise from summing the vector and axial vector contributions.
The coefficients $V_{\alpha\beta}$ denote the entries of the Cabibbo-Kobayashi-Maskawa (CKM) matrix
\begin{align}
V_\mathrm{CKM}=\begin{pmatrix}
V_{ud} & V_{us} & V_{ub}\\
V_{cd} & V_{cs} & V_{cb}\\
V_{td} & V_{ts} & V_{tb}
\end{pmatrix} \, ,
\end{align}
and describe the size of the flavour mixing.

At LO the symmetric SIDIS $\CC$ structure functions are therefore given by
\begin{alignat*}{2}
& \Big[ \Fcal_{T}^{h,W^-},~ \Fcal_{T}^{h,W^+} \Big] &&= 2 \sum_{\phantom| \alpha\in \mathfrak{u} \phantom|} \sum_{\phantom| \beta  \in \mathfrak{d} \phantom| } |V_{\alpha\beta}|^2 \Big[ f_\alpha^N \, D_\beta^h + f_{\bar{\beta}}^N \, D_{\bar{\alpha}}^h ,~ f_\beta^N \, D_\alpha^h + f_{\bar{\alpha}}^N \, D_{\bar{\beta}}^h \Big] \, , \\
& \Big[ \Fcal_{L}^{h,W^-},~ \Fcal_{L}^{h,W^+} \Big] &&= 0 \, . \numberthis
\end{alignat*}

At NLO there exist the quark-initiated contributions
\begin{align*}
\Ccal^{\{T,L\},W^{-},(1)}_{\beta \alpha} &= 2\left|V_{\alpha\beta}\right|^2 C^{\{T,L\},(1)}_{qq} \,  , \\
\Ccal^{\{T,L\},W^{-},(1)}_{g\alpha} &= 2\sum_\beta \left|V_{\alpha\beta}\right|^2 C^{\{T,L\},(1)}_{gq} \,  , \numberthis
\end{align*}
the anti-quark-initiated contributions
\begin{align*}
\Ccal^{\{T,L\},W^{-},(1)}_{\bar{\alpha}\bar{\beta}} &= C^{\{T,L\},W^{-},(1)}_{\beta\alpha}  \,  , \\
\Ccal^{\{T,L\},W^{-},(1)}_{g\bar{\beta}} & = 2 \sum_\alpha \left|V_{\alpha\beta}\right|^2 C^{\{T,L\},(1)}_{gq} \, ,\numberthis
\end{align*}
and the gluon-initiated contributions
\begin{align*}
\Ccal^{\{T,L\},W^{-},(1)}_{\beta g} &=  2\sum_\alpha \left|V_{\alpha\beta}\right|^2 C^{\{T,L\},(1)}_{qg} \,  , \\
\Ccal^{\{T,L\},W^{-},(1)}_{\bar{\alpha}g}  & =  2\sum_\beta \left|V_{\alpha\beta}\right|^2 C^{\{T,L\},(1)}_{qg} \, . \numberthis
\end{align*}

At NNLO we have the quark-initiated contributions
\begin{align*}
C^{\{T,L\},W^-,(2)}_{\beta\alpha} & =2\left|V_{\alpha\beta} \right|^2 C^{\{T,L\},(2)}_{qq}  \,  , \\
C^{\{T,L\},W^-,(2)}_{qq,\Fcon} &= 2\sum_{\alpha,\beta}  \left|V_{\alpha\beta} \right|^2  C^{\{T,L\},(2)}_{qq, \Fcon,2} \,  , \\
C^{\{T,L\},W^-,(2)}_{\alpha\alpha,\Fcon} &= 2\sum_\beta \left|V_{\alpha\beta} \right|^2 C^{\{T,L\},(2)}_{qq, \Fcon,1}   \,  , \\
C^{\{T,L\},W^-,(2)}_{\bar{\beta} \alpha} & = 2\left|V_{\alpha\beta} \right|^2 C^{\{T,L\},(2)}_{\qb q}  \,  , \\
C^{\{T,L\},W^-,(2)}_{\bar{\alpha} \alpha, \Fcon} &= 2\sum_\beta \left|V_{\alpha\beta} \right|^2 C^{\{T,L\},(2)}_{\qb q, \Fcon}  \,  , \\
C^{\{T,L\},W^-,(2)}_{\qp \alpha} & = 2\sum_\beta \left|V_{\alpha\beta}\right|^2 C^{\{T,L\},(2)}_{\qp q, 1}  \,  , \\
C^{\{T,L\},W^-,(2)}_{\beta' q} & =  2\sum_\alpha \left|V_{\alpha\beta}\right|^2 C^{\{T,L\},(2)}_{\qp q, 2} \,  , \\
C^{\{T,L\},W^-,(2)}_{\qbp \alpha} & = 2\sum_\beta \left|V_{\alpha\beta}\right|^2  C^{\{T,L\},(2)}_{\qp q, 1}  \,  , \\
C^{\{T,L\},W^-,(2)}_{\bar{\alpha}' q} & = 2\sum_\beta \left|V_{\alpha\beta}\right|^2 C^{\{T,L\},(2)}_{\qp q, 2} \,  , \\
C^{\{T,L\},W^-,(2)}_{g\alpha} &=  2\sum_\beta \left|V_{\alpha\beta}\right|^2 C^{\{T,L\},(2)}_{gq} \, , \numberthis
\end{align*}
with $q\in\mathfrak{u}\cup\mathfrak{d}$ denoting a quark of any active flavour; the anti-quark-initiated contributions
 \begin{align*}
C^{\{T,L\},W^-,(2)}_{\bar{\alpha}\bar{\beta}} & = C^{\{T,L\},W^-,(2)}_{\beta\alpha}  \,  , \\
C^{\{T,L\},W^-,(2)}_{\bar{q}\bar{q},\Fcon} &= 2\sum_{\alpha,\beta}  \left|V_{\alpha\beta} \right|^2  C^{\{T,L\},(2)}_{qq, \Fcon, 2}   \,  , \\
C^{\{T,L\},W^-,(2)}_{\bar{\beta}\bar{\beta},\Fcon} &= 2\sum_\alpha \left|V_{\alpha\beta} \right|^2 C^{\{T,L\},(2)}_{qq, \Fcon, 1}  \,  , \\
C^{\{T,L\},W^-,(2)}_{\alpha\bar{\beta}} & = C^{\{T,L\},W^-,(2)}_{\bar{\beta} \alpha} \,  , \\
C^{\{T,L\},W^-,(2)}_{\beta\bar{\beta}, \Fcon} &= 2\sum_\alpha \left|V_{\alpha\beta} \right|^2 C^{\{T,L\},(2)}_{\qb q, \Fcon}  \,  , \\
C^{\{T,L\},W^-,(2)}_{\qp \bar{\beta}} & = 2\sum_\alpha \left|V_{\alpha\beta}\right|^2 C^{\{T,L\},(2)}_{\qp q, 1}  \,  , \\
C^{\{T,L\},W^-,(2)}_{\beta \bar{q}} & =  2\sum_\alpha \left|V_{\alpha\beta}\right|^2 C^{\{T,L\},(2)}_{\qp q, 2} \,  , \\
C^{\{T,L\},W^-,(2)}_{\qbp \bar{\beta}} & = 2\sum_\alpha \left|V_{\alpha\beta}\right|^2  C^{\{T,L\},(2)}_{\qp q, 1}  \,  , \\
C^{\{T,L\},W^-,(2)}_{\bar{\alpha}' \bar{q}} & = 2\sum_\beta \left|V_{\alpha\beta}\right|^2 C^{\{T,L\},(2)}_{\qp q, 2} \,  , \\
C^{\{T,L\},W^-,(2)}_{g\bar{\beta}} &=  2\sum_\alpha \left|V_{\alpha\beta}\right|^2 C^{\{T,L\},(2)}_{gq} \,  , \numberthis
\end{align*}
and the gluon-initiated contributions
\begin{align*}
C^{\{T,L\},W^-,(2)}_{\beta g} &= 2\sum_\alpha \left|V_{\alpha\beta}\right|^2 C^{\{T,L\},(2)}_{qg}  \, ,  \\
C^{\{T,L\},W^-,(2)}_{\bar{\alpha}g} &= 2\sum_\beta \left|V_{\alpha\beta}\right|^2 C^{\{T,L\},(2)}_{qg} \, ,  \\
C^{\{T,L\},W^-,(2)}_{gg} &= 2\sum_{\alpha,\beta} \left|V_{\alpha\beta}\right|^2 C^{\{T,L\},(2)}_{gg} \, . \numberthis
\end{align*}
In order to obtain a coefficient function from the channel decomposition, all the above contributions which admit the initial and final state particle need to be summed.

The coefficient functions $\Ccal^{\{T,L\},W^+,(j)}_{p'p}$ describing the exchange of a $W^+$ boson are obtained for $\alpha\in \mathfrak{d}$ and $\beta\in \mathfrak{u}$.

\subsubsection{Antisymmetric Coefficient Functions}\label{sec:CCAntisymmCF}
At LO the antisymmetric CC coefficient functions are
\begin{align*}
\Ccal^{3,W^{-},(0)}_{\beta\alpha} &= 2 \left|V_{\alpha\beta}\right|^2 C^{3,(0)}_{qq} \,  ,  \\
\Ccal^{3,W^{-},(0)}_{\bar{\alpha}\bar{\beta}} &=- 2 \left|V_{\alpha\beta}\right|^2 C^{3,(0)}_{qq} \,  , \numberthis
\end{align*}
where $\alpha\in \mathfrak{u}$ and  $\beta\in \mathfrak{d}$.
The factors of $2$ arise from our choice of normalization of $C^{3,(0)}_{qq} = \delta(1-\xh) \delta(1-\zh)$, like for NC.

The leading order $\Fcal_3^h$ structure function is therefore given by
\begin{alignat*}{3}
& \Big[ \Fcal_{3}^{h,W^-},~ \Fcal_{3}^{h,W^+} \Big] &&= 2 \sum_{\phantom| \alpha\in \mathfrak{u} \phantom|} \sum_{\phantom| \beta\in \mathfrak{d} \phantom| } |V_{\alpha\beta}|^2 \Big[ f_\alpha^N \, D_\beta^h - f_{\bar{\beta}}^N \, D_{\bar{\alpha}}^h ,~ f_\beta^N \, D_\alpha^h - f_{\bar{\alpha}}^N \, D_{\bar{\beta}}^h \Big] \, . \numberthis
\end{alignat*}

At NLO the quark-initiated coefficient functions are given by
\begin{align*}
\Ccal^{3,W^{-},(1)}_{\beta\alpha} &= 2 \left|V_{\alpha\beta}\right|^2 C^{3,(1)}_{qq} \,  , \\
\Ccal^{3,W^{-},(1)}_{g\alpha} &= 2 \sum_\beta \left|V_{\alpha\beta}\right|^2 C^{3,(1)}_{gq}  \numberthis \,  ,
\end{align*}
the anti-quark-initiated coefficient functions by
\begin{align*}
\Ccal^{3,W^{-},(1)}_{\bar{\alpha}\bar{\beta}} &= -C^{3,W^{-},(1)}_{\beta\alpha}  \,  , \\
\Ccal^{3,W^{-},(1)}_{g\bar{\beta}} & = -2\sum_\alpha \left|V_{\alpha\beta}\right|^2 C^{3,(1)}_{gq} \, , \numberthis
\end{align*}
and the gluon-initiated coefficient functions by
\begin{align*}
\Ccal^{3,W^{-},(1)}_{\beta g} &= 2 \sum_\alpha \left|V_{\alpha\beta}\right|^2 C^{3,(1)}_{qg} \,  , \\
\Ccal^{3,W^{-},(1)}_{\bar{\alpha}g}  & =  - 2 \sum_\beta \left|V_{\alpha\beta}\right|^2 C^{3,(1)}_{qg} \numberthis \, .
\end{align*}

At NNLO we have the quark-initiated contributions
\begin{align*}
C^{3,W^-,(2)}_{\beta\alpha} & = 2 \left|V_{\alpha\beta} \right|^2 C^{3,(2)}_{qq}  \,  , \\
C^{3,W^-,(2)}_{\alpha\alpha,\Fcon} &= 2 \sum_\beta \left|V_{\alpha\beta} \right|^2 C^{3,(2)}_{qq, \Fcon, 1}  \,  , \\
C^{3,W^-,(2)}_{\bar{\beta} \alpha} & = 2 \left|V_{\alpha\beta} \right|^2 C^{\{T,L\},(2)}_{\qb q}  \,  , \\
C^{3,W^-,(2)}_{\bar{\alpha} \alpha, \Fcon} &= 2 \sum_\beta \left|V_{\alpha\beta} \right|^2 C^{3,(2)}_{\qb q, \Fcon}  \,  , \\
C^{3,W^-,(2)}_{\qp \alpha} & = 2 \sum_\beta \left|V_{\alpha\beta}\right|^2 C^{3,(2)}_{\qp q, 1}  \,  , \\
C^{3,W^-,(2)}_{\beta' q} & = 2 \sum_\alpha \left|V_{\alpha\beta}\right|^2 C^{3,(2)}_{\qp q, 2} \,  , \\
C^{3,W^-,(2)}_{\qbp \alpha} & = 2 \sum_\beta \left|V_{\alpha\beta}\right|^2  C^{3,(2)}_{\qp q, 1}  \,  , \\
C^{3,W^-,(2)}_{\bar{\alpha}' q} & = -2 \sum_\beta \left|V_{\alpha\beta}\right|^2 C^{3,(2)}_{\qp q, 2} \,  , \\
C^{3,W^-,(2)}_{g\alpha} &= 2 \sum_\beta \left|V_{\alpha\beta}\right|^2 C^{3,(2)}_{gq} \, , \numberthis
\end{align*}
with $q\in\mathfrak{u}\cup\mathfrak{d}$; the anti-quark-initiated contributions are given by
 \begin{align*}
C^{3,W^-,(2)}_{\bar{\alpha}\bar{\beta}} & = -C^{3,W^-,(2)}_{\beta\alpha}  \,  , \\
C^{3,W^-,(2)}_{\bar{\beta}\bar{\beta},\Fcon} &= -2 \sum_\alpha \left|V_{\alpha\beta} \right|^2 C^{3,(2)}_{qq, \Fcon, 1} \,  , \\
C^{3,W^-,(2)}_{\bar{\alpha}\beta} & = -C^{3,W^-,(2)}_{\bar{\beta} \alpha} \,  , \\
C^{3,W^-,(2)}_{\beta\bar{\beta}, \Fcon} &=-2 \sum_\alpha \left|V_{\alpha\beta} \right|^2 C^{3,(2)}_{\qb q, \Fcon}  \,  , \\
C^{3,W^-,(2)}_{\qp \bar{\beta}} & = -2 \sum_\alpha \left|V_{\alpha\beta}\right|^2 C^{3,(2)}_{\qp q, 1}  \,  , \\
C^{3,W^-,(2)}_{\beta \bar{q}} & =  2 \sum_\alpha \left|V_{\alpha\beta}\right|^2 C^{3,(2)}_{\qp q, 2} \,  , \\
C^{3,W^-,(2)}_{\qbp \bar{\beta}} & = -2 \sum_\alpha \left|V_{\alpha\beta}\right|^2  C^{3,(2)}_{\qp q, 1}  \,  , \\
C^{3,W^-,(2)}_{\bar{\alpha}' \bar{q}} & = -2 \sum_\beta \left|V_{\alpha\beta}\right|^2 C^{3,(2)}_{\qp q,2} \,  , \\
C^{3,W^-,(2)}_{g\bar{\beta}} &=  -2 \sum_\alpha \left|V_{\alpha\beta}\right|^2 C^{3,(2)}_{gq}  \numberthis \, ,
\end{align*}
and the gluon initiated contributions read
\begin{align*}
C^{3,W^-,(2)}_{\beta g} &= 2\sum_\alpha \left|V_{\alpha\beta}\right|^2 C^{3,(2)}_{qg} \, ,  \\
C^{3,W^-,(2)}_{\bar{\alpha}g} &= -2 \sum_\beta \left|V_{\alpha\beta}\right|^2 C^{3,(2)}_{qg} \, , \\
C^{3,W^-,(2)}_{gg} &=0 \,  . \numberthis
\end{align*}
For $W^+$-exchange one instead has to consider $\alpha\in \mathfrak{d}$ and $\beta\in \mathfrak{u}$.

\section{Analytical Calculation of the Coefficient Functions}
\label{sec:calc}
In order to calculate the NNLO corrections to neutral- and charged-current SIDIS we use our established toolchain \cite{Bonino:2024qbh,Bonino:2024wgg}.
The diagrams are generated with \texttt{QGRAF}~\cite{Nogueira:1991ex}, and their
algebraic evaluation is then performed in
 \texttt{FORM}~\cite{Vermaseren:2000nd} and \texttt{Mathematica}.

We apply projectors to obtain the structure functions $\mathcal{F}_T^h$, $\mathcal{F}_L^h$, and $\mathcal{F}_3^h$ from Eq.~\eqref{eq:had_tens}.
In our coefficient functions we encounter up to two $\gamma_5$ matrices from the axial vector vertex couplings
at the weak gauge boson vertices. We use the Larin scheme~\cite{Larin:1993tq} to
treat $\gamma_5$ in dimensional regularisation, which consists in replacing
\begin{align}
\gamma_\mu \gamma_5 = \frac{i}{3!} \varepsilon_{\mu\nu\rho\sigma} \gamma^\nu \gamma^\rho \gamma^\sigma,
\label{eq:Larin_g5}
\end{align}
and contracting the two resulting  Levi-Civita tensors into $d$-dimensional scalar products,
\begin{align}
\varepsilon_{\mu\nu\rho\sigma} \, \varepsilon_{\alpha\beta\gamma\delta} = \operatorname{det}
\begin{pmatrix}
g_{\mu \alpha} & g_{\mu \beta} & g_{\mu \gamma} & g_{\mu \delta} \\
g_{\nu \alpha} & g_{\nu \beta} & g_{\nu \gamma} & g_{\nu \delta} \\
g_{\rho \alpha} & g_{\rho \beta} & g_{\rho \gamma} & g_{\rho \delta} \\
g_{\sigma \alpha} & g_{\sigma \beta} & g_{\sigma \gamma} & g_{\sigma \delta}
\end{pmatrix}.
\label{eq:Larin_epstens}
\end{align}

At NNLO, there are three classes of contributing
subprocesses: the real-real (RR) contributions, describing the additional radiation of two partons; the real-virtual (RV) contributions, describing the radiation of a single parton with an additional loop; and the virtual-virtual contributions with no extra radiation, but two additional loops, all with respect to the leading order.

The calculation of the tree-level RR matrix elements is straightforward.
Using \texttt{REDUZE}~\cite{vonManteuffel:2012np}, the loop-integrals in the RV and VV contributions are mapped onto integral families and reduced to master integrals with integration-by-parts identities~\cite{Chetyrkin:1981qh,Laporta:2000dsw}.
For the RV we reduce the loop integrals to the 1-loop bubble and box integral~\cite{Gehrmann:1999as}.
The integrals appearing in the VV are reduced to the four master integrals in~\cite{Gehrmann:2005pd}.

To obtain the coefficient functions, the contributions need to be integrated over the corresponding phase spaces.
For the RR, we employ reverse unitarity \cite{Anastasiou:2002yz} to map the integrals onto 14 integral families containing a set of 8 independent propagators, 3 constraints from reverse unitarity, and the SIDIS phase space constraint introducing the fragmentation variable $z$, as described in~\cite{Bonino:2024adk}.
The phase space integrals are then reduced to a set of 21 known master
integrals~\cite{Bonino:2024adk,Ahmed:2024owh}.

For the RV, the phase space integration is fully constrained by $x$ and $z$, such that only an
expansion of the unintegrated RV contributions in distributions in $(1-x)$ and $(1-z)$ needs to be performed.
These expansions are described in detail in~\cite{Gehrmann:2022cih}, and
must be performed in the appropriate kinematical regions requiring analytical continuations of
the one-loop bubble and box master integrals~\cite{Gehrmann:2022cih,Haug:2022hkr}.

The resulting bare coefficient functions
 are first expressed in terms of harmonic polylogarithms (HPLs) \cite{Remiddi:1999ew,Gehrmann:2000zt} using the Mathematica packages \texttt{HPL} \cite{Maitre:2005uu} and \texttt{PolyLogTools}~\cite{Duhr:2019tlz}, and subsequently converted into logarithms and polylogarithms.

The ultraviolet poles are removed by renormalizing the QCD coupling constant in the usual manner, and applying the divergent renormalization on the axial vector current \cite{Larin:1993tq}.
Both renormalizations are multiplicative and independent of the kinematics, and yield the UV-finite coefficient functions in the Larin scheme.
We then apply mass factorization to cancel the IR poles associated with PDFs and fragmentation functions.
In a final step, we apply the multiplicative finite renormalization of the axial current \cite{Larin:1993tq} to restore the Ward identities, and obtain the $\MSbar$ SIDIS coefficient functions.

Our results are expressed with the full scale dependence on $(\mu_A,\mu_F,\mu_R)$.
They are distributed inside an ancillary file with the \texttt{arXiv} submission.
Our notation inside the ancillary file is explained in Appendix~\ref{app:ancillary}.

To validate our results we have carried out the following checks:
Our results fulfill the scale dependence predicted by the renormalization group equations in $(\mu_F, \mu_A, \mu_R)$.
We have cross-checked our results for $\mathcal{F}_T^h$ and $\mathcal{F}_L^h$ against previous results for photon exchange~\cite{Goyal:2023zdi,Bonino:2024qbh,Goyal:2024emo}.
We find full agreement for the flavour decomposed vector coefficient functions, and agreement for the axial vector coefficient functions if the axial vector couplings are attached to a common quark line.
An equality for the axial vector couplings attaching to different quark lines is not required by theory, and also not realized.

\section{Phenomenological Results}
\label{sec:num}

\begin{figure}[tb]
\centering
\includegraphics[width=0.49\linewidth]{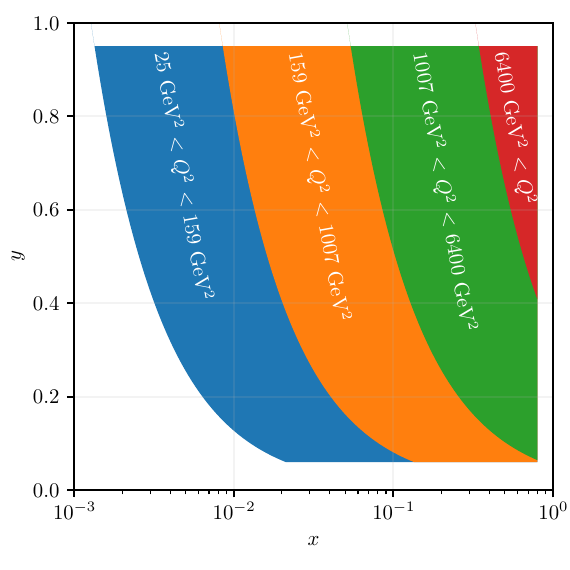}\,
\includegraphics[width=0.49\linewidth]{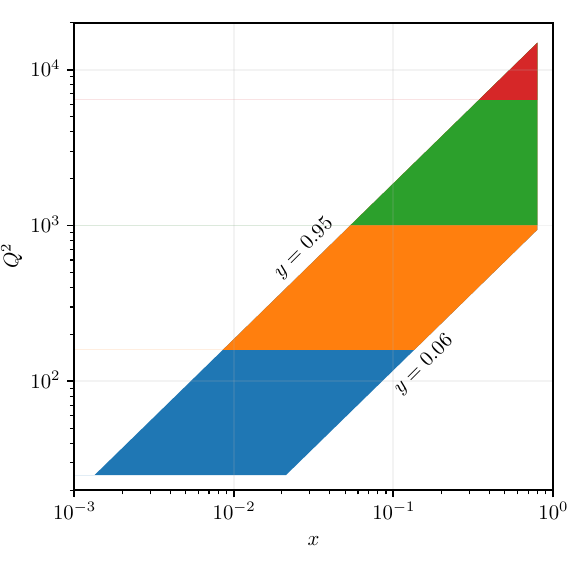}
\caption{Fiducial phase space covered by the four $Q^2$ bins of~\eqref{eq:cutsQ2} with kinematic cuts~\eqref{eq:cutsxyz} in the $(x,y)$ plane (left) or $(x,Q^2)$ plane (right).}
\label{fig:ps}
\end{figure}

In the following we study the phenomenological impact of the electroweak effects in neutral and charged current SIDIS at the BNL~EIC.
We consider electron-initiated $\pi^{\pm}$ production at a centre-of-mass energy $\sqrt{s} = 140 \, \,\mathrm{GeV}$.
We impose
\begin{align}\label{eq:cutsxyz}
& 0.06 < y < 0.95 \, , && x < 0.8 \, , && 0.2 < z < 0.85
\end{align}
as kinematic cuts on the SIDIS variables, restricting ourselves to vector boson virtualities $Q^2 > 25\,\GeV^2$ divided in four logarithmically equispaced bins, which we denote as
\begin{alignat}{7}\label{eq:cutsQ2}
&\text{Low-}Q^2:~& \,  25\,\GeV^2 &< Q^2 < &~ 159 \,\GeV^2 & \, ,
&&\quad &&
\text{Mid-}Q^2:~& \, 159 \,\GeV^2 &< Q^2 < &~ 1007\,\GeV^2 & \, ,
\nonumber \\
&\text{High-}Q^2:~& \,1007\,\GeV^2 &< Q^2 < &~ 6400\,\GeV^2 & \, ,
&&\quad &&
\text{Extreme-}Q^2:~& \, 6400\,\GeV^2 &< Q^2 \, . \hspace{-.2em}
\end{alignat}
The kinematic range of these four bins in the $(x,y)$ and $(x,Q^2)$ planes is shown in Fig.~\ref{fig:ps}.
The variable $z$ is not constrained by the virtuality.

We employ the FF set \code{BDSSV22}~\cite{Borsa:2022vvp} and the PDF set \code{NNPDF40}~\cite{NNPDF:2021njg} at NNLO throughout.
Interpolation grids for \code{NNPDF40\_nnlo\_as\_01180} and the evolution of $\alpha_s$ from the corresponding set are provided via \code{LHAPDF}~\cite{Buckley:2014ana}.
We evaluate the cross sections for $\nf=5$ massless flavours, consistent with the adopted range in $Q^2$.
Uncertainty bands for cross sections are obtained from 7-point scale variation of $(\mu_R^2, \mu_F^2=\mu_A^2)$ around the central scale $Q^2$, discarding extreme variations.
Additional uncertainties stemming from the PDFs and FFs are not taken into account.
We provide results for the fiducial SIDIS cross section differential in $Q^2$, $x$ or $z$.
All quantities displayed in plots are integrated over the respective bin.
Numerical fluctuations visible in plots of cross section ratios are due to Monte Carlo integration uncertainties.

As electroweak parameters, we adopt $G_F = 1.166379\times 10^{-5}~\mathrm{GeV}^{-2}$, $M_W = 80.369 \, \mathrm{GeV}$, $M_Z = 91.188~\mathrm{GeV}$, from which we derive $\alpha \simeq 1/132.10$ and $\sin^2 \theta_\mathrm{W} \simeq 0.2232$ with EW tree-level relations.
In the predictions for the charged current cross section, we keep the full CKM matrix, with numerical values for the CKM entries equal to~\cite{ParticleDataGroup:2024cfk}:
\begin{align}
  |V_{ud}| &= 0.97435\,, & |V_{us}| &= 0.22500\,, & |V_{ub}| &= 0.00369\,, \nonumber \\
  |V_{cd}| &= 0.22486\,, & |V_{cs}| &= 0.97349\,, & |V_{cb}| &= 0.04182\,.
\end{align}

\subsection{Neutral Current}

We first present predictions for the helicity-averaged cross section $\dd \sigma^{h,\NC} (\lambda_\ell = 0)$ corresponding to the $\lambda_\ell$-even pieces in eq.~\eqref{eq:xsNCCC}.
To this end we define
\begin{align}
\dd \sigma^{h,\NC}(\lambda_\ell=\lambda) = \frac{1+\lambda}{2} \, \dd \sigma^{h,\NC}(\lambda_\ell=+1) + \frac{1-\lambda}{2} \, \dd \sigma^{h,\NC}(\lambda_\ell=-1)
\end{align}
for the lepton beam polarisation $-1\leq \lambda \leq 1$.
We additionally provide predictions for the structure function decomposition of the cross section,
\begin{align}\label{eq:xsNCCC_dec}
  \dd \sigma^{h,\NC} = \dd \sigma^{h,\gamma\gamma, T} + \dd \sigma^{h,\gamma\gamma, L}
  +\dd \sigma^{h,(\gamma Z+ZZ), T} + \dd \sigma^{h,(\gamma Z+ZZ), L} + \dd \sigma^{h,(\gamma Z+ZZ), 3} \,.
\end{align}
Here we have introduced
\begin{equation}
  \dd \sigma^{h,(\gamma Z+ZZ), i}=\dd \sigma^{h,\gamma Z, i} + \dd \sigma^{h,ZZ, i} \,,  \quad i = T,L,3
\end{equation}
to isolate effects of electroweak origin.
Moreover, instead of presenting results for the NC cross section with definite lepton helicity states ($\lambda_\ell=\pm 1$),
we show results for the \emph{neutral current asymmetry} defined as
\begin{align}\label{eq:A_NC}
A^h_\NC = \frac{\dd \sigma^{h,\NC} (\lambda_{\ell}=1) - \dd \sigma^{h,\NC} (\lambda_{\ell}=-1)}{\dd \sigma^{h,\NC} (\lambda_{\ell}=1) + \dd \sigma^{h,\NC} (\lambda_{\ell}=-1)} \, .
\end{align}
This observable quantifies the $\lambda_\ell\,$-odd component in eq.~\eqref{eq:xsNCCC} for a helicity eigenstate $\lambda_\ell=\pm 1$.
The cross sections entering eq.~\eqref{eq:A_NC} are subjected to the same kinematic cuts of eq.~\eqref{eq:cutsxyz} and \eqref{eq:cutsQ2}.
By constructing \eqref{eq:A_NC} from cross sections single-differential in $Q^2$, $x$ or $z$, we obtain the asymmetry as
function of either of these variables.
Theory uncertainties are obtained with a 31-point uncorrelated scale variation between numerator and denominator.
Note that the cross section for any lepton beam polarisation $-1 \leq \lambda \leq 1$ is then obtained by
\begin{align}
\dd \sigma^{h,\NC} (\lambda_\ell = \lambda) = \big( 1 + \lambda \, A^h_\NC \big) \cdot \dd \sigma^{h,\NC} (\lambda_\ell =0) \, .
\end{align}

For certain observables the cross section can be reconstructed with reasonable accuracy from the knowledge of the leading order ratio $\dd \sigma^{h,\NC,(0)}/\dd \sigma^{h,\gamma\gamma,(0)}$ and the $n^\text{th}$ order photon exchange cross section.
For these observables we will display the ratio
\begin{align*}
  R^{h, (n)} = \dd \sigma^{h,\NC,(n)} \Bigg/ \left( {\dd \sigma^{h,\gamma\gamma,(n)}}
  \frac{\dd \sigma^{h,\NC,(0)}}{\dd \sigma^{h,\gamma\gamma,(0)}} \right)
  = \frac{\dd \sigma^{h,\NC,(n)}}{\dd \sigma^{h,\gamma\gamma,(n)}} \frac{\dd \sigma^{h,\gamma\gamma,(0)}}{\dd \sigma^{h,\NC,(0)}} \, , \numberthis
\end{align*}
describing the accuracy of the prediction of the total cross section based on dressing the $n^\text{th}$ order photon exchange cross section with the ratio between the LO $\NC$ and the LO photon exchange cross sections.
The error bands for $R^{h,(n)}$ are determined by a correlated $7$-point scale variation of the order-dependent ratio ${\dd \sigma^{h,\NC,(n)}} / {\dd \sigma^{h,\gamma\gamma,(n)}}$.

\begin{figure}[tb]
\centering
\includegraphics[width=\linewidth]{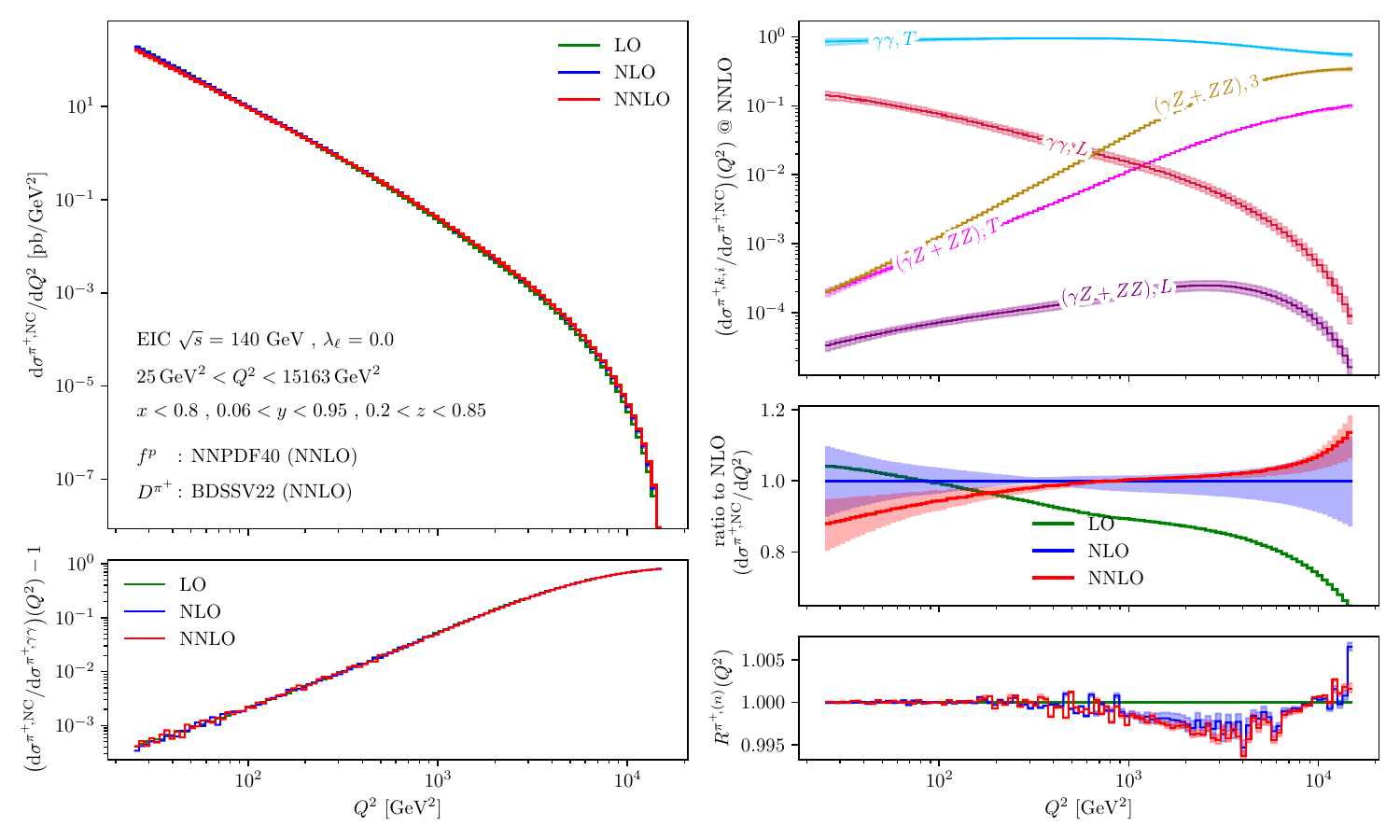}
\caption{
$\NC$ $Q^2$-distribution for $\lambda_\ell$-averaged $\pi^+$ production.
Top left panel: total $\NC$ cross section.
Bottom left panel: ratio $(\dd\sigma^{\pi^+,\NC}/\dd\sigma^{\pi^+,\gamma\gamma})(Q^2)-1$.
Top right panel: structure function contributions to the total cross section.
Centre right panel: $\NC$ Ratio to NLO.
Bottom right panel: $R^{\pi^+,(n)} (Q^2)$.
}
\label{fig:Q2_0.0_NC_pip}
\end{figure}
\begin{figure}[tb]
\centering
\includegraphics[width=\linewidth]{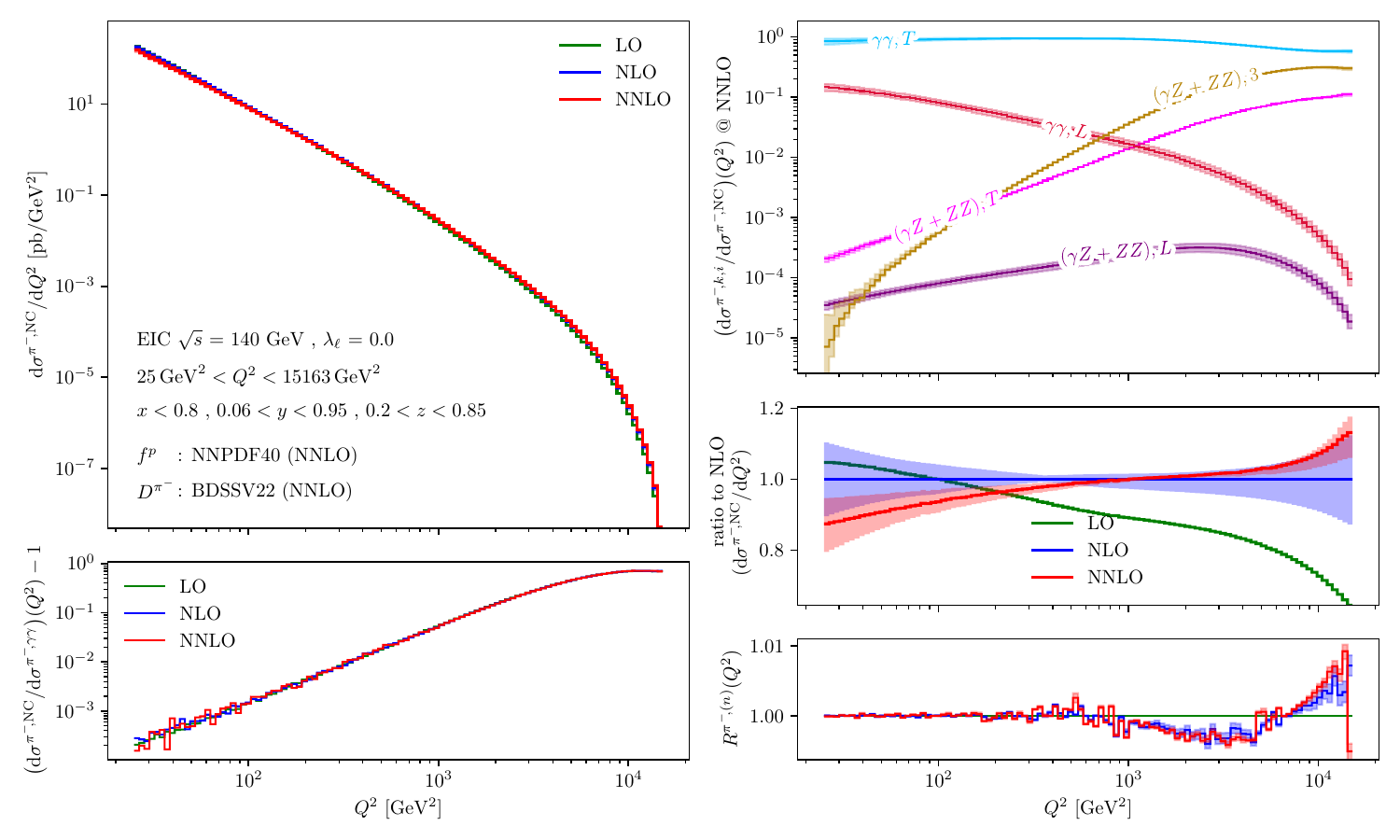}
\caption{
$\NC$ $Q^2$-distribution for helicity-averaged $\pi^-$ production.
Top left panel: total $\NC$ cross section.
Bottom left panel: ratio $(\dd\sigma^{\pi^-,\NC}/\dd\sigma^{\pi^-,\gamma\gamma})(Q^2)-1$.
Top right panel: structure function contributions to the total cross section.
Centre right panel: $\NC$ Ratio to NLO.
Bottom right panel: $R^{\pi^-,(n)}(Q^2)$.
}
\label{fig:Q2_0.0_NC_pim}
\end{figure}

\subsubsection[\texorpdfstring{$Q^2$-distributions}{\$Q\^2\$-distributions}]{\boldmath $Q^2$-distributions}

In this section we discuss the impact of electroweak effects in $\NC$ SIDIS in different energetic regimes.
Fig.~\ref{fig:Q2_0.0_NC_pip} shows the $Q^2$-distribution for the helicity-averaged cross section $\dd \sigma^{\pi^+,\NC}/\dd Q^2\, (\lambda_\ell = 0)$.
In the top left panel, the $\NC$ cross section comprised of $\gamma\gamma$, $\gamma Z$ and $ZZ$ exchanges is shown in the entire allowed kinematic range.
In the considered kinematic regime the $\NC$ cross section largely follows the photon-exchange cross section.
For virtualities above $6000\, \GeV^2$ the kinematical constraints, in particular on the allowed range in $x$ lead to a sharply decreasing cross section due to the depletion of PDFs at high-$x$.
Electroweak effects become visible in the ratio $(\dd\sigma^{\pi^+,\NC}/\dd\sigma^{\pi^+,\gamma\gamma}) (Q^2) -1$,
plotted in the bottom left panel of Fig.~\ref{fig:Q2_0.0_NC_pip}.
Their magnitude is approximately determined by $\eta_{\gamma Z}$ and $\eta_{ZZ}$:
as a consequence they are suppressed at low $Q^2$, but become relevant at high $Q^2$.
For $Q^2 < 10^2\,\GeV^2$ they are almost negligible, for $10^2\,\GeV^2< Q^2 < 10^3\,\GeV^2$ they constitute a few percent effect, above $10^3\,\GeV^2$ they increase monotonically to about 60\% at $Q^2 \sim 10^4\,\GeV^2$ of the photon-exchange contributions.
An important message emerging from the bottom left panel of Fig.~\ref{fig:Q2_0.0_NC_pip} is that
the ratio $(\dd\sigma^{\pi^+,\mathrm{NC}}/\dd\sigma^{\pi^+,\gamma\gamma})(Q^2)$ is almost insensitive to the perturbative order.
This is quantified in the bottom right panel, which displays the ratio $R^{\pi^+,(n)}(Q^2)$.
It displays that predictions of the total cross section and its uncertainties based on dressing the $n^\text{th}$ order photon exchange cross section with the ratio between the LO $\NC$ and the LO photon exchange cross sections are accurate at sub-percent-level
for all values of $Q^2$.

The top right panel of Fig.~\ref{fig:Q2_0.0_NC_pip} gives insights on the size of individual structure function contributions to the total NNLO cross section according to the decomposition in eq.~\eqref{eq:xsNCCC_dec}.
The dominant contribution to the cross section throughout the studied range of $Q^2$ is given by $\dd\sigma^{\pi^+,\gamma\gamma,T}/\dd Q^2$.
Especially at lower $Q^2$ longitudinally polarized photons provide a significant fraction of about $10\,\%$ at $Q^2\sim 25\,\GeV^2$ to the total cross section.
The longitudinal contribution rapidly declines to percent level and below starting from $Q^2 > 600\,\GeV^2$.
While for low $Q^2$ the electroweak effects in $\Fcal^{\pi^+}_T$ and $\Fcal^{\pi^+}_3$, denoted as $(\gamma Z + ZZ)$, are of similarly small size, the $\Fcal^{\pi^+}_3$ contribution decreases only moderately, becoming larger than $\Fcal^{\pi^+}_L$ at around $Q^2 \sim 600\, \GeV^2$.
It eventually approaches the magnitude of the photon contributions in $\Fcal^{\pi^+}_T$ at high $Q^2>5000\,\GeV^2$.
The electroweak effects from $\Fcal^{\pi^+}_T$ also decrease more slowly than the photon contribution, but faster than those of $\Fcal^{\pi^+}_3$.
Electroweak effects in $\Fcal^{\pi^+}_L$ are negligible in the kinematic region where $\Fcal^{\pi^+}_L$ is relevant.
As a consequence, $\Fcal^{\pi^+}_3$ is the primary origin of electroweak effects.

Finally, in the centre right panel of Fig.~\ref{fig:Q2_0.0_NC_pip} we show the size of the $Q^2$-differential cross section $\dd\sigma^{\pi^+,\NC,(n)}/\dd Q^2$ at different orders in perturbation theory relative to the NLO prediction.
The NNLO enhances the trend of the NLO corrections with respect to the LO, decreasing the cross section at low $Q^2$ and increasing it at large $Q^2$.
While the scale uncertainty bands of the NLO and NNLO corrections overlap in the entire $Q^2$ range, a slower perturbative convergence is seen in the low $Q^2$ regime.
The convergence improves significantly for $Q^2 \gtrsim 100\,\GeV^2$, with the NNLO predictions well within the NLO uncertainty bands and with significantly smaller scale variation uncertainties.
In particular in the range between $100\,\GeV^2$ and $4000\,\GeV^2$ the predictions remain very stable under higher order corrections, with small NNLO corrections.
For $Q^2 \gtrsim 4000\,\GeV^2$, NNLO corrections become larger, but remain within the NLO envelope.

The $Q^2$-differential total cross section for $\pi^-$ production in Fig.~\ref{fig:Q2_0.0_NC_pim} is very similar to $\pi^+$ production in size and shape.
A difference is visible for $Q^2 \lesssim 70\,\GeV^2$ in the structure function decomposition in the top right panel, where the cross section contribution $\dd \sigma^{\pi^-,(\gamma Z + ZZ),3}/\dd Q^2$ falls off steeply and is associated with a rather large scale variation uncertainty.
In this Low-$Q^2$ region the SIDIS differential cross section becomes highly sensitive to unfavoured $\pi^-$ fragmentation functions especially at Low-$Q^2$, cf.\ Fig.~\ref{fig:z_NC_channel_pim} below.
These unfavoured fragmentation functions are only poorly constrained from other experimental observables.
In $\pi^-$ production these unfavoured channels become relevant for two reasons.
One reason is the relative suppression of the flavour-favoured $d\to d$ channel, which is valence content of both proton and $\pi^-$.
This suppression is due to the electromagnetic charge $e_d^2=\frac{1}{4}e_u^2$ and the smaller valence distribution of the $d$.
The second reason is the range of $x$ probed at Low-$Q^2$, in which the proton's gluon and sea quark distributions grow sizeable.

\begin{figure}[tb]
\centering
\includegraphics[width=0.6\linewidth]{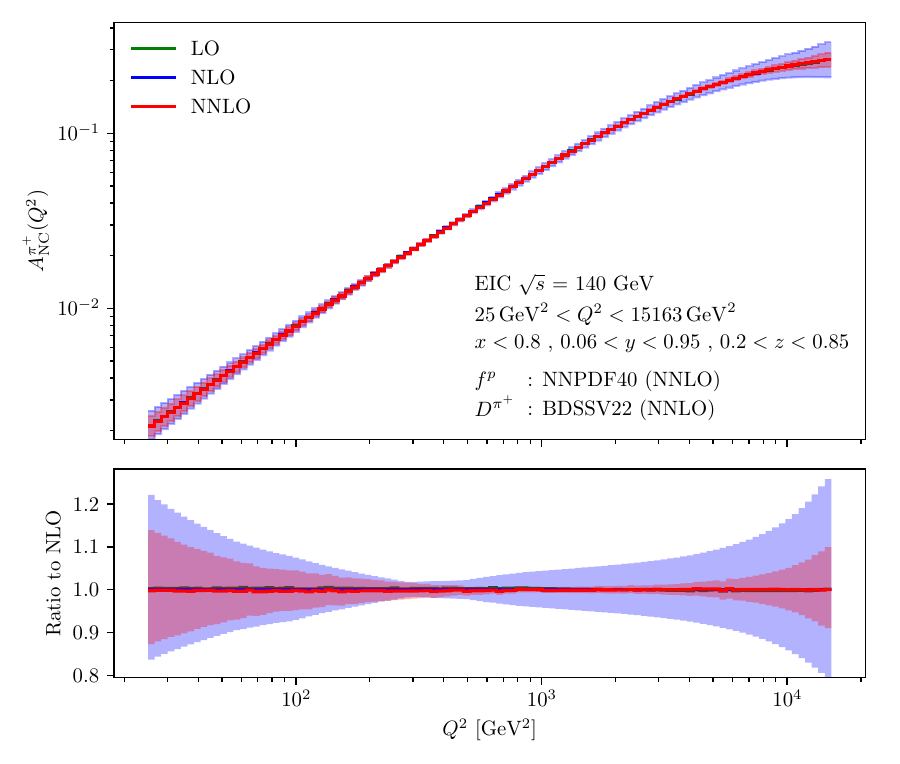}
\caption{
NC
asymmetry
 $A_\mathrm{NC}^{\pi^+}$ as function of $Q^2$.
Top panel: total asymmetry.
Bottom panel: Ratio to NLO.
}
\label{fig:Q2_A_NC}
\end{figure}

In Fig.~\ref{fig:Q2_A_NC} we plot the neutral current asymmetry $A^{\pi^+}_\NC(Q^2)$, defined in eq.~\eqref{eq:A_NC} as a function of $Q^2$.
It is designed to assess the relative size of the $\lambda_\ell\,$-odd parts of the cross section with respect to the helicity-averaged cross section.
Inspecting the top panel of Fig.~\ref{fig:Q2_A_NC}, the $\lambda_\ell$-even electroweak effects appear to be small at low $Q^2$,
reaching a percent level magnitude at $Q^2\sim 200\, \GeV^2$.
Then $A^{\pi^+}_\NC(Q^2)$ increases to $\sim 8\,\%$ at $Q^2\sim 1000\,\GeV^2$ and further,
slowly starting to saturate towards $\sim 30 \,\%$ at the highest considered $Q^2$.
Only above approximately $1500\,\GeV^2$ the $\lambda_\ell$-odd electroweak contributions are surpassed in size by the $\lambda_\ell$-even electroweak effects $\dd \sigma^{\pi^+,(\gamma Z+ ZZ)}/\dd Q^2$, which can be inferred from comparing Fig.~\ref{fig:Q2_A_NC} with the bottom left panel in Fig.~\ref{fig:Q2_0.0_NC_pip}.
In the bottom panel, we plot the $K$-factor for the asymmetry.
While $A^{\pi^+}_\NC(Q^2)$ displays a noticeable shrinking of NNLO uncertainty bands over the entire $Q^2$ range,
the perturbative corrections are very small, at sub-percent-level throughout the $Q^2$ spectrum.
The $\NC$ asymmetry $A_\NC^{\pi^-}(Q^2)$ for $\pi^-$ production is indistinguishable from $A_\NC^{\pi^+}(Q^2)$,
and is thus not displayed separately.

\subsubsection[\texorpdfstring{$z$-distribution}{z-distribution}]{\boldmath $z$-distribution}

\begin{figure}[tbp]
\centering
\begin{subfigure}{0.49\textwidth}
\centering
\hspace{-7pt}
\includegraphics[width=\linewidth]{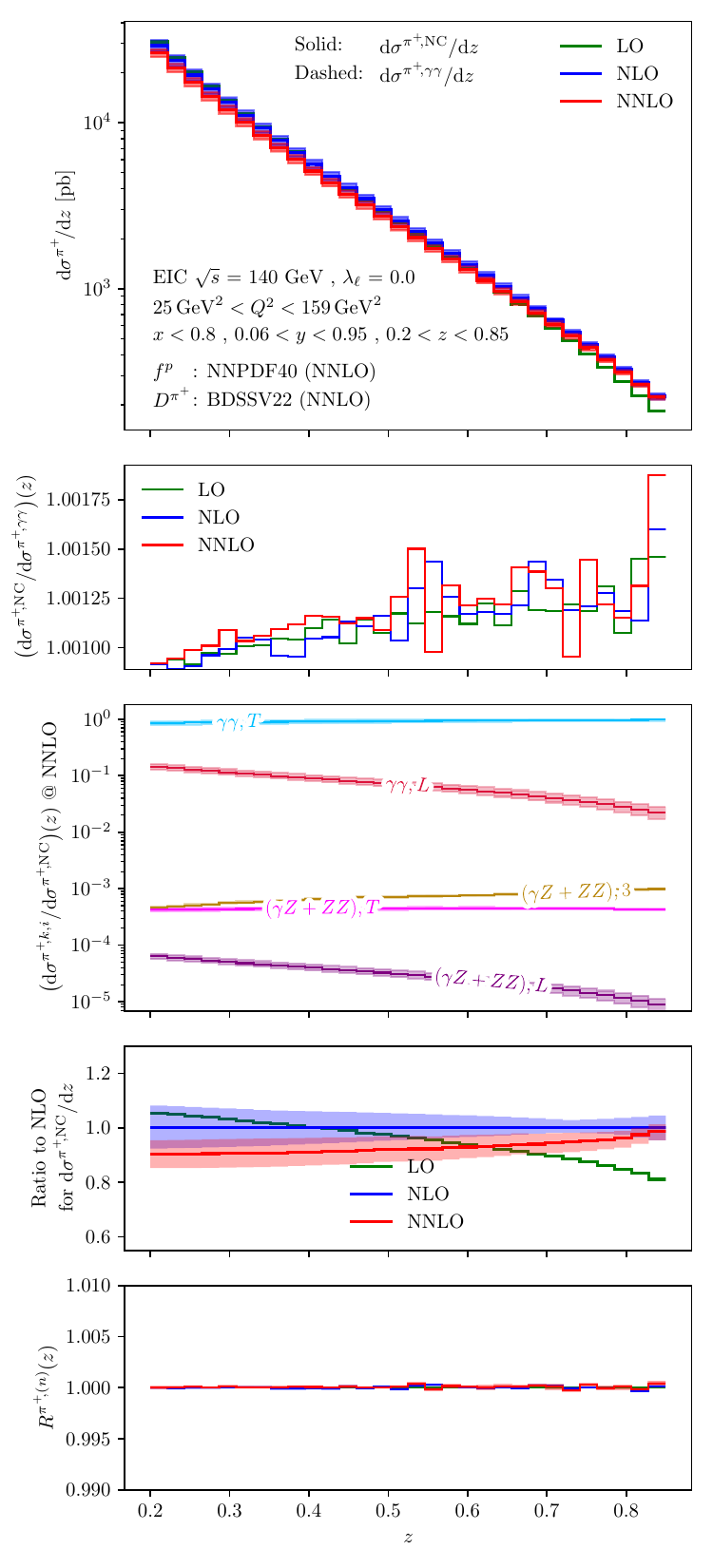}
\caption{Low-$Q^2$}
\end{subfigure}
\begin{subfigure}{0.49\textwidth}
\centering
\includegraphics[width=1.0\linewidth]{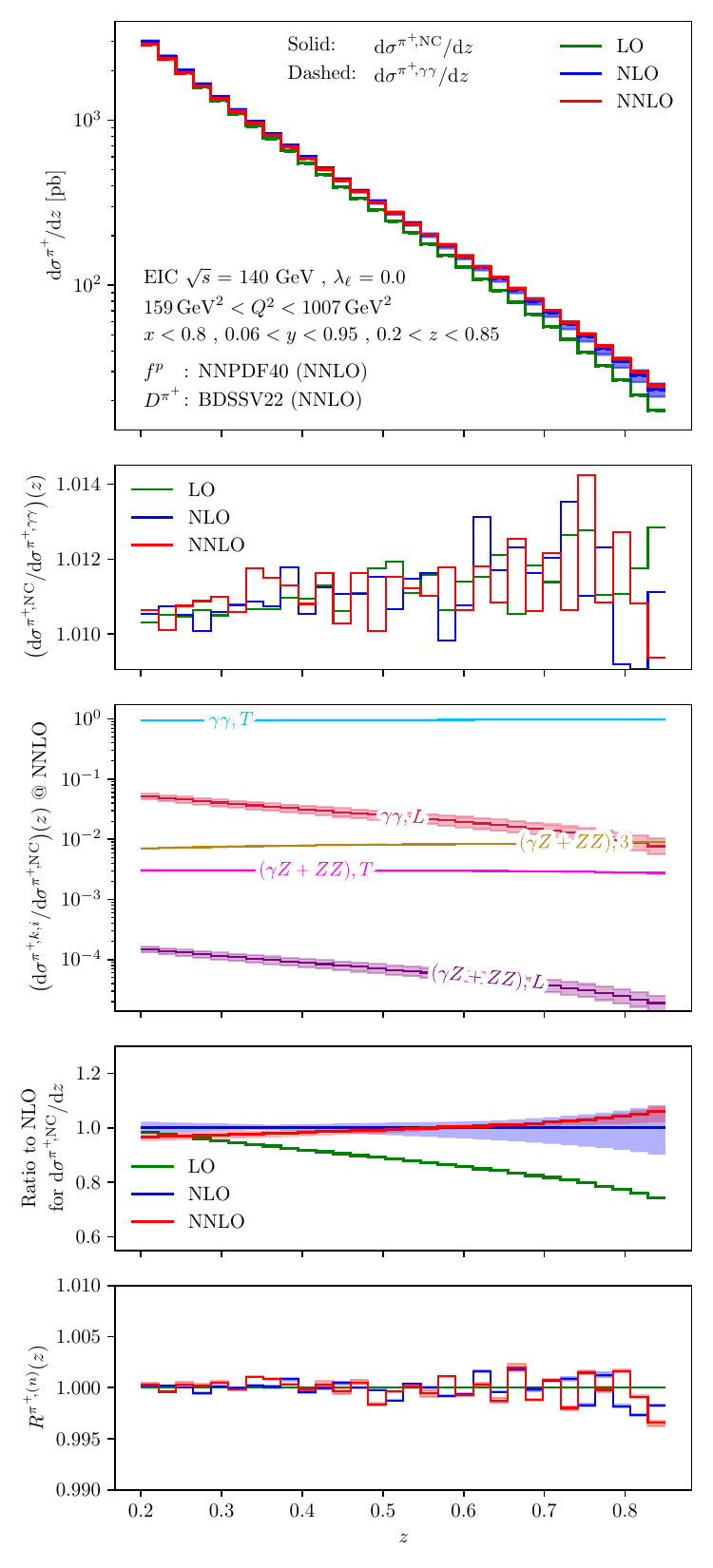}
\caption{Mid-$Q^2$}
\end{subfigure}

\caption{NC $z$-distributions for $\lambda_\ell$-averaged $\pi^+$ production at low and intermediate $Q^2$.
First panels: total cross section.
Second panels: ratio $(\dd\sigma^{\pi^+,\NC}/\dd\sigma^{\pi^+,\gamma\gamma})(z)$.
Third panels: structure function contributions to the total cross section.
Fourth panels: $\NC$ Ratio to NLO.
Fifth Panels: $R^{\pi^+,(n)}$ ratio as function of $z$.
}
\label{fig:Z_0.0_NC_lowQ2_pip}

\end{figure}

\begin{figure}[tbp]
\centering
\begin{subfigure}{0.49\textwidth}
\centering
\includegraphics[width=.995\linewidth]{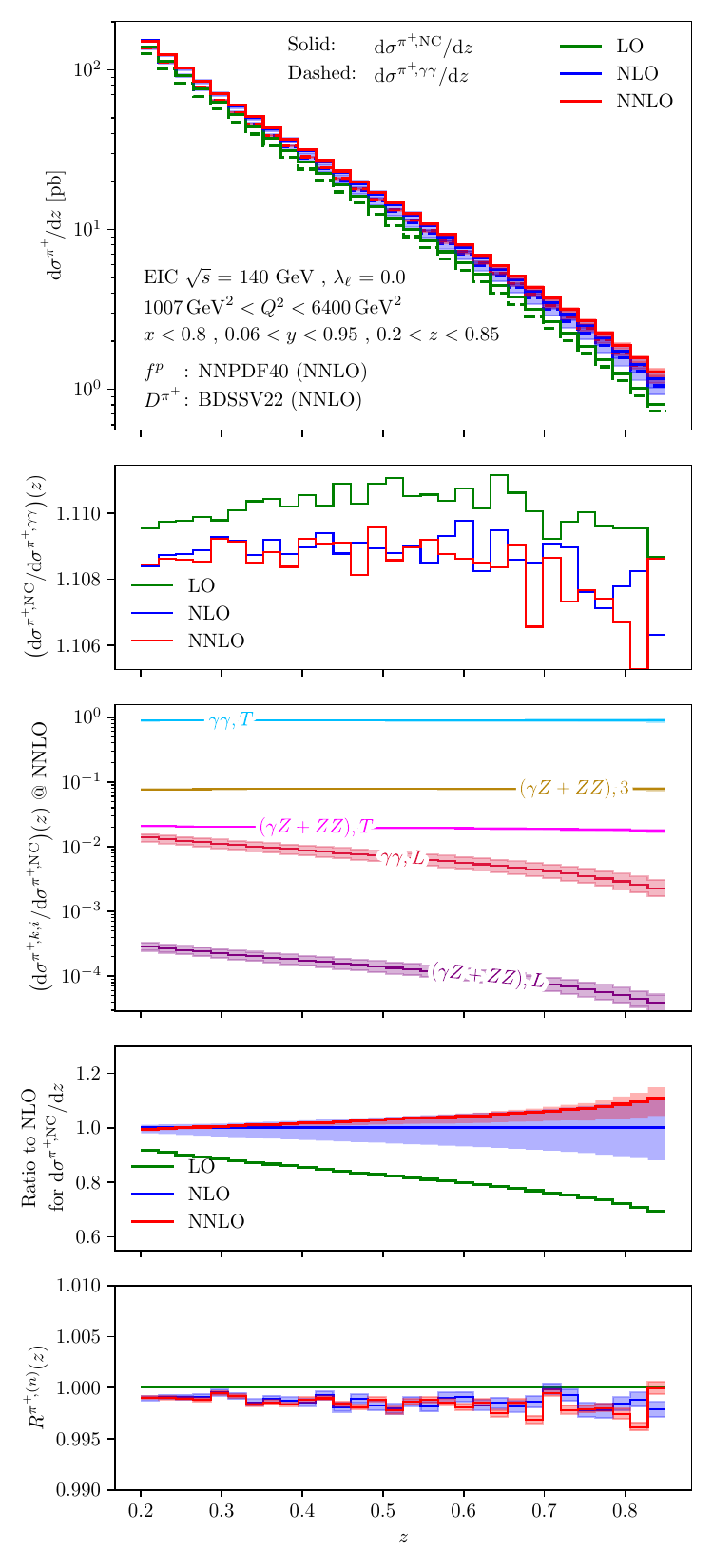}
\caption{High-$Q^2$}
\end{subfigure}
\begin{subfigure}{0.49\textwidth}
\centering
\includegraphics[width=\linewidth]{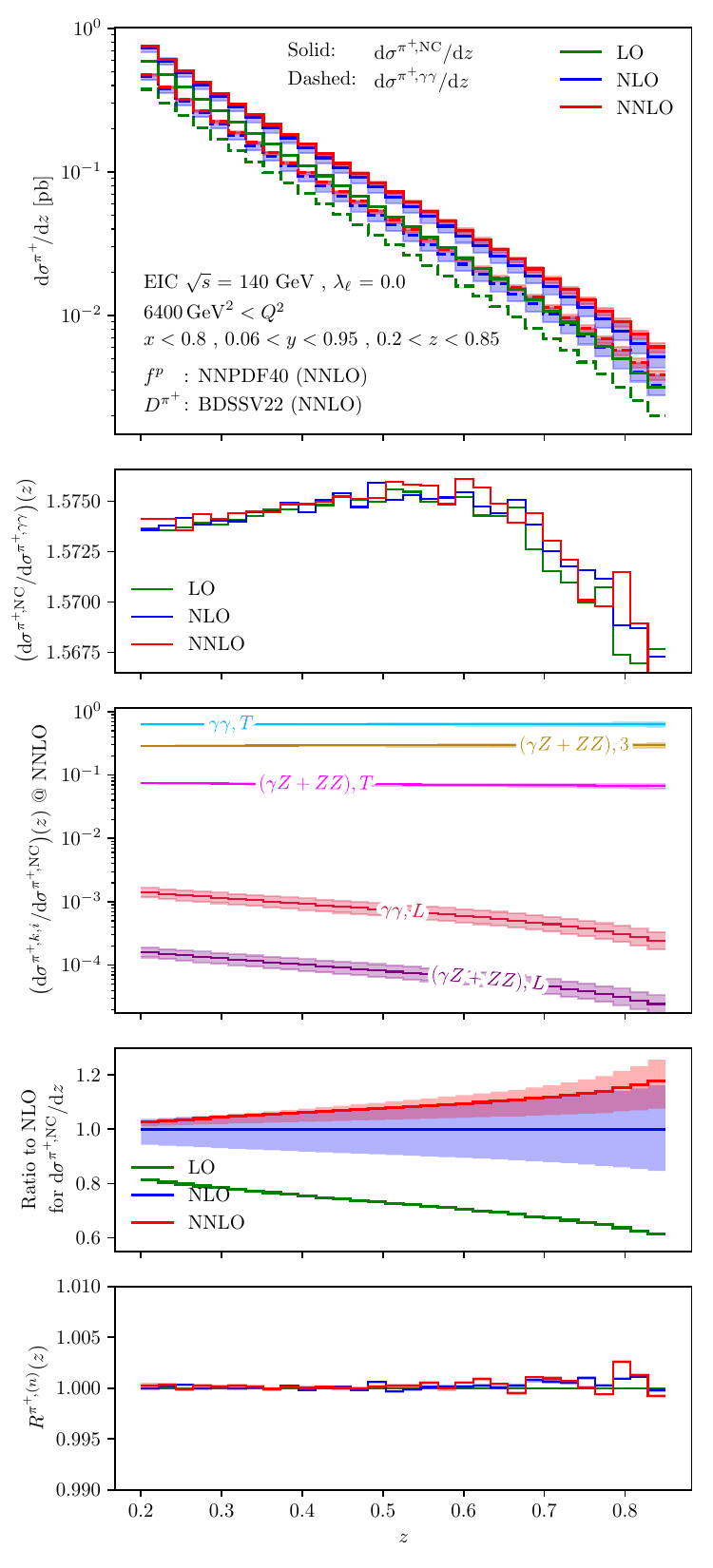}
\caption{Extreme-$Q^2$}
\end{subfigure}

\caption{
NC $z$-distributions for $\lambda_\ell$-averaged $\pi^+$ production at high $Q^2$.
First panels: total cross section. Second panels: ratio $(\dd\sigma^{\pi^+,\NC}/\dd\sigma^{\pi^+,\gamma\gamma})(z)$.
Third panels: structure function contributions to the total cross section.
Fourth panels: $\NC$ Ratio to NLO.
Fifth Panels: $R^{\pi^+,(n)}$ ratio as function of $z$.
}
\label{fig:Z_0.0_NC_highQ2_pip}

\end{figure}

In Fig.~\ref{fig:Z_0.0_NC_lowQ2_pip} and~\ref{fig:Z_0.0_NC_highQ2_pip} the NC distributions $\dd \sigma^{\pi^+}/\dd z$ for the four ranges in $Q^2$ defined in eq.~\eqref{eq:cutsQ2} are depicted.
The $z$-differential cross section decreases with increasing $Q^2$ from Low-$Q^2$ to Extreme-$Q^2$ based on the DIS scaling $1/Q^2$.
Between the $Q^2$ ranges we observe moderate changes in the downward sloping shape with increasing $z$.
The ratio $(\dd\sigma^{\pi^+,\NC}/\dd\sigma^{\pi^+,\gamma\gamma})(z)$, as displayed in the second panel of each plot, is largely determined by the lower bound of $Q^2$ through the propagator ratio $\eta_{\gamma Z}$, reaching a maximum value of $\sim 1.6$ in the Extreme-$Q^2$ bin.
A $z$-dependence in this ratio is almost absent, with moderate per-mille-level dependence only in the highest $Q^2$ region.
The $n^\text{th}$-order corrections to the $\NC$ cross section $\dd \sigma^{\pi^+,\NC,(n)}/\dd z$ and their scale variation uncertainties are predicted at per-mille level accuracy by rescaling the photon exchange cross section $\dd\sigma^{\pi^+,\gamma\gamma,(n)}/\dd z$ with the LO ratio $(\dd\sigma^{\pi^+,\NC,(0)}/\dd \sigma^{\pi^+,\gamma\gamma,(0)})(z)$, as visible from $R^{\pi^+,(n)}(z)$ in the fifth panel.
In the fourth panel, we observe good perturbative convergence, especially upon increasing the lower bound on $Q^2$.

In the third panel in the figures we display the decomposition in structure function contributions.
In the Low-$Q^2$ and Mid-$Q^2$ regions, the contribution $\dd\sigma^{\pi^+,\gamma\gamma,L}/\dd z$ from the $\Fcal^{\pi^+}_L$ structure function is the second most dominant after $\dd\sigma^{\pi^+,\gamma\gamma,T}/\dd z$.
The large relative uncertainties in $\dd\sigma^{\pi^+,\gamma\gamma,L}/\dd z$ compared to the other contributions are due to the vanishing LO of $\Fcal_L^h$.
The $\dd\sigma^{\pi^+,(\gamma Z + ZZ),3}/\dd z$ contribution from the $\Fcal^{\pi^+}_3$ structure function increases with $Q^2$ to become the second largest in the High-$Q^2$ and Extreme-$Q^2$ regions, where it reaches a magnitude of $10\,\%$ to $50 \, \%$ of $\dd\sigma^{\pi^+,\gamma\gamma,T}/\dd z$.
The EW effects are dominated by $\dd\sigma^{\pi^+,(\gamma Z + ZZ),3}/\dd z$, which is around twice the size of the EW effects $\dd\sigma^{\pi^+,(\gamma Z + ZZ),T}/\dd z$ in the Low-$Q^2$ and Mid-$Q^2$ regions, and three to four times their size in the High-$Q^2$ and Extreme-$Q^2$ regions.
The $z$-dependence of $\dd\sigma^{\pi^+,\gamma\gamma,T}/\dd z$ and $\dd\sigma^{\pi^+,(\gamma Z + ZZ),3}/\dd z$ is similar, resulting in a negligible change in the $z$-dependence when adding EW effects.

\begin{figure}[tbp]
\centering
\begin{subfigure}{0.49\textwidth}
\centering
\hspace{-7pt}
\includegraphics[width=\linewidth]{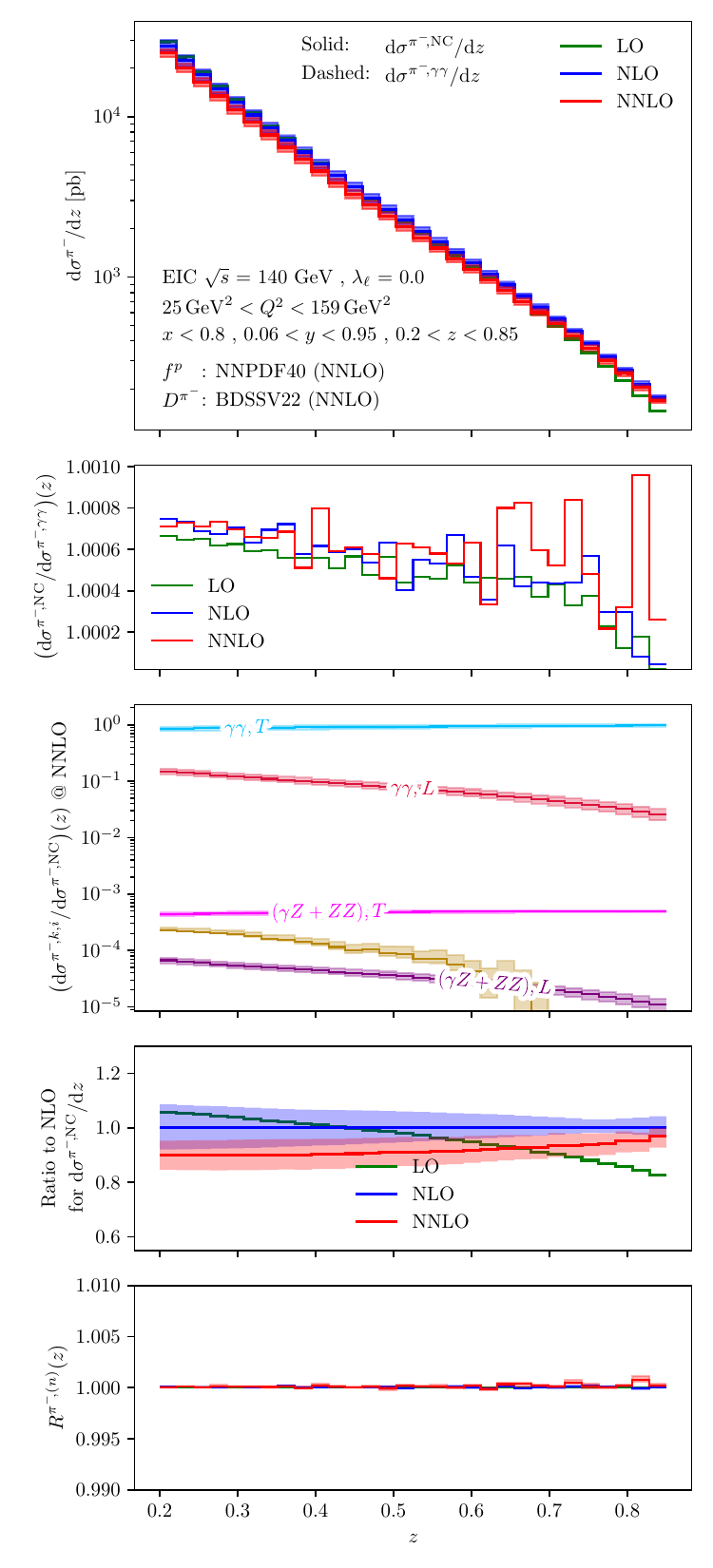}
\caption{Low-$Q^2$}
\end{subfigure}
\begin{subfigure}{0.49\textwidth}
\centering
\includegraphics[width=1.0\linewidth]{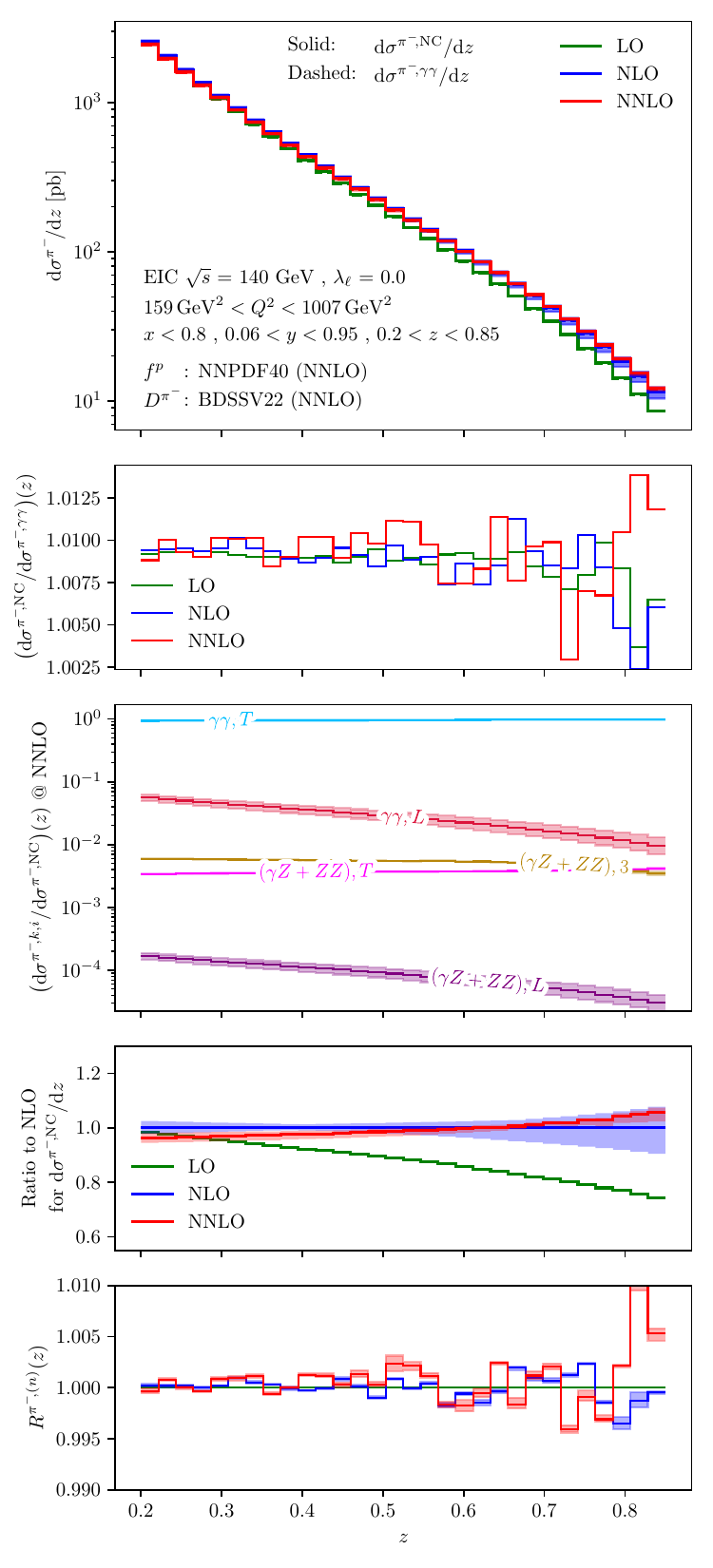}
\caption{Mid-$Q^2$}
\end{subfigure}

\caption{
NC $z$-distributions for $\lambda_\ell$-averaged $\pi^-$ production at low and intermediate $Q^2$.
First panels: total cross section.
Second panels: ratio $(\dd\sigma^{\pi^-,\NC}/\dd\sigma^{\pi^-,\gamma\gamma})(z)$.
Third panels: structure function contributions to the total cross section.
Fourth panels: Ratio to NLO.
Fifth Panels: $R^{\pi^-,(n)}$ ratio as function of $z$.
}
\label{fig:Z_0.0_NC_lowQ2_pim}

\end{figure}

\begin{figure}[tbp]
\centering
\begin{subfigure}{0.49\textwidth}
\centering
\includegraphics[width=\linewidth]{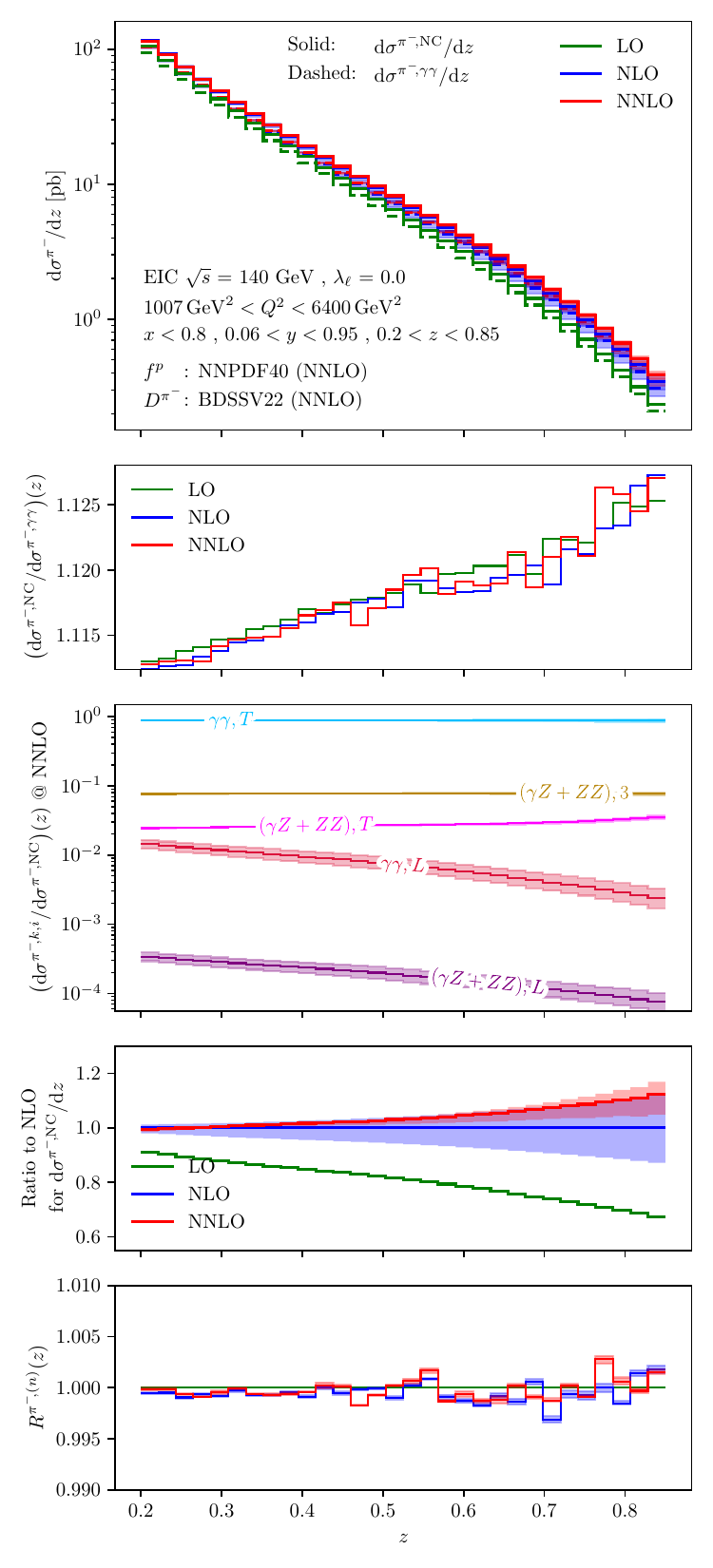}
\caption{High-$Q^2$}
\end{subfigure}
\begin{subfigure}{0.49\textwidth}
\centering
\includegraphics[width=\linewidth]{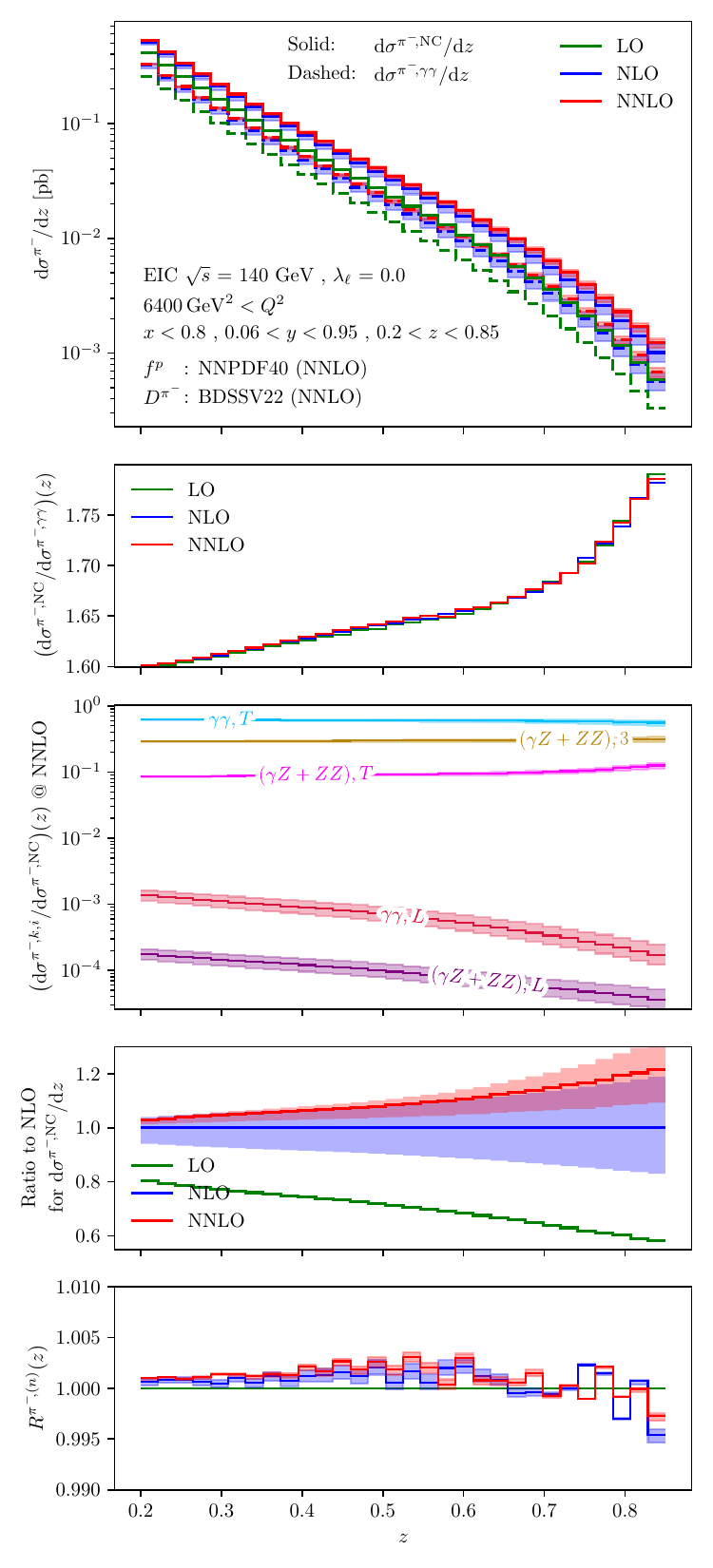}
\caption{Extreme-$Q^2$}
\end{subfigure}

\caption{
NC $z$-distributions for $\lambda_\ell$-averaged $\pi^-$ production at high $Q^2$.
First panels: total cross section.
Second panels: ratio $(\dd\sigma^{\pi^-,\NC}/\dd\sigma^{\pi^-,\gamma\gamma})(z)$.
Third panels: structure function contributions to the total cross section.
Fourth panels: Ratio to NLO.
Fifth Panels: $R^{\pi^-,(n)}$ ratio as function of $z$.}
\label{fig:Z_0.0_NC_highQ2_pim}

\end{figure}

The $z$-differential cross sections for $\pi^-$ production in Fig.~\ref{fig:Z_0.0_NC_lowQ2_pim} and~\ref{fig:Z_0.0_NC_highQ2_pim} are insignificantly smaller for low and intermediate values of $z$, but are falling off slightly faster at high $z$.
Only in the Extreme-$Q^2$ region the $\pi^-$ production cross section reduces to merely half the size of their $\pi^+$ counterparts, while at the same time showing larger EW effects.
In this kinematic region $\dd \sigma^{\pi^-,(\gamma Z + ZZ)}/\dd z$ also exhibits a different $z$-dependence compared to the pure photon-exchange.
Noteworthy is the greatly reduced size of $\dd\sigma^{\pi^-,(\gamma Z+ZZ),3}/\dd z$ and its remarkable dependence on $z$ at Low-$Q^2$.
The $\Fcal^{\pi^-}_3$ contribution essentially vanishes above $z\sim 0.7$, while the contributions from the other structure functions remain unchanged.
For the Low-$Q^2$ region we have already noticed this effect in our discussion of $\dd \sigma^{\pi^-,\NC}/\dd Q^2$.

\begin{figure}[tb]
\centering
\begin{subfigure}{0.45\textwidth}
\centering
\includegraphics[width=1.0\linewidth]{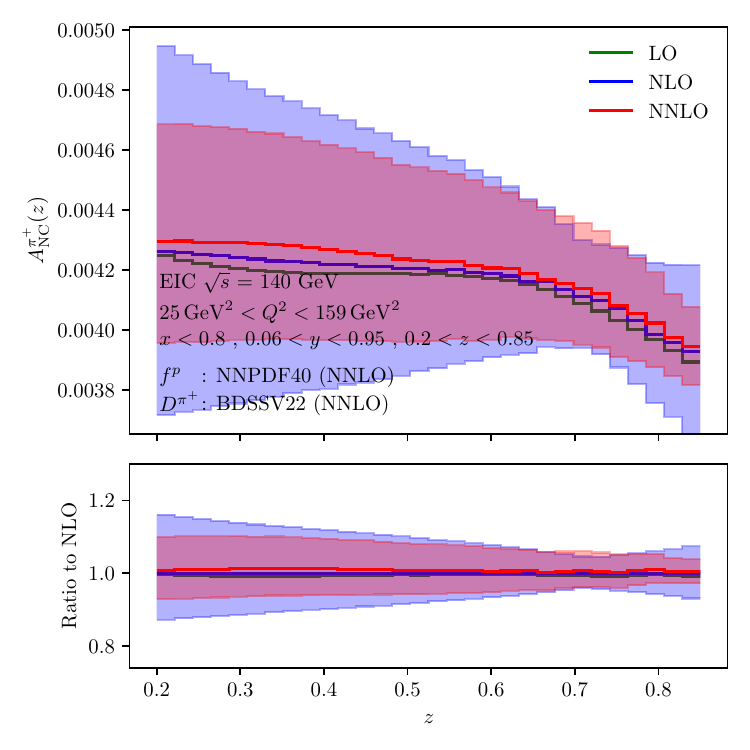}
\caption{Low-$Q^2$}
\end{subfigure}
\begin{subfigure}{0.45\textwidth}
\centering
\includegraphics[width=.995\linewidth]{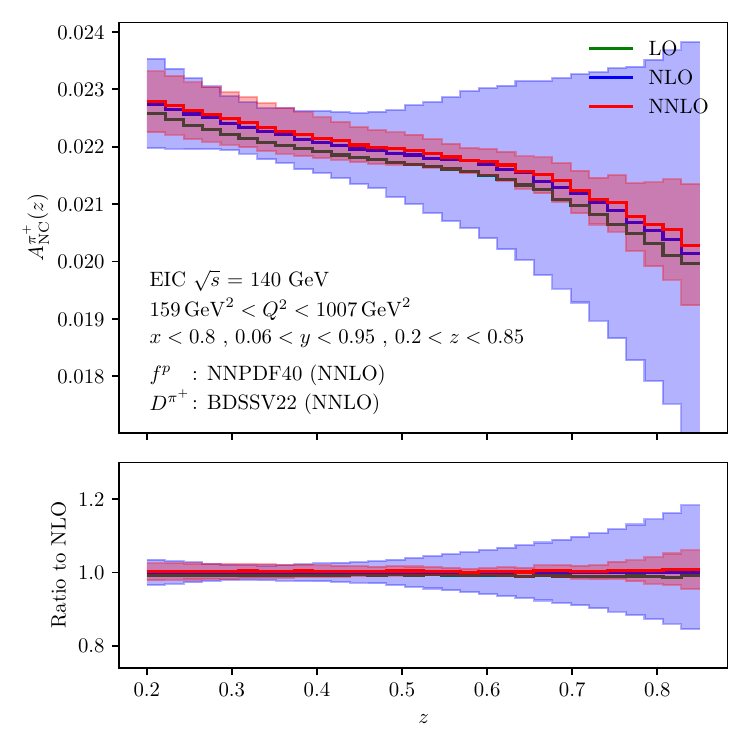}
\caption{Mid-$Q^2$}
\end{subfigure}
\begin{subfigure}{0.45\textwidth}
\centering
\includegraphics[width=1.0\linewidth]{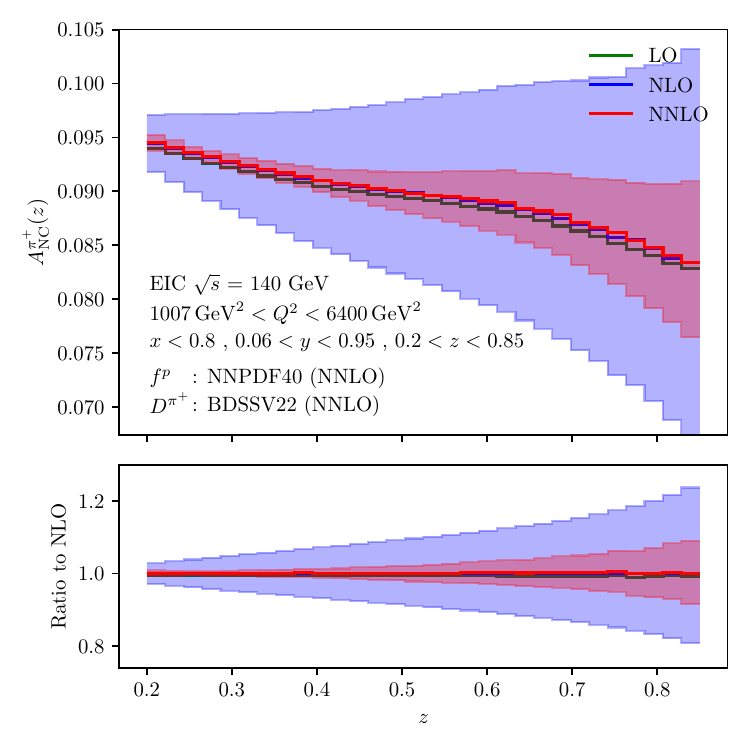}
\caption{High-$Q^2$}
\end{subfigure}
\begin{subfigure}{0.45\textwidth}
\centering
\includegraphics[width=.99\linewidth]{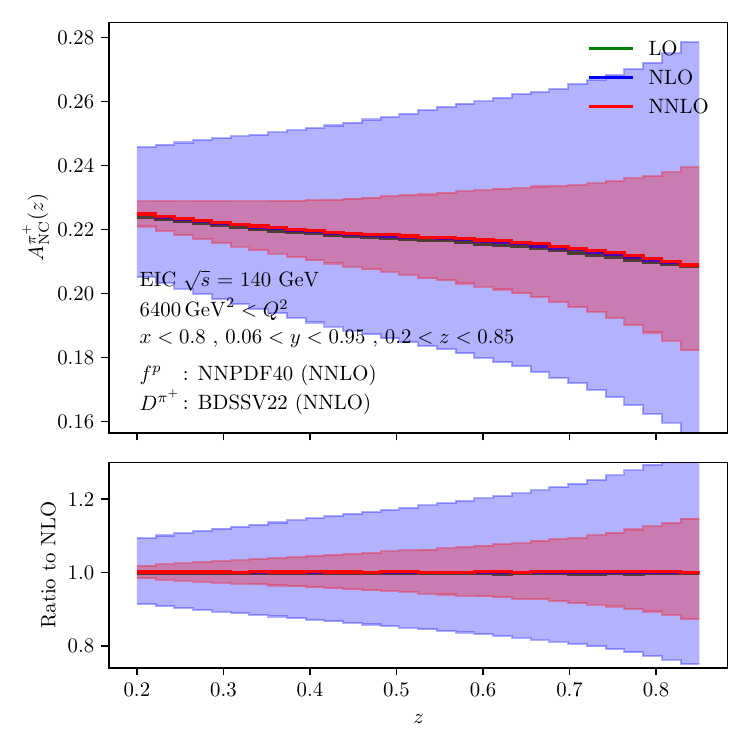}
\caption{Extreme-$Q^2$}
\end{subfigure}

\caption{
NC asymmetry $A_\NC^{\pi^+}$ as a function of $z$.
Top panels: total cross section.
Bottom panels: Ratio to NLO.
}
\label{fig:Z_A_NC_pip}

\end{figure}

In Fig.~\ref{fig:Z_A_NC_pip} we show the neutral current asymmetry for $\pi^+$ production as a function of $z$.
$A^{\pi^+}_{\NC}(z)$ carries a moderate $z$-dependence, but is rather sensitive to the average $Q^2$:
its absolute value increases significantly with $Q^2$, ranging from sub-percent-level in the Low-$Q^2$ region to around twenty percent in Extreme-$Q^2$ region.
The scale uncertainties are large in the Low-$Q^2$ region, but with stable central values between the different orders.
At higher average $Q^2$ we observe a greatly improved perturbative convergence.

\begin{figure}[tb]
\centering
\begin{subfigure}{0.45\textwidth}
\centering
\includegraphics[width=1.0\linewidth]{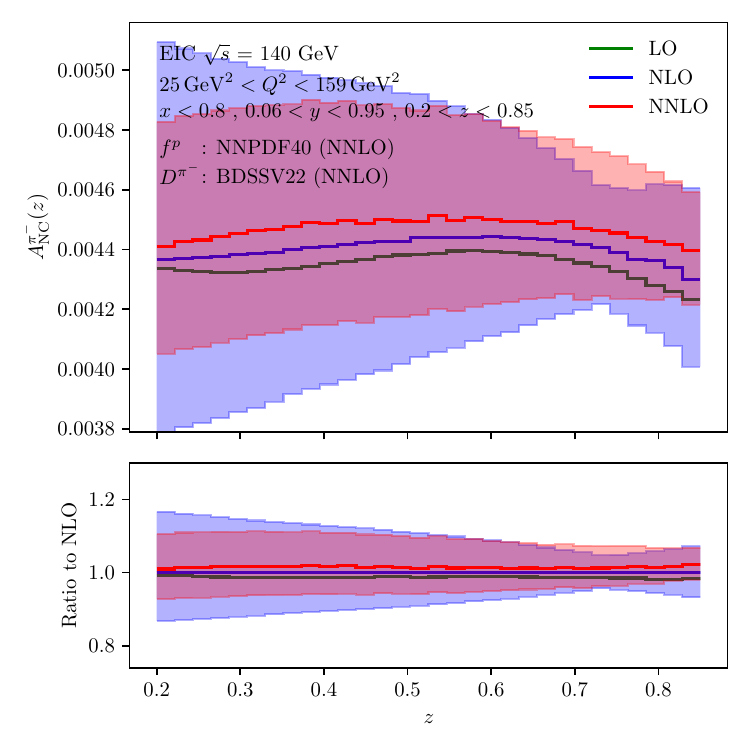}
\caption{Low-$Q^2$}
\end{subfigure}
\begin{subfigure}{0.45\textwidth}
\centering
\includegraphics[width=1.0\linewidth]{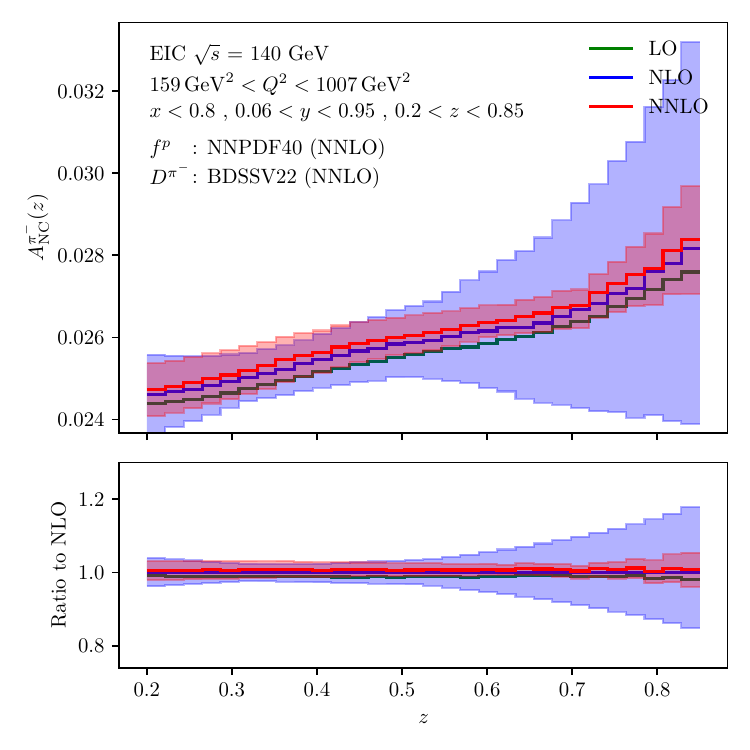}
\caption{Mid-$Q^2$}
\end{subfigure}
\begin{subfigure}{0.45\textwidth}
\centering
\includegraphics[width=1.0\linewidth]{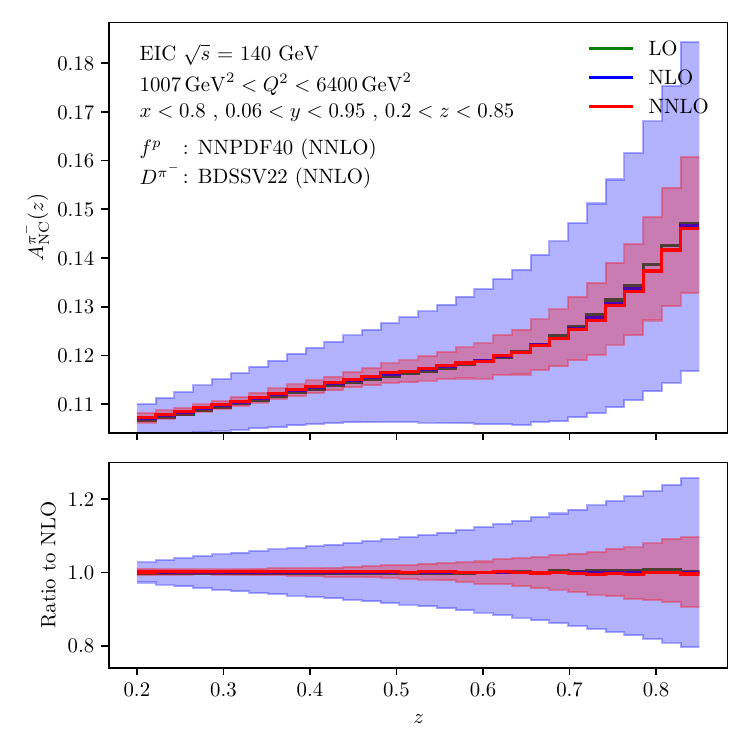}
\caption{High-$Q^2$}
\end{subfigure}
\begin{subfigure}{0.45\textwidth}
\centering
\includegraphics[width=1.0\linewidth]{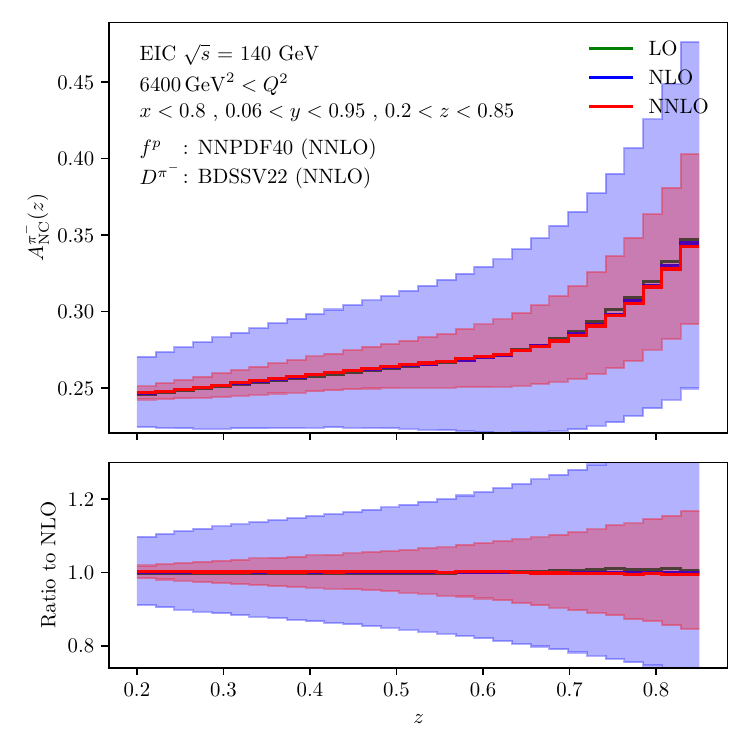}
\caption{Extreme-$Q^2$}
\end{subfigure}

\caption{
NC asymmetry $A_\NC^{\pi^-}$ as a function of $z$.
Top panels: total cross section.
Bottom panels: Ratio to NLO.
}
\label{fig:Z_A_NC_pim}

\end{figure}

The $\NC$ asymmetry $A_\NC^{\pi^-}(z)$ for $\pi^-$ production is shown in Fig.~\ref{fig:Z_A_NC_pim}.
The perturbative convergence is slower than for $\pi^+$ production, hinting at a non-trivial interplay between different partonic channels for the production of the final state hadron.

Both $A^{\pi^{+}}_\NC(z)$ and $A^{\pi^{-}}_\NC(z)$ appear to be largely insensitive to the perturbative order.
This allows for a reasonable estimate of the $\lambda_\ell$-odd cross section by dressing the leading order with photon-exchange K-factors for the fragmentation-sensitive $z$-distributions.

\begin{figure}[p]
\centering
\includegraphics[width=\textwidth]{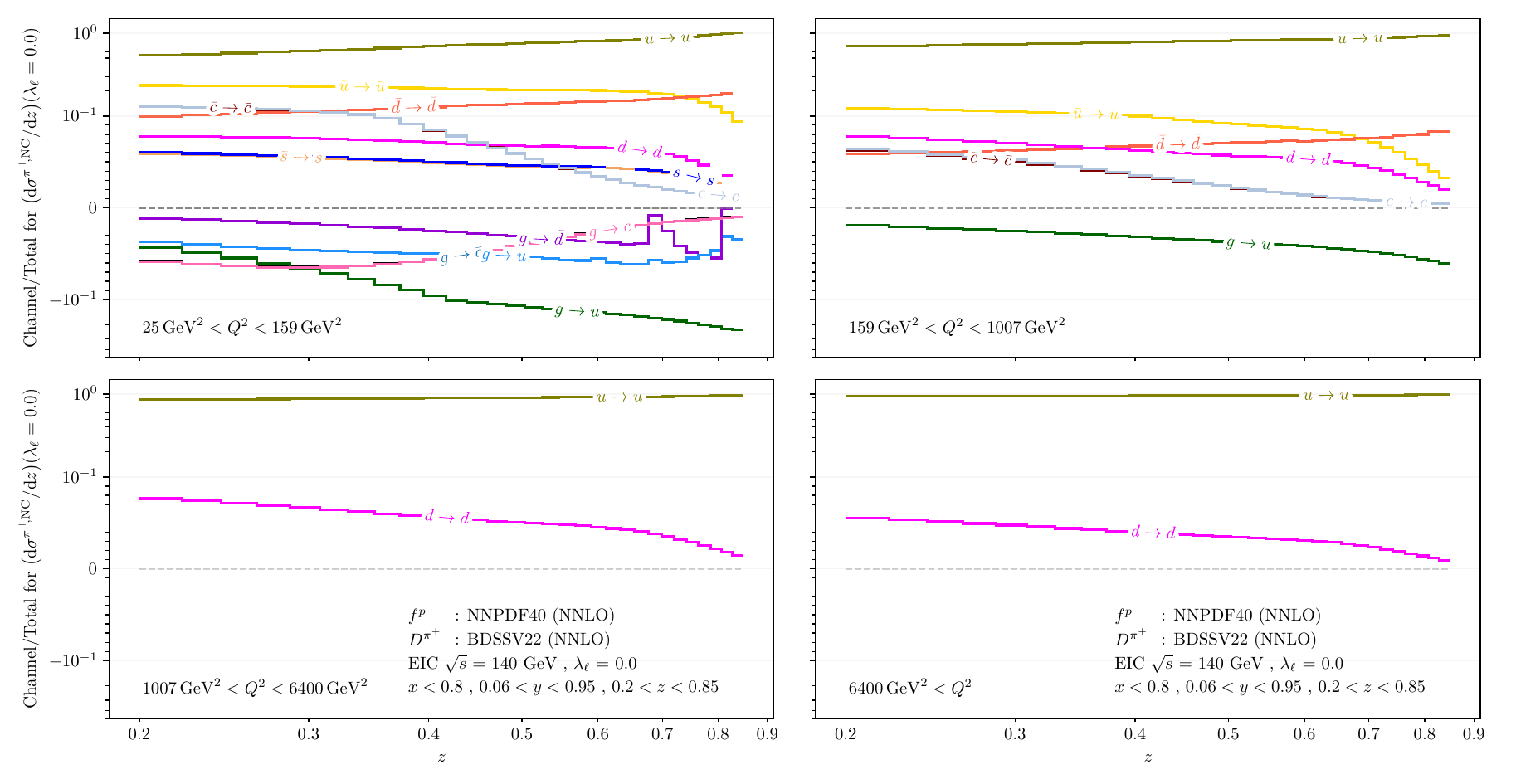}
\caption{
Channel decomposition for $\dd \sigma^{\pi^+,\NC}$ as a function of $z$.
Only channels contributing more than $4\,\%$ in any bin are displayed.
The range $(-0.1, 0.1)$ on the vertical axis is linear, the ranges above and below are plotted logarithmically.
}
\label{fig:z_NC_channel_pip}
\end{figure}

In Fig.~\ref{fig:z_NC_channel_pip} the channel decomposition for $\dd \sigma^{\pi^+,\NC}/\dd z$ is displayed.
In the following, we describe the features of the channel decomposition by decreasing virtuality.
At High- and Extreme-$Q^2$ the $u\to u$ channel accounts for almost the entire production cross section.
Due to the range of $x$ probed at High- and Extreme-$Q^2$, only $f^p_d$ is comparable in size to $f^p_u$.
At Mid-$Q^2$, sea-initiated channels start to become sizeable.
In particular the channel $\bar{d}\to \bar{d}$, which is preferred by the $\pi^+$, and $\bar{u}\to \bar{u}$, which has a larger electromagnetic coupling, as well as the charge- and fragmentation-enhanced gluonic channel $g\to u$ start to become sizeable.
At Low-$Q^2$ the sea quark contributions in the proton become yet more significant, and also unfavoured gluon channels become larger.
The gluonic channels always give negative contributions to the cross sections.
Note that channels appearing for the first time at NNLO are not shown in the plot as they contribute less than $4\%$ everywhere.

\begin{figure}[p]
\centering
\includegraphics[width=\textwidth]{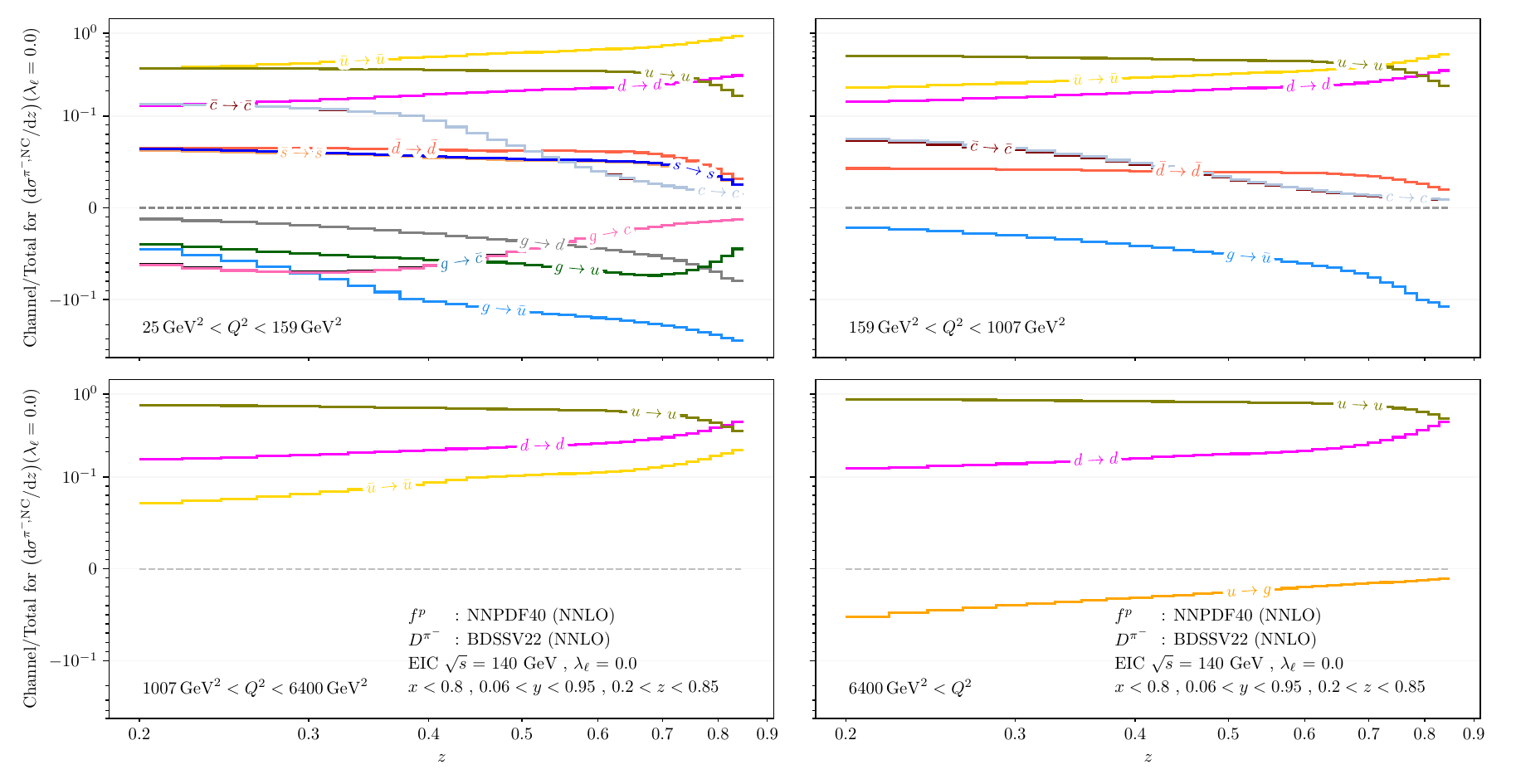}
\caption{
Channel decomposition for $\dd \sigma^{\pi^-,\NC}$ as a function of $z$.
Only channels contributing more than $4\,\%$ in any bin are displayed.
The range $(-0.1, 0.1)$ on the vertical axis is linear, the ranges above and below are plotted logarithmically.
}
\label{fig:z_NC_channel_pim}
\end{figure}

As shown in Fig.~\ref{fig:z_NC_channel_pim}, also for $\pi^-$ the predominant production channel is $u\to u$ in the valence regime (High- and Extreme-$Q^2$), due to the abundance of up-type quarks in the proton and the stronger electromagnetic coupling.
At higher $z$ the contribution from  $d\to d$ (which is the favoured fragmentation channel for $\pi^-$ production)
becomes sizable, since
the $u\to u$ channel is disfavoured by the fragmentation functions at high $z$.
The same effect is visible in $\pi^+$ production, with $u\to u$ becoming yet more pronounced as production mode for $\pi^+$ at high-$z$ and $d\to d$ depleting.
Conversely at small-$z$, $d$ and $u$ quarks fragment into $\pi^+$ and $\pi^-$ at almost equal rates since the resulting hadron retains only a small momentum fraction of the parent parton.

\begin{figure}[p]
\centering
\includegraphics[width=\textwidth]{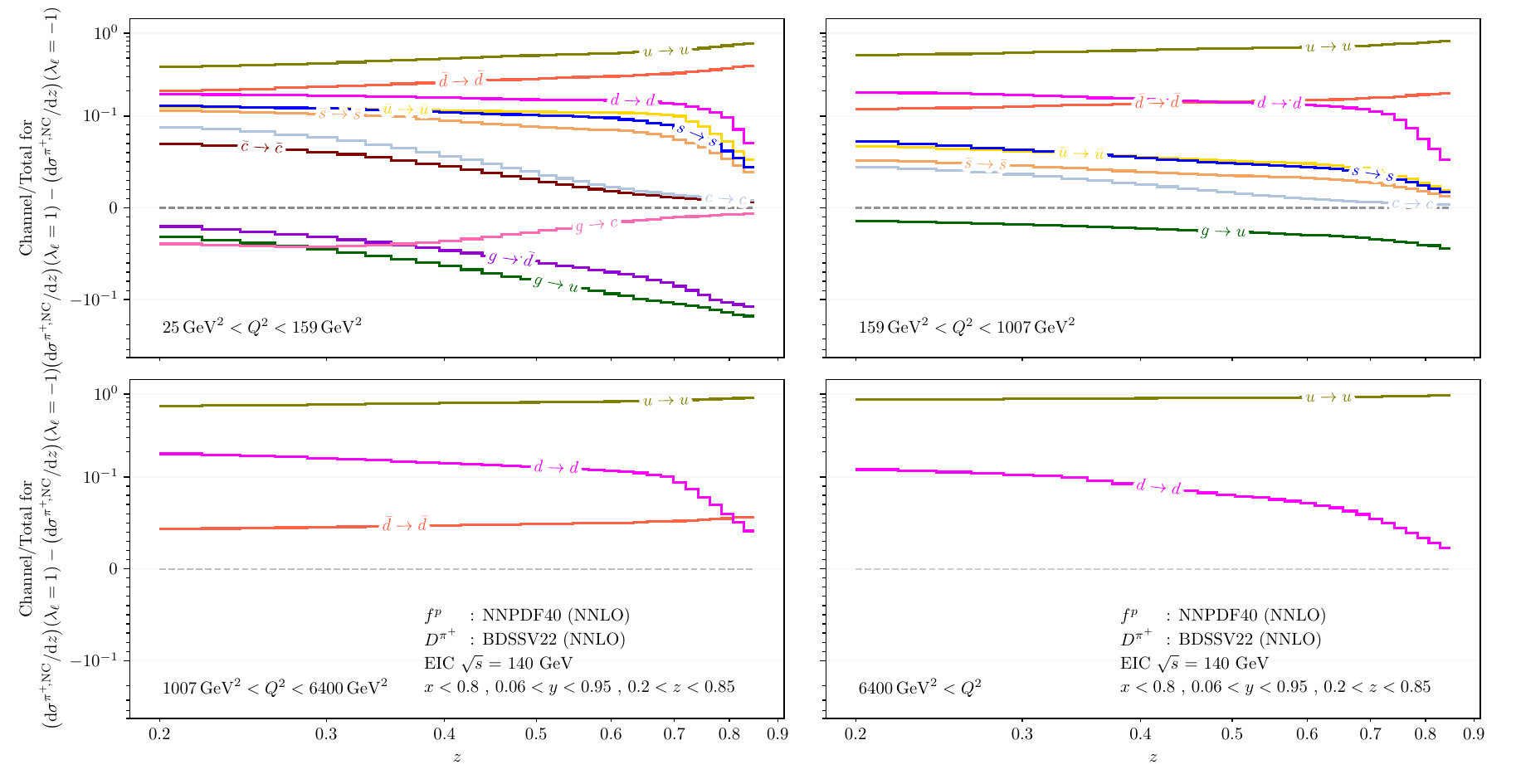}
\caption{
Channel decomposition for $\lambda_\ell$-odd cross section $\dd \sigma^{\pi^+,\NC}(\lambda_\ell=+1) - \dd \sigma^{\pi^+,\NC}(\lambda_\ell=-1)$ as a function of $z$.
Only channels contributing more than $4\,\%$ in any bin are displayed.
The range $(-0.1, 0.1)$ on the vertical axis is linear, the ranges above and below are plotted logarithmically.
}
\label{fig:z_NC_AS_channel_pip}
\end{figure}

\begin{figure}[p]
\centering
\includegraphics[width=\textwidth]{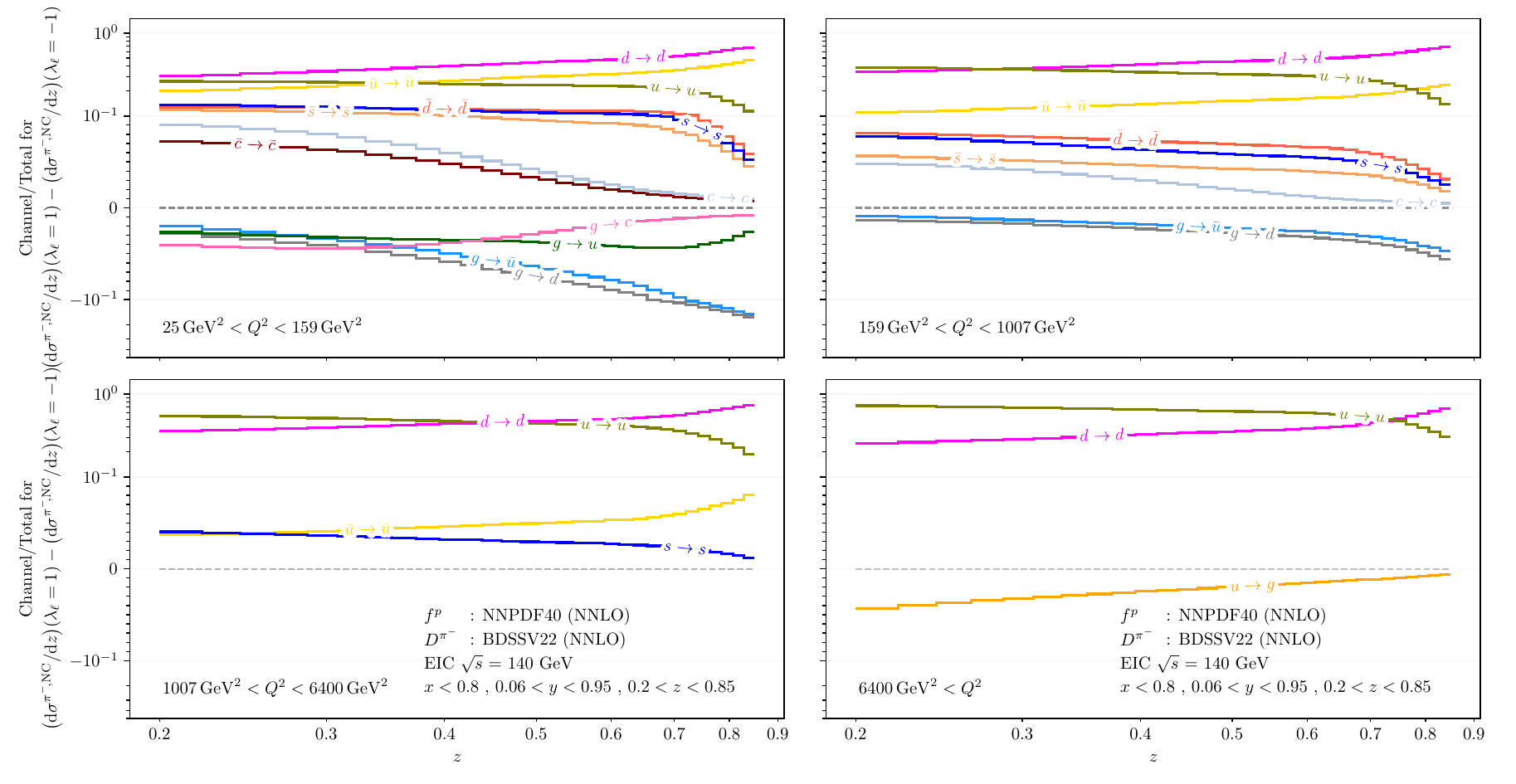}
\caption{
Channel decomposition for $\lambda_\ell$-odd cross section $\dd \sigma^{\pi^-,\NC}(\lambda_\ell=+1) - \dd \sigma^{\pi^-,\NC}(\lambda_\ell=-1)$ as a function of $z$.
Only channels contributing more than $4\,\%$ in any bin are displayed.
The range $(-0.1, 0.1)$ on the vertical axis is linear, the ranges above and below are plotted logarithmically.
}
\label{fig:z_NC_AS_channel_pim}
\end{figure}

In Fig.~\ref{fig:z_NC_AS_channel_pip} and~\ref{fig:z_NC_AS_channel_pim} the channel decomposition for the $z$-distributions of the $\lambda_\ell$-odd cross section differences $\dd \sigma^{\pi^\pm,\NC}/\dd z\,(\lambda_\ell=+1) - \dd \sigma^{\pi^\pm,\NC}/\dd z\,(\lambda_\ell=-1)$ are shown.
We observe drastic differences in the size and hierarchy of the individual contributing channels compared to the $\lambda_\ell$-even channel decomposition, Fig.~\ref{fig:z_NC_channel_pip} and~\ref{fig:z_NC_channel_pim}.
As a consequence the $\lambda_\ell$-odd channel decomposition cannot be directly predicted from the $\lambda_\ell$-even channel decomposition.

\clearpage

\subsubsection[\texorpdfstring{$x$-distribution}{x-distribution}]{\boldmath $x$-distribution}

\begin{figure}[p]
\centering

\begin{subfigure}{0.49\textwidth}
\centering
\includegraphics[width=\linewidth]{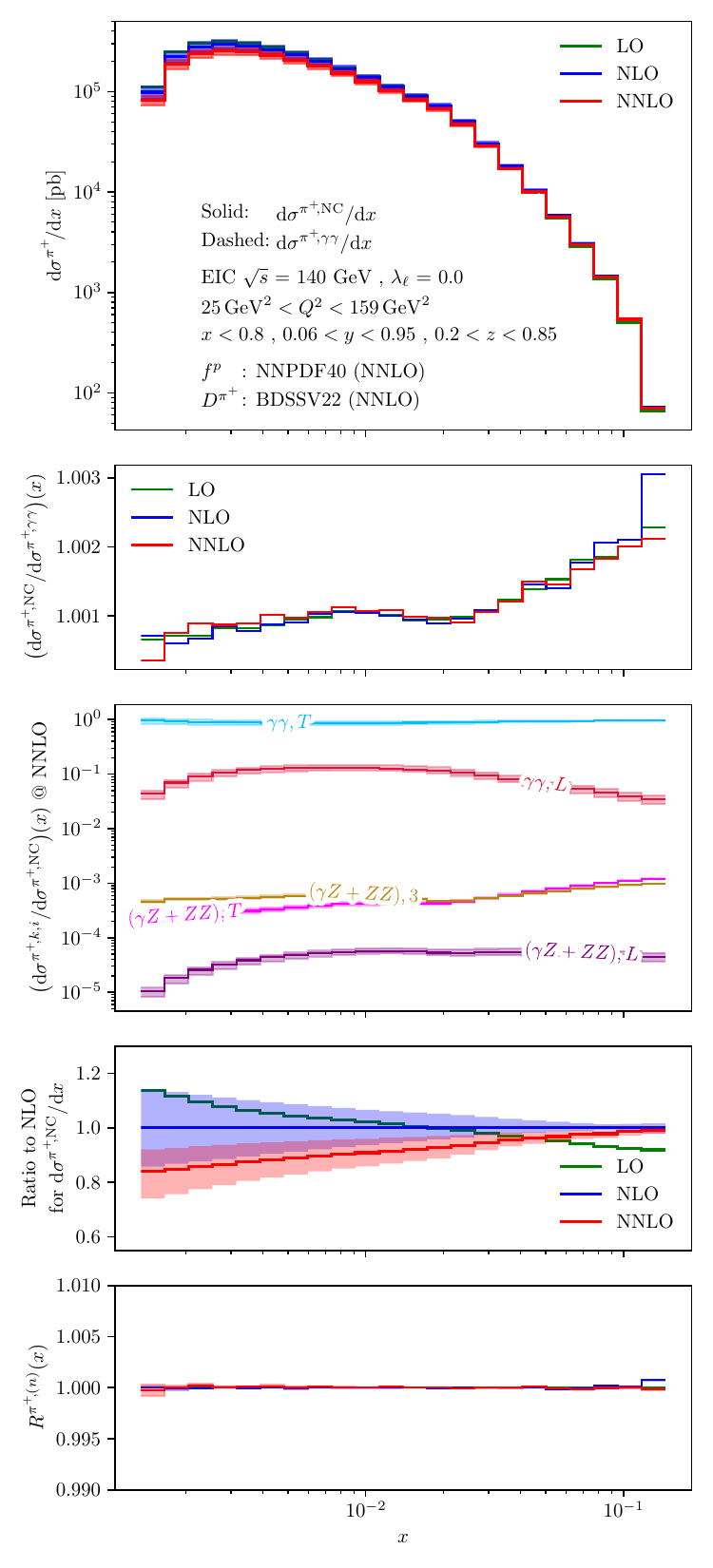}
\caption{Low-$Q^2$}
\end{subfigure}
\begin{subfigure}{0.49\textwidth}
\centering
\includegraphics[width=\linewidth]{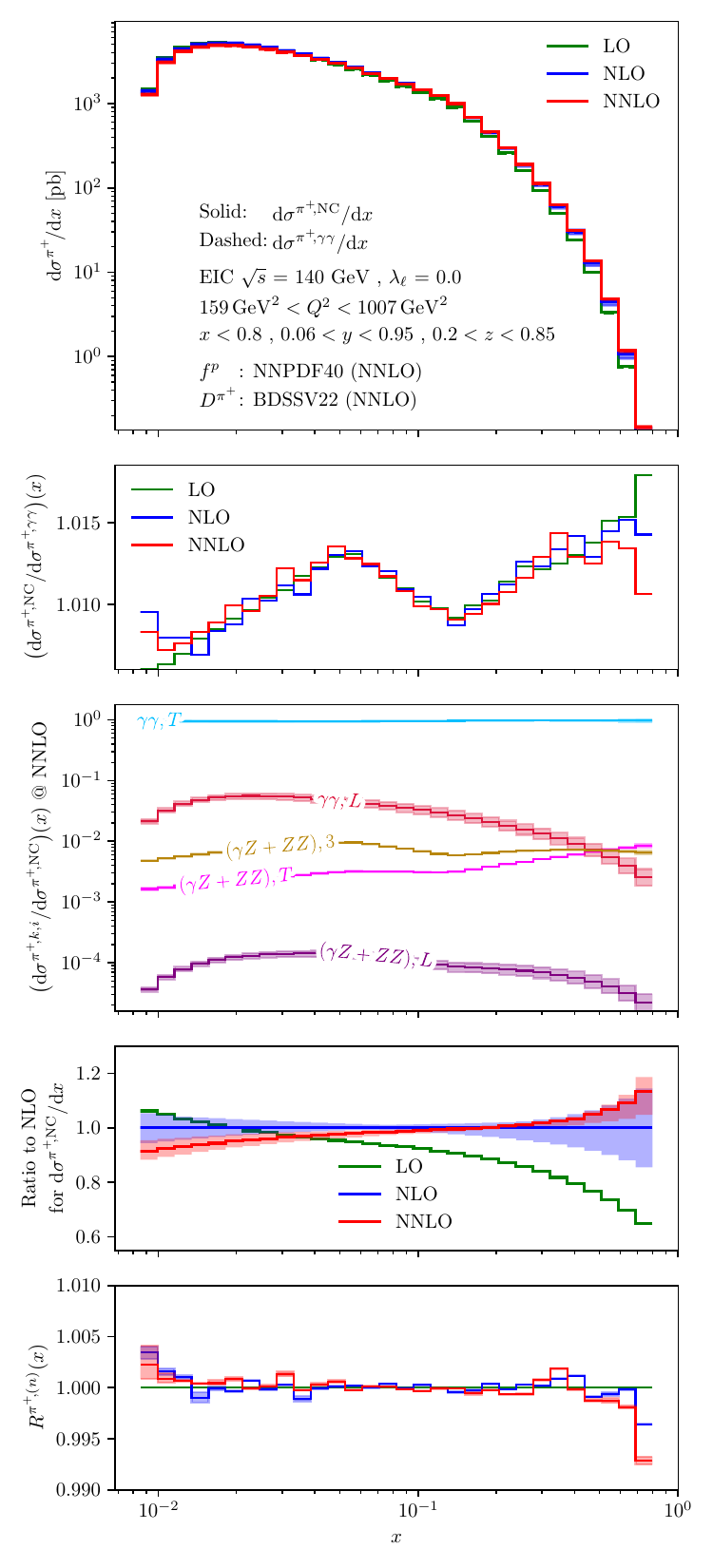}
\caption{Mid-$Q^2$}
\end{subfigure}

\caption{
NC $x$-distributions for $\lambda_\ell$-averaged $\pi^+$ production at Low-$Q^2$ and Mid-$Q^2$.
First panels: total cross section.
Second panels: ratio $(\dd\sigma^{\pi^+,\mathrm{NC}}/\dd\sigma^{\pi^+,\gamma\gamma})(x)$.
Third panels: structure function contributions to the total cross section.
Fourth panels: $\NC$ Ratio to NLO.
Fifth panels: $R^{\pi^+,(n)}(x)$ ratio.
}
\label{fig:X_0.0_NC_lowQ2_pip}

\end{figure}

\begin{figure}[p]
\centering

\begin{subfigure}{0.49\textwidth}
\centering
\includegraphics[width=1.0\linewidth]{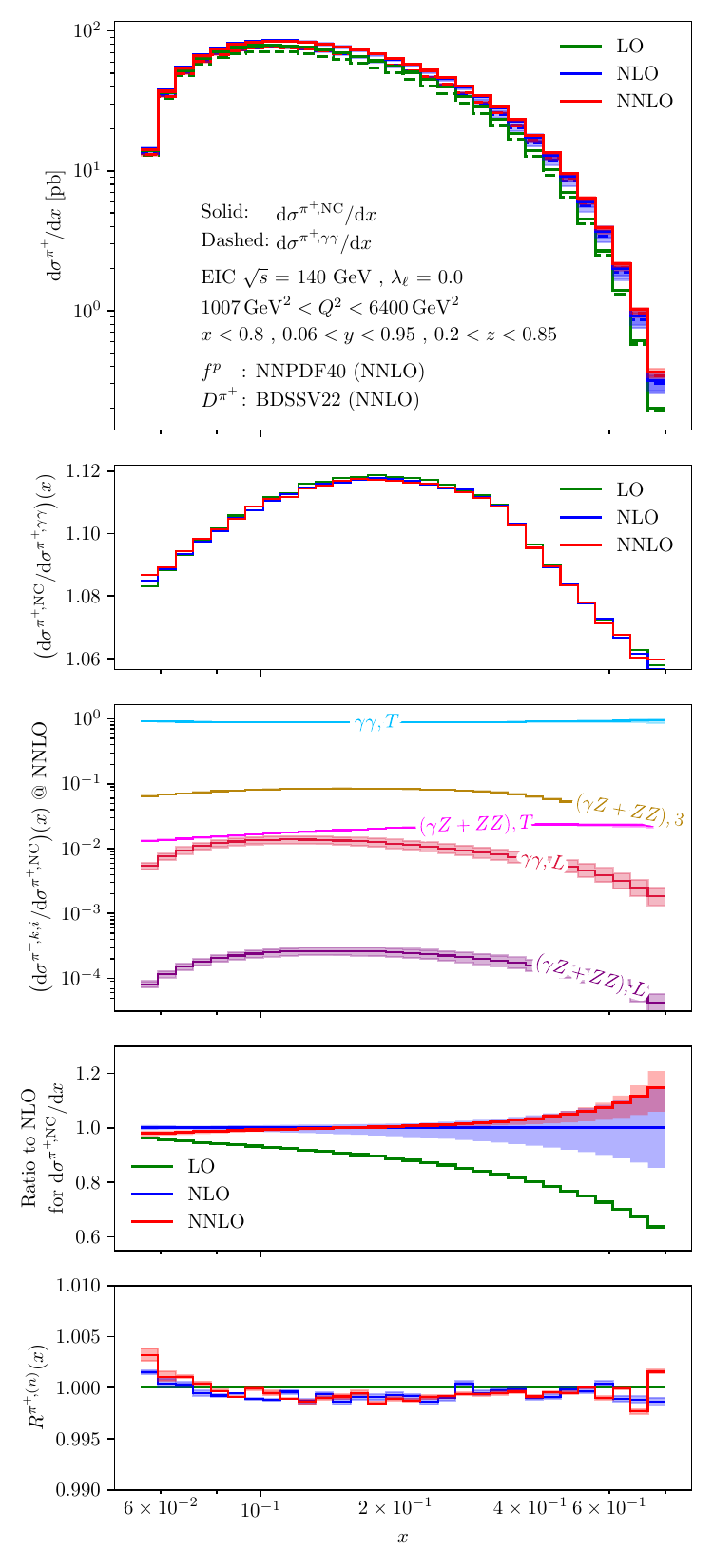}
\caption{High-$Q^2$}
\end{subfigure}
\begin{subfigure}{0.49\textwidth}
\centering
\includegraphics[width=1.0\linewidth]{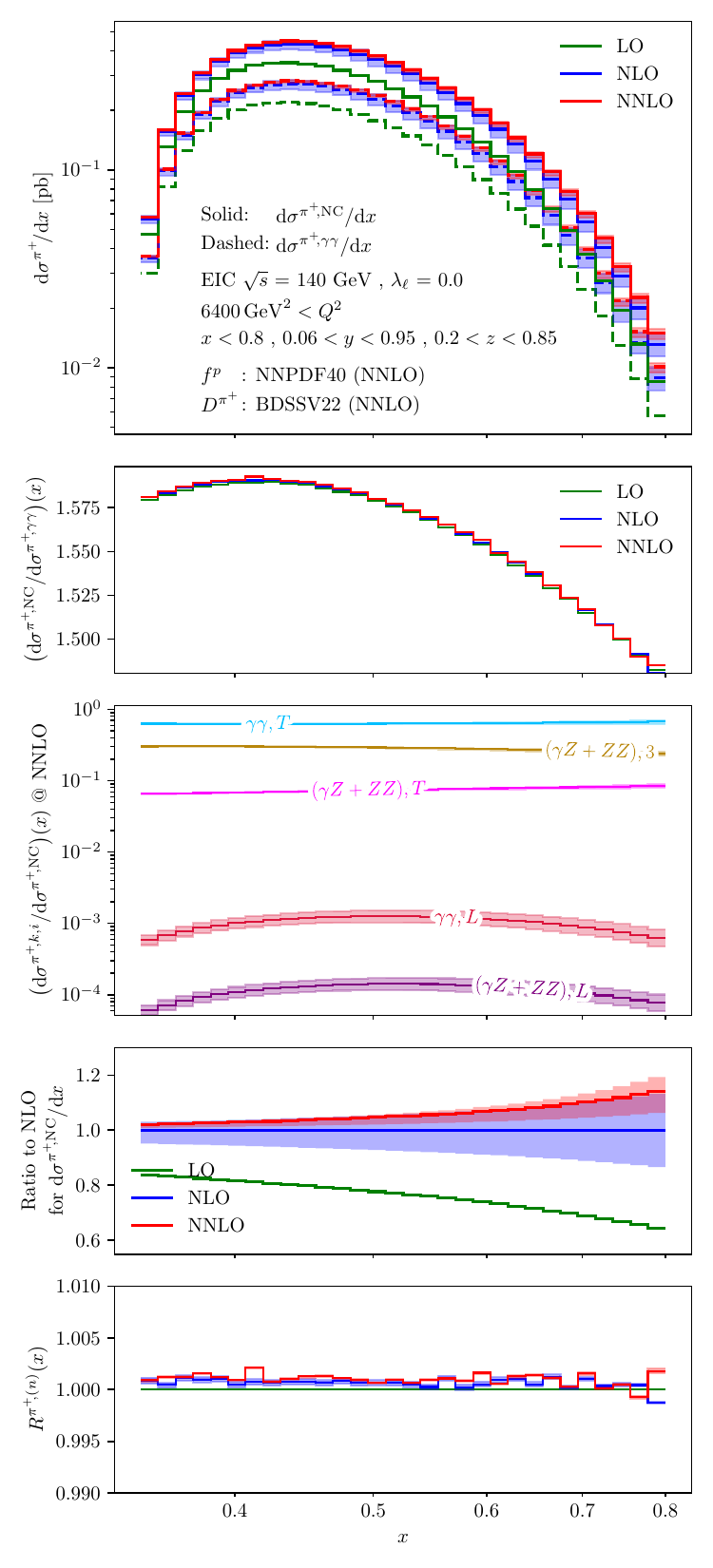}
\caption{Extreme-$Q^2$}
\end{subfigure}

\caption{
NC $x$-distributions for $\lambda_\ell$-averaged $\pi^+$ production at High-$Q^2$ and Extreme-$Q^2$.
First panels: total cross section.
Second panels: ratio $(\dd\sigma^{\pi^+,\mathrm{NC}}/\dd\sigma^{\pi^+,\gamma\gamma})(x)$.
Third panels: structure function contributions to the total cross section.
Fourth panels: $\NC$ Ratio to NLO.
Fifth panels: $R^{\pi^+,(n)}(x)$ ratios.
}
\label{fig:X_0.0_NC_highQ2_pip}

\end{figure}

In Fig.~\ref{fig:X_0.0_NC_lowQ2_pip} and~\ref{fig:X_0.0_NC_highQ2_pip} we display the helicity-averaged NC $x$-differential cross sections $\dd \sigma^{\pi^+}/\dd x$ for the four regions in $Q^2$, cf.\ eq.~\eqref{eq:cutsQ2}.
The range in $x$ is determined by $Q^2=xys$ and restricted to $x<0.8$.
As the range in $y$ spans only one order of magnitude, the $x$ range is greatly affected by the chosen $Q^2$, see Fig.~\ref{fig:ps}.
The $x$-differential cross sections in the different $Q^2$ ranges display a similar shape, although with different allowed intervals in $x$.
The total cross section rises in the first few $x$ bins, then it attains its maximum, after which at first moderately decreases before rapidly falling off.
For Mid-$Q^2$, High-$Q^2$, and Extreme-$Q^2$ an accelerated decline is visible, matching the depletion of the quark PDFs at high $x$.
The differential cross section overall decreases with increasing average $Q^2$ ($Q^2_\mathrm{avg}$), as predicted in the $Q^2$ differential distribution in Fig.~\ref{fig:Q2_0.0_NC_pip}.
At higher $Q^2_\mathrm{avg}$ the cross section falls less rapidly upon increasing $x$ than in the lower $Q^2$ ranges.

The magnitude of the EW effects $(\dd\sigma^{\pi^+,\mathrm{NC}}/\dd\sigma^{\pi^+,\gamma\gamma})(x)$ as presented in the second panels can be approximately determined by $\eta_{\gamma Z}(Q^2_\mathrm{min})$, with $Q^2_\mathrm{min}$ the respective lower bound of $Q^2$.
The functional shape of the ratio $(\dd\sigma^{\pi^+,\mathrm{NC}}/\dd\sigma^{\pi^+,\gamma\gamma})(x)-1$ appears to be moderately dependent on $x$.
In the Low- and Mid-$Q^2$ region, an up-down-up trend is visible in the ratio $(\dd\sigma^{\pi^+,\mathrm{NC}}/\dd\sigma^{\pi^+,\gamma\gamma})(x)$.
This behaviour can be attributed to the $y$-cuts becoming active and discarding events in the different regions, as depicted in Fig.~\ref{fig:ps} on the right: the first incline is due to the upper $y=0.95$ cut, in the central declining region the full range in $y$ contributes, and the subsequent increase is due to the lower cut in $y$ rejecting events.
Modifying the cuts in $y$ also modifies the shape of the $x$-differential cross section, thus the shape is heavily impacted by the imposed fiducial cuts.
In the higher $Q^2$ bins the $x$-dependence of the ratio is more pronounced and peaks in the region where the differential cross section is largest.

We observe a good perturbative convergence, especially for small and moderate values of $x$ above Low-$Q^2$.
In the Low-$Q^2$ region the $\dd\sigma^{\pi^+,\gamma\gamma,L}$ contribution from $\Fcal^{\pi^+}_L$ is the second most dominant, and appears to have a larger relative theoretical uncertainty than the other two structure functions as it vanishes at LO.
The $\dd\sigma^{\pi^+,(\gamma\gamma+\gamma Z)\,3}/\dd x$ contribution from $\Fcal^{\pi^+}_3$ becomes the second largest contribution starting from the High-$Q^2$ region, and has the same order of magnitude as the $\dd\sigma^{\pi^+,(\gamma\gamma+\gamma Z)\,T}/\dd x$ contribution from $\Fcal^{\pi^+}_T$ in the High-$Q^2$ region.

The overall shape and magnitude of the ratio $(\dd\sigma^{\pi^+,\mathrm{NC}}/\dd\sigma^{\pi^+,\gamma\gamma})(x)$ in the Low-$Q^2$ region is in agreement with the inclusive predictions in the top right panel of Fig.~1 in~\cite{Borsa:2022cap}.
Comparing the Low-$Q^2$ and Mid-$Q^2$ regions, it is visible that the cross section in~\cite{Borsa:2022cap} is dominated by the lower $Q^2$ values in the allowed range, except for the highest bins in $x$.

The $R^{\pi^+,(n)}(x)$ ratio plotted in the fifth panel again shows that the NC cross section and error bands are reproduced at the per-mille level by the NC leading order dressed with the K-factors for pure photon exchange.

\begin{figure}[p]
\centering

\begin{subfigure}{0.49\textwidth}
\centering
\includegraphics[width=\linewidth]{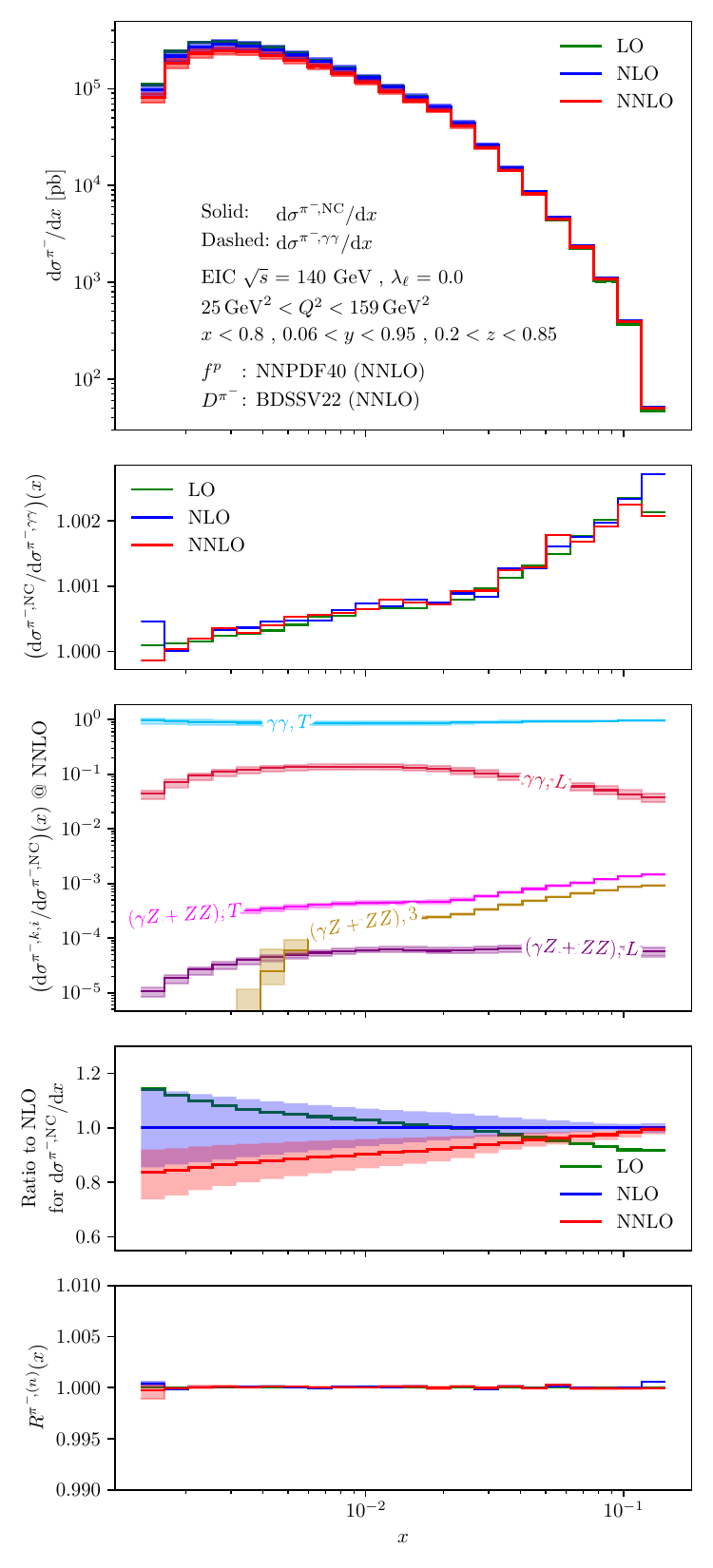}
\caption{Low-$Q^2$}
\end{subfigure}
\begin{subfigure}{0.49\textwidth}
\centering
\includegraphics[width=\linewidth]{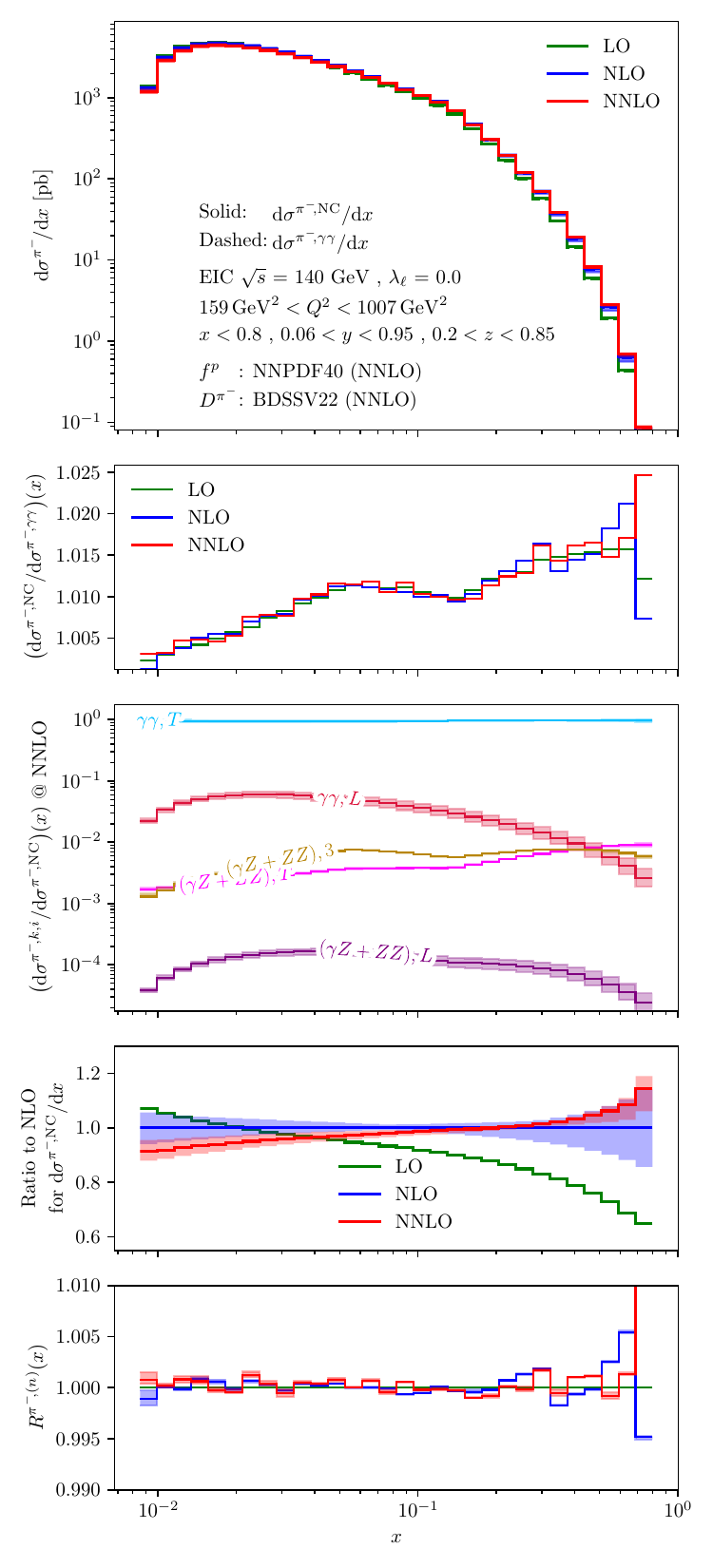}
\caption{Mid-$Q^2$}
\end{subfigure}

\caption{
NC $x$-distributions for $\lambda_\ell$-averaged $\pi^-$ production at low and intermediate $Q^2$.
First panels: total cross section.
Second panels: ratio $(\dd\sigma^{\pi^+,\mathrm{NC}}/\dd\sigma^{\pi^-,\gamma\gamma})(x)$.
Third panels: structure function contributions to the total cross section.
Fourth panels: Ratio to NLO.
Fifth panels: $R^{\pi^-,(n)}(x)$ ratio.
}
\label{fig:X_0.0_NC_lowQ2_pim}

\end{figure}

\begin{figure}[p]
\centering

\begin{subfigure}{0.49\textwidth}
\centering
\includegraphics[width=1.0\linewidth]{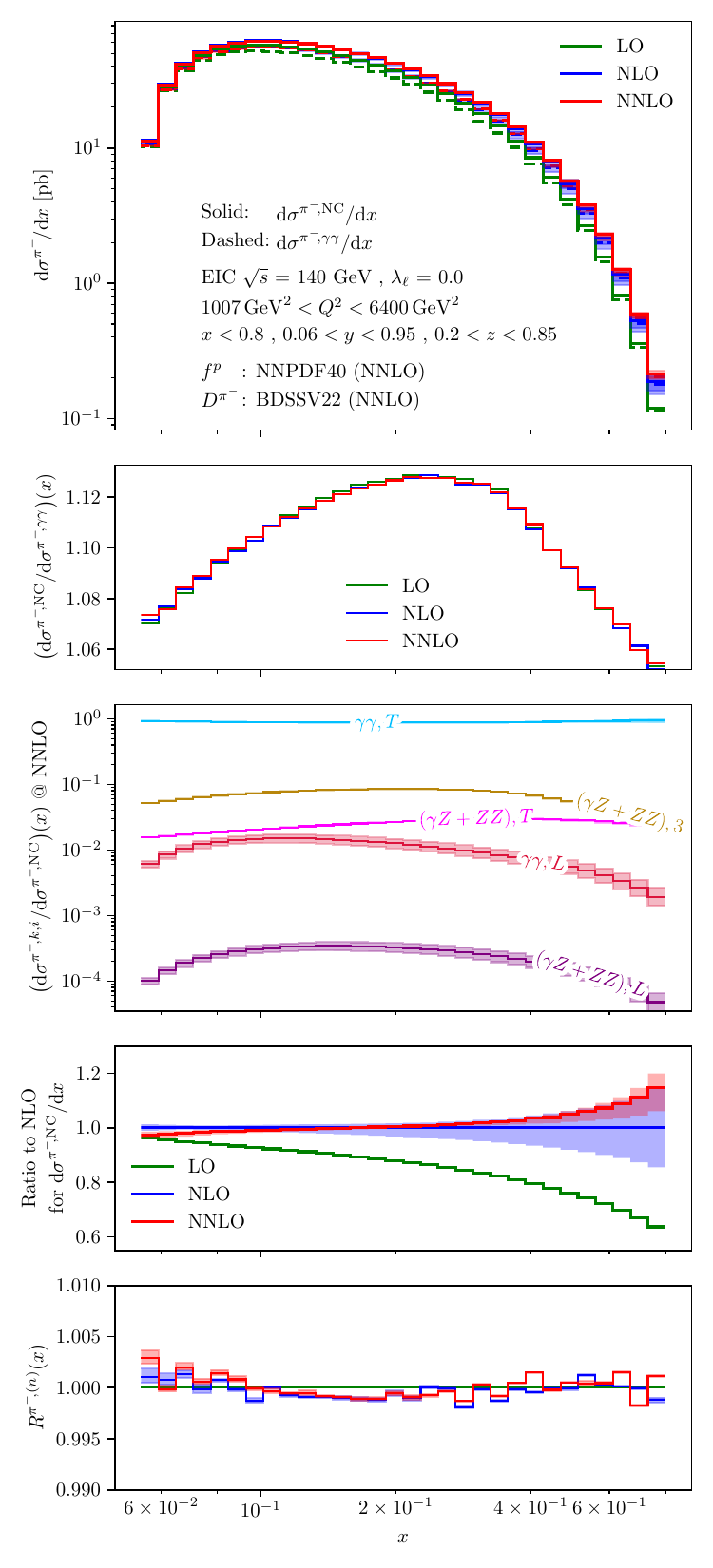}
\caption{High-$Q^2$}
\end{subfigure}
\begin{subfigure}{0.49\textwidth}
\centering
\includegraphics[width=1.0\linewidth]{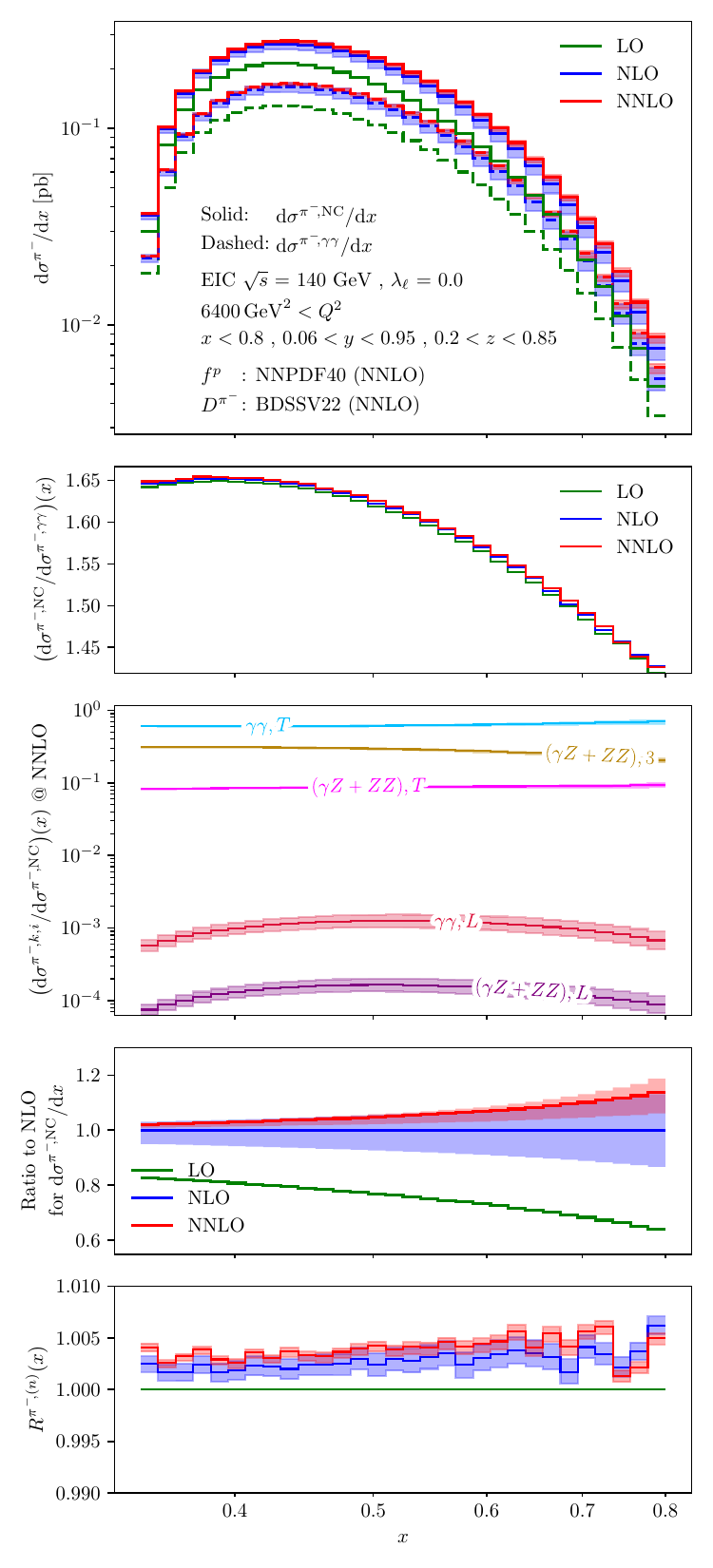}
\caption{Extreme-$Q^2$}
\end{subfigure}

\caption{
NC $x$-distributions for $\lambda_\ell$-averaged $\pi^-$ production at high $Q^2$.
First panels: total cross section.
Second panels: ratio $(\dd\sigma^{\pi^+,\mathrm{NC}}/\dd\sigma^{\pi^-,\gamma\gamma})(x)$.
Third panels: structure function contributions to the total cross section.
Fourth panels: Ratio to NLO.
Fifth panels: $R^{\pi^+,(n)}(x)$ ratios.
}
\label{fig:X_0.0_NC_highQ2_pim}

\end{figure}

For $\pi^-$ production the $x$-differential NC cross sections are shown in Fig.~\ref{fig:X_0.0_NC_lowQ2_pim} and~\ref{fig:X_0.0_NC_highQ2_pim}.
In each of the virtuality ranges the behaviour of the cross section for $\pi^+$ and $\pi^-$ production is very similar.
The structure function decomposition shows however that the contribution $\dd\sigma^{\pi^-,(\gamma Z + ZZ),3}/\dd x$ depletes for $x<0.007$ at Low-$Q^2$, and that the same contribution is slightly smaller than $\dd\sigma^{\pi^+,(\gamma Z + ZZ),3}/\dd x$ at Mid-$Q^2$.
At higher $Q^2$ the $\pi^+$ cross section is overall enhanced by around $50\,\%$ over the $\pi^-$ cross section due to the larger up-quark distribution at the higher $x$ probed at these virtualities.
At Extreme-$Q^2$ the electroweak effects are found to be $10\,\%$ larger for $\pi^-$ than for $\pi^+$.

\begin{figure}[tb]
\centering

\begin{subfigure}{0.45\textwidth}
\centering
\includegraphics[width=1.0\linewidth]{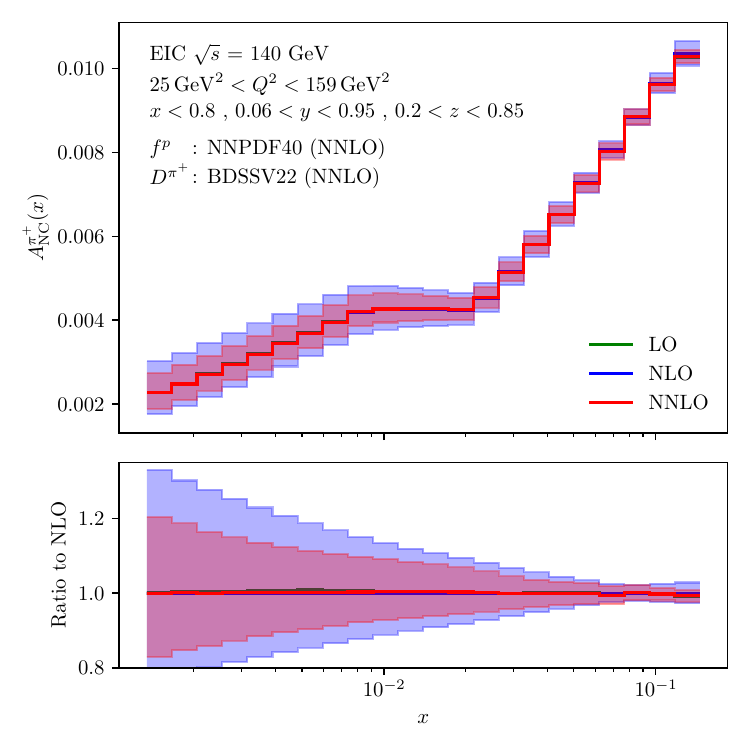}
\caption{Low-$Q^2$}
\end{subfigure}
\begin{subfigure}{0.45\textwidth}
\centering
\includegraphics[width=1.0\linewidth]{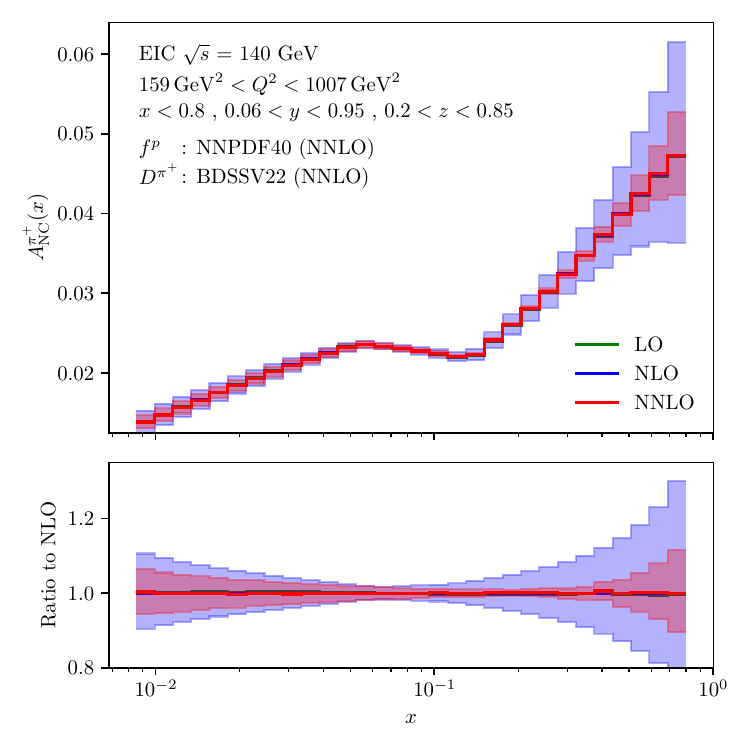}
\caption{Mid-$Q^2$}
\end{subfigure}
\begin{subfigure}{0.45\textwidth}
\centering
\includegraphics[width=1.0\linewidth]{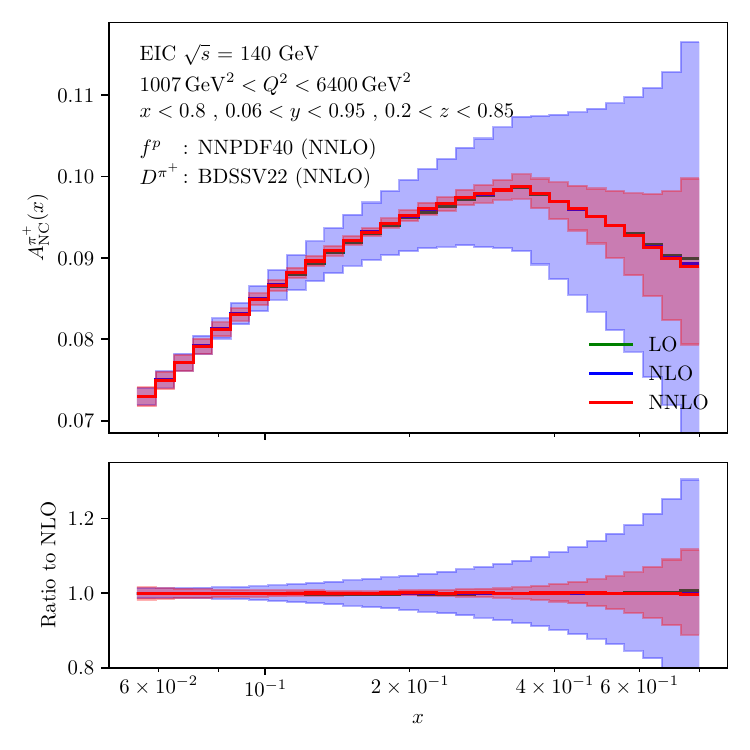}
\caption{High-$Q^2$}
\end{subfigure}
\begin{subfigure}{0.45\textwidth}
\centering
\includegraphics[width=1.0\linewidth]{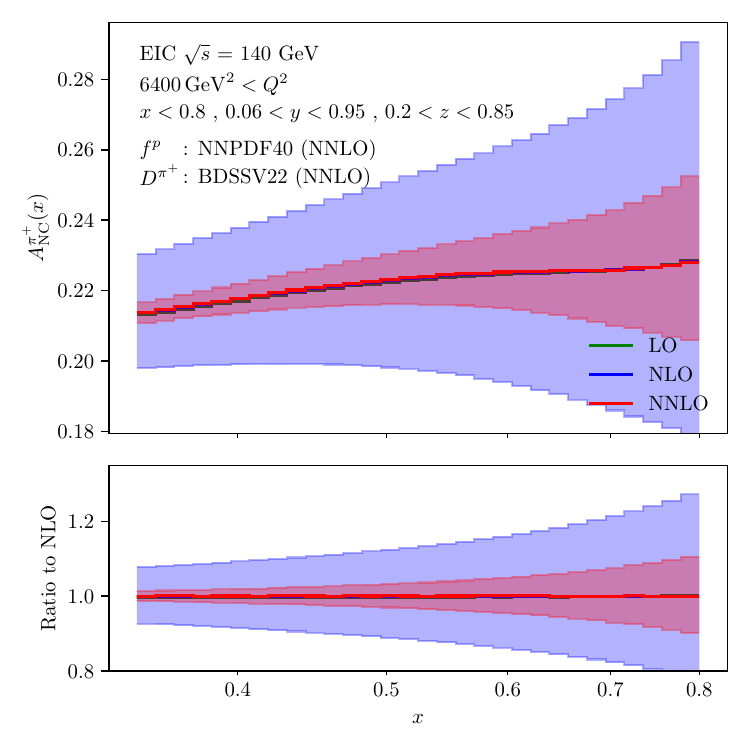}
\caption{Extreme-$Q^2$}
\end{subfigure}

\caption{
NC asymmetry $A_\mathrm{NC}^{\pi^+}$ in $\pi^+$ production as a function of $x$.
Top panels: total cross section.
Bottom panels: Ratio to NLO.
}
\label{fig:X_A_NC_pip}
\end{figure}

\begin{figure}[tb]
\centering

\begin{subfigure}{0.45\textwidth}
\centering
\includegraphics[width=1.0\linewidth]{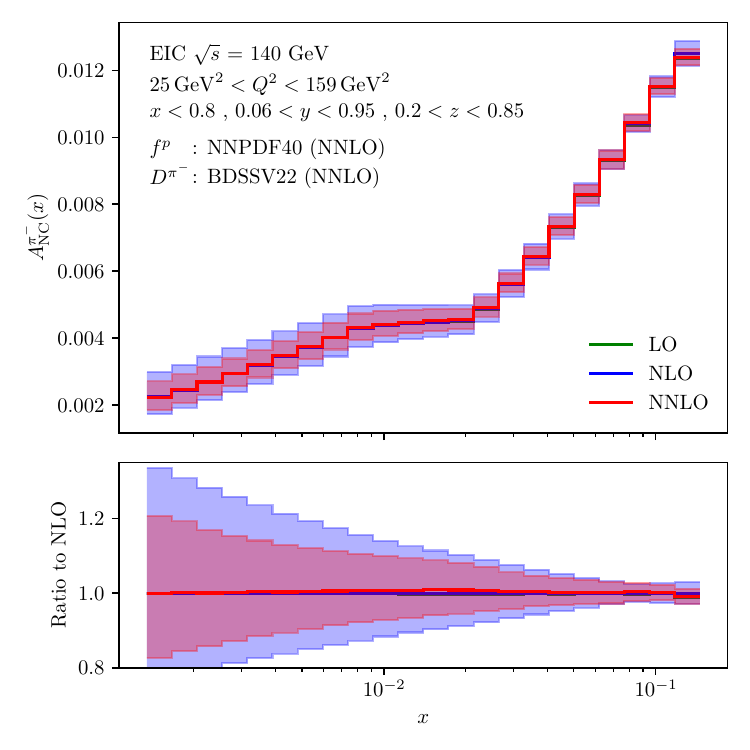}
\caption{Low-$Q^2$}
\end{subfigure}
\begin{subfigure}{0.45\textwidth}
\centering
\includegraphics[width=1.0\linewidth]{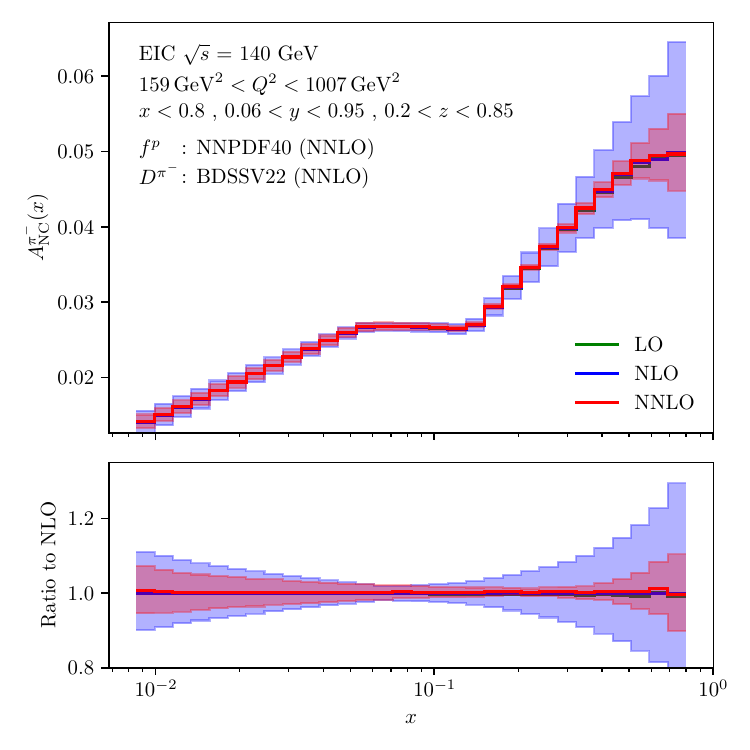}
\caption{Mid-$Q^2$}
\end{subfigure}
\begin{subfigure}{0.45\textwidth}
\centering
\includegraphics[width=1.0\linewidth]{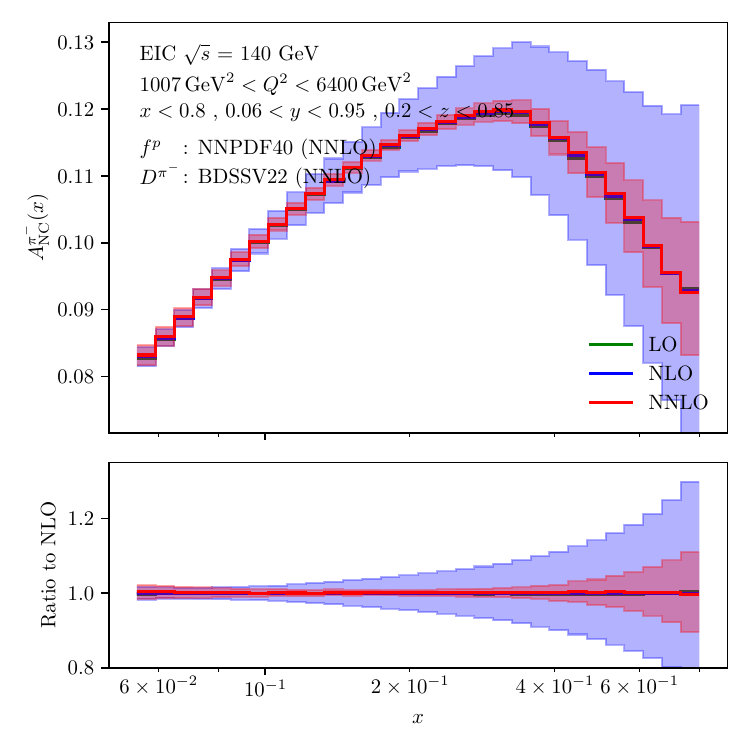}
\caption{High-$Q^2$}
\end{subfigure}
\begin{subfigure}{0.45\textwidth}
\centering
\includegraphics[width=1.0\linewidth]{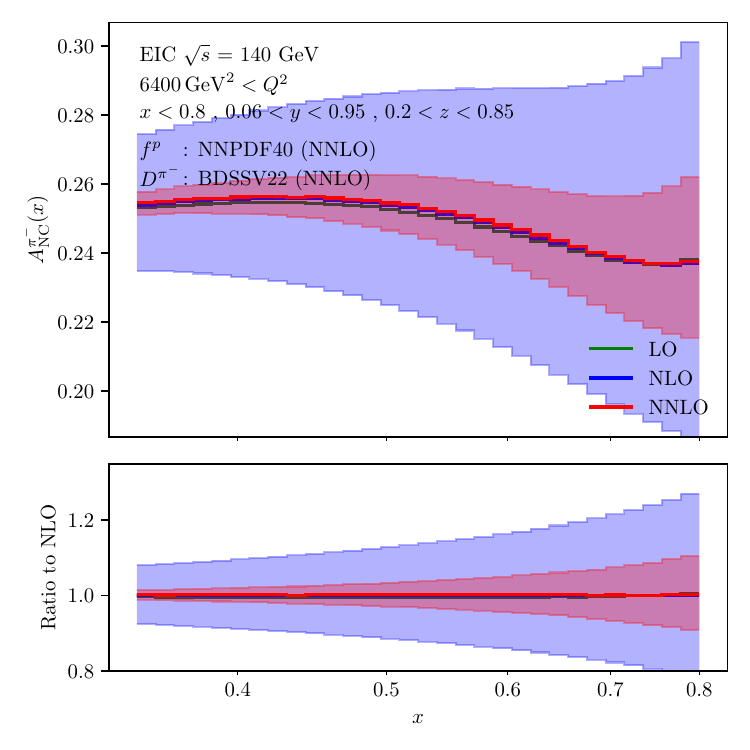}
\caption{Extreme-$Q^2$}
\end{subfigure}

\caption{
NC asymmetry $A_\mathrm{NC}^{\pi^-}$ in $\pi^-$ production as a function of $x$.
Top panels: total cross section.
Bottom panels: Ratio to NLO.
}
\label{fig:X_A_NC_pim}
\end{figure}

In Fig.~\ref{fig:X_A_NC_pip} we show the $x$-dependency of the neutral current asymmetry as defined in eq.~\eqref{eq:A_NC}.
The absolute values of $A^{\pi^+}_\NC(x)$ grow with increasing $Q^2_\mathrm{avg}$.
In the Low-$Q^2$ region the $\lambda_\ell$-odd terms yield sub-percent-level effects in the total cross section.
At Mid-$Q^2$ they become a five-percent-level effect, and increase further in the High-$Q^2$ region to attain around $10\%$.
$A^{\pi^+}_\NC(x)$ displays a moderate growth in $x$, interrupted by a plateau, which shifts towards higher $x$ as
$Q^2$
is increased.
This plateau also has its origin in the $y$-cut and is present in the same range as the downward region in the ratio $(\dd \sigma^{\pi^+,\NC}/\dd \sigma^{\pi^+,\gamma\gamma})(x)$ in Fig.~\ref{fig:X_0.0_NC_lowQ2_pip}.
We observe a significant reduction of scale uncertainties especially upon increasing $Q^2$, and an overall trend for the asymmetry to increase with $x$.
The K-factors for $A^{\pi^+}_\NC(x)$ appear to be very stable in $Q^2$ and $x$ and do not depend on the perturbative order. We observe a substantially  reduced scale uncertainty at NNLO.
In $\pi^-$ production the NC asymmetry $A_\NC^{\pi^-}(x)$ is of similar size and shape, with a tendency of falling off slightly faster with increasing $x$ than for $\pi^+$ in the higher virtuality ranges.
This is shown in Fig.~\ref{fig:X_A_NC_pim}.

\subsection{Charged Current}

Moving to the charged current case, we first consider cross sections for left-handed electrons $\dd\sigma^{\pi^{\pm},\CC} (\lambda_{\ell}=-1)$.
Consequently, only charged current processes with a $W^-$ exchange from the lepton to the quark line are probed.
In order to study the charged current contribution even at low $Q^2$,
we define the \emph{charged-neutral current asymmetry}
\begin{align}\label{eq:A_CCNC}
A^h_\mathrm{CC/NC} = \frac{\dd \sigma^{h,\CC} (\lambda_{\ell}=1) - \dd\sigma^{h,\CC} (\lambda_{\ell}=-1)}{\dd \sigma^{h,\NC} (\lambda_{\ell}=1) - \dd \sigma^{h,\NC} (\lambda_{\ell}=-1)} \, .
\end{align}
The cross section differences in $A^h_\mathrm{CC/NC}$ select the $\lambda_{\ell}\,$-odd components of the differential cross section,
cf.\ eq.~\eqref{eq:xsNCCC}, and remove the dominant photon exchange contribution (which is $\lambda_{\ell}$-even) from the denominator.
Note that always one of the two lepton helicity states in the numerator vanishes in $\CC$ SIDIS.

Furthermore, the differences in $A^h_\mathrm{CC/NC}$ allow for experimental systematic uncertainties to cancel.
Experimentally, charged-lepton--proton scattering $\CC$ events are distinguished from $\NC$ events by the absence of a highly-energetic lepton in the final state.
If the lepton is not detected or misidentified in a $\NC$ interaction it can be mislabelled as a $\CC$ event.
However, the misidentification is agnostic to the helicity state of the incoming lepton.
The helicity difference in the numerator therefore removes the misidentified events which originate from the $\lambda_\ell$-even $\NC$ process,
which constitutes the largest contribution to the scattering cross section.
In addition, in the denominator of $A^h_\mathrm{CC/NC}$ systematic uncertainties related to the lepton beam polarisation cancel.

Similarly to eq.~\eqref{eq:A_NC}, the cross sections entering eq.~\eqref{eq:A_CCNC} are subjected to the kinematic cuts of eq.~\eqref{eq:cutsxyz} and \eqref{eq:cutsQ2}
and theory uncertainties are obtained with a 31-point uncorrelated scale variation.

\subsubsection[\texorpdfstring{$Q^2$-distribution}{\$Q\^2\$-distribution}]{\boldmath $Q^2$-distribution}

\begin{figure}[tb]
\centering
\begin{subfigure}{.45\textwidth}
\centering
\includegraphics[width=\textwidth]{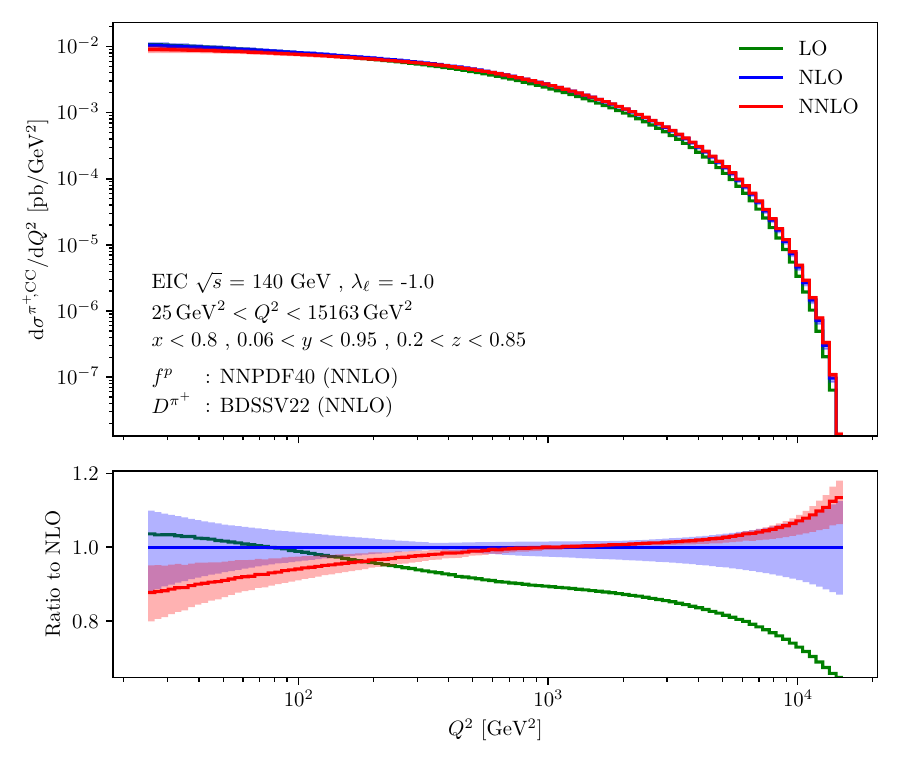}
\end{subfigure}
\begin{subfigure}{.45\textwidth}
\centering
\includegraphics[width=\textwidth]{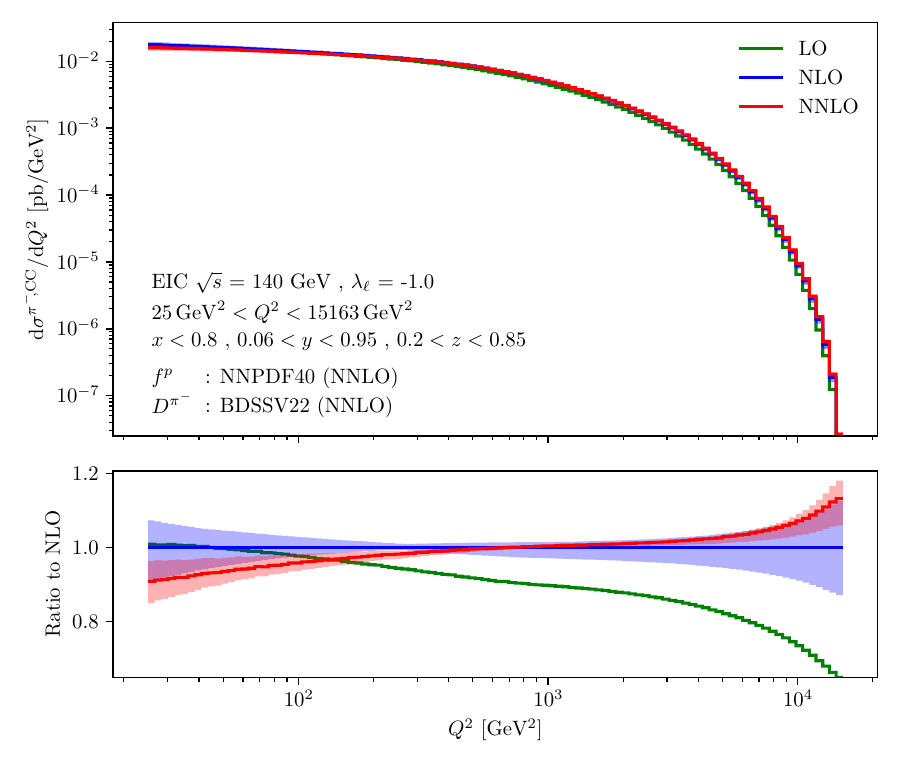}
\end{subfigure}
\caption{
Electron-initiated CC $Q^2$-distribution for $\pi^+$ and $\pi^-$ production.
Top panels: total cross section.
Bottom panels: Ratio to NLO.
}
\label{fig:Q2_CH}
\end{figure}

In Fig.~\ref{fig:Q2_CH} the electron-initiated $Q^2$-differential cross sections for $\pi^\pm$ production in CC SIDIS are shown.
For low virtuality $Q^2$, the differential cross sections decrease only very slowly with increasing $Q^2$,
up to about $Q^2\sim 7000\,\GeV^2$, when the decrease of the PDFs and
the onset of propagator effects starts reducing the cross section more
significantly.
We observe sizeable NLO and NNLO corrections and uncertainty bands at low $Q^2$ and very high $Q^2$.
The relative size of the perturbative corrections is very similar to the NC case, cf. Fig.~\ref{fig:Q2_0.0_NC_pip}.
Together with the NNLO corrections the scale variation bands shrink considerably between  $100 \,\GeV^2 \lesssim Q^2 \lesssim 4000\,\GeV^2$.
At higher virtualities,  the NNLO corrections and associated uncertainties become sizeable again.
The NLO and NNLO predictions display overlapping theory uncertainty bands in the entire kinematic range.
The cross section for the production of a $\pi^-$ in $W^-$ exchange is around $60\,\%$ larger compared to $\pi^+$ production.
The enhancement of $\pi^-$ in $W^-$ exchange is due to the size of the $f_u^p$ distribution, which at LO exclusively transitions into an outgoing $d$ quark.
The $d$ in turn prefers to fragment into a $\pi^-$ over a $\pi^+$ at intermediate and high $z$, reflecting the valence content of the pion.
The production of $\pi^+$ is instead suppressed as the corresponding weak-isospin-conjugated transition $d\to u$ is forbidden at LO and NLO in $W^-$ exchange due to charge conservation.
We will therefore refer to the two production modes as \emph{flavour-favoured} ($\pi^-$) and \emph{flavour-unfavoured} ($\pi^+$).
The perturbative corrections are overall similar for $\pi^+$ and $\pi^-$ production, with slightly smaller perturbative corrections in the latter for $Q^2<100\,\GeV^2$.

\begin{figure}[tb]
\centering
\begin{subfigure}{.45\textwidth}
\centering
\includegraphics[width=\textwidth]{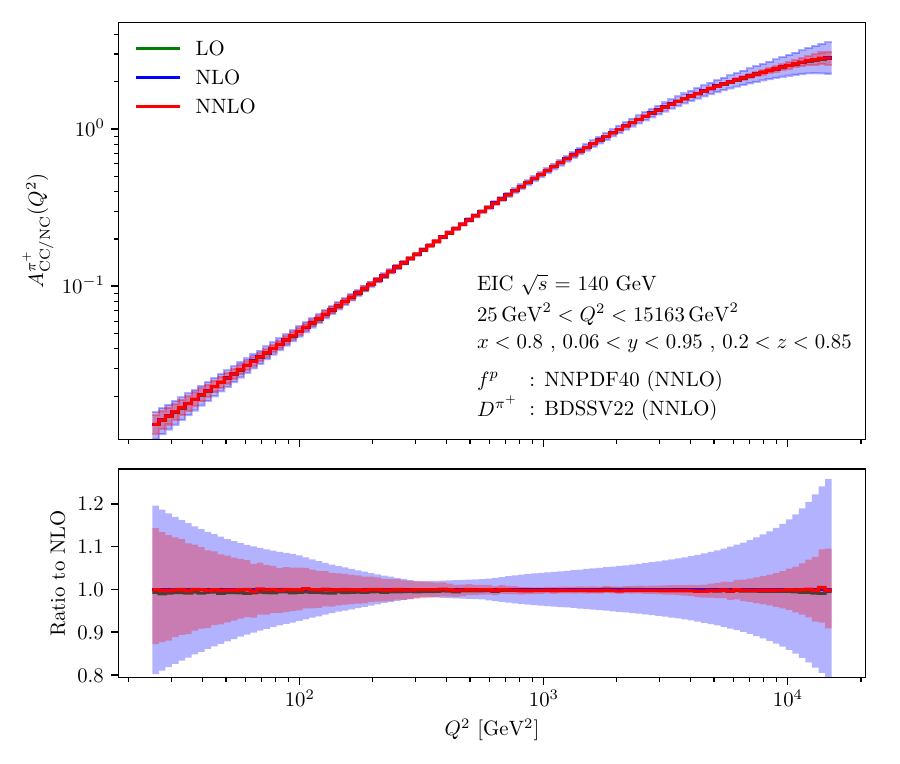}
\end{subfigure}
\begin{subfigure}{.45\textwidth}
\centering
\includegraphics[width=\textwidth]{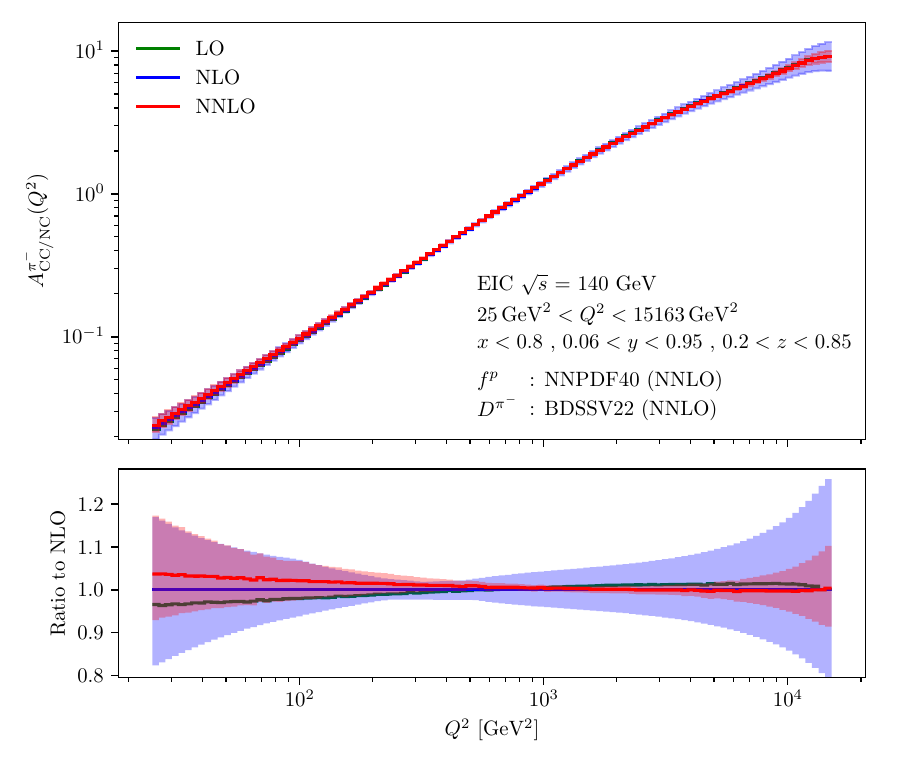}
\end{subfigure}

\caption{
Electron-initiated charged-neutral current asymmetry $A^{\pi^\pm}_\mathrm{CC/NC}$ as a function of $Q^2$ in $\pi^+$ and $\pi^-$ production.
Top panels: total asymmetry.
Bottom panels: Ratio to NLO.
}
\label{fig:Q2_A_CCNC}
\end{figure}

In Fig.~\ref{fig:Q2_A_CCNC} the charged-neutral current asymmetry $A^{\pi^\pm}_\mathrm{CC/NC}(Q^2)$ of eq.~\eqref{eq:A_CCNC} is displayed as function of $Q^2$ for $\pi^+$ and $\pi^-$ production.
A strong dependence on the virtuality $Q^2$ is observed.
Provided a reliable reconstruction of CC events is achieved, the asymmetries are capable of
probing CC contributions in particular at low $Q^2$ to a few percent of the $\lambda_\ell$-odd cross section difference.
In the Mid-$Q^2$ region the asymmetries quickly surpass the level of $10\,\%$.
At higher $Q^2>1000\,\GeV^2$ the asymmetries even exceed the size of the $\lambda_\ell$-odd parts of the NC cross section before levelling off.
Below $Q^2 \sim 2000 \, \GeV^2$ the flavour-favoured asymmetry for $\pi^-$ production is roughly twice the size of its unfavoured ($\pi^+$) counterpart.
Above, the flavour-favoured production mode levels off less quickly, resulting in a three times as large asymmetry at the highest values of $Q^2$.
In $\pi^+$ production the charged-neutral current asymmetry appears to be insensitive to the perturbative order.
This is largely also the case for $\pi^-$, where some minor dependence on the order becomes visible in the Low-$Q^2$ and Extreme-$Q^2$ domains.
In the Mid-$Q^2$ and High-$Q^2$ regions we observe NNLO corrections of below $3\,\%$ to the NLO.
In these regions also the scale uncertainties shrink drastically going from NLO to NNLO.

\subsubsection[\texorpdfstring{$z$-distribution}{z-distribution}]{\boldmath $z$-distribution}

\begin{figure}[tb]
\centering

\begin{subfigure}{0.45\textwidth}
\centering
\includegraphics[width=1.0\linewidth]{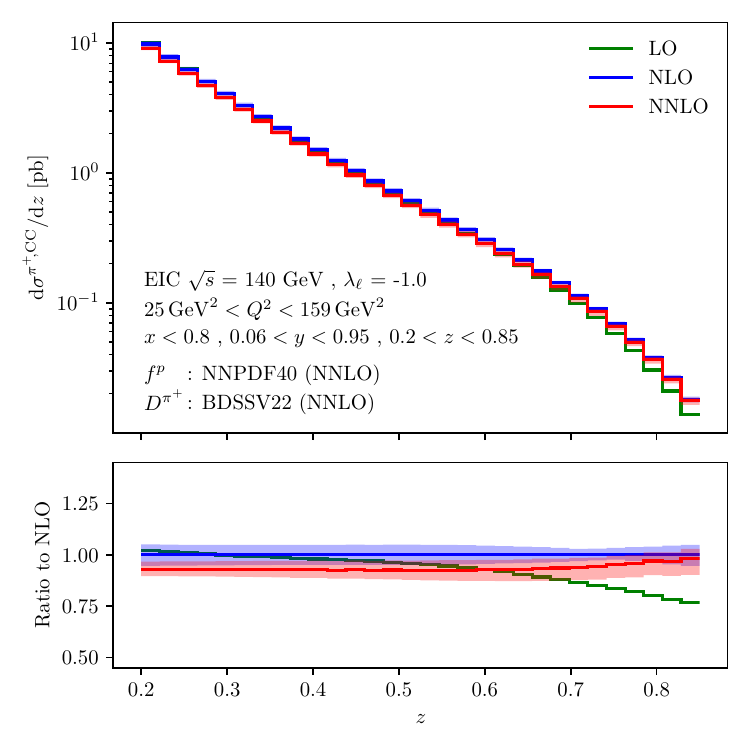}
\caption{Low-$Q^2$}
\end{subfigure}
\begin{subfigure}{0.45\textwidth}
\centering
\includegraphics[width=1.0\linewidth]{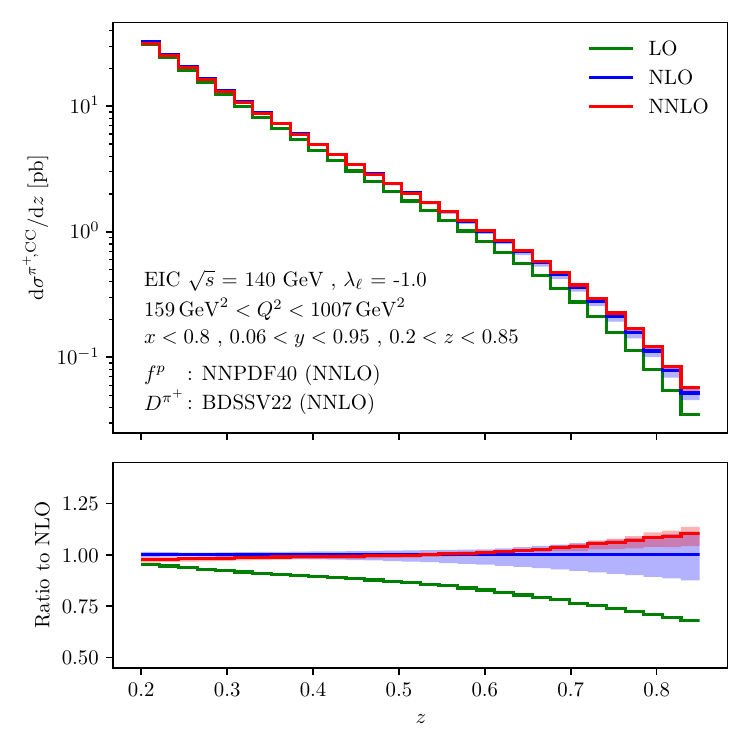}
\caption{Mid-$Q^2$}
\end{subfigure}
\begin{subfigure}{0.45\textwidth}
\centering
\includegraphics[width=1.0\linewidth]{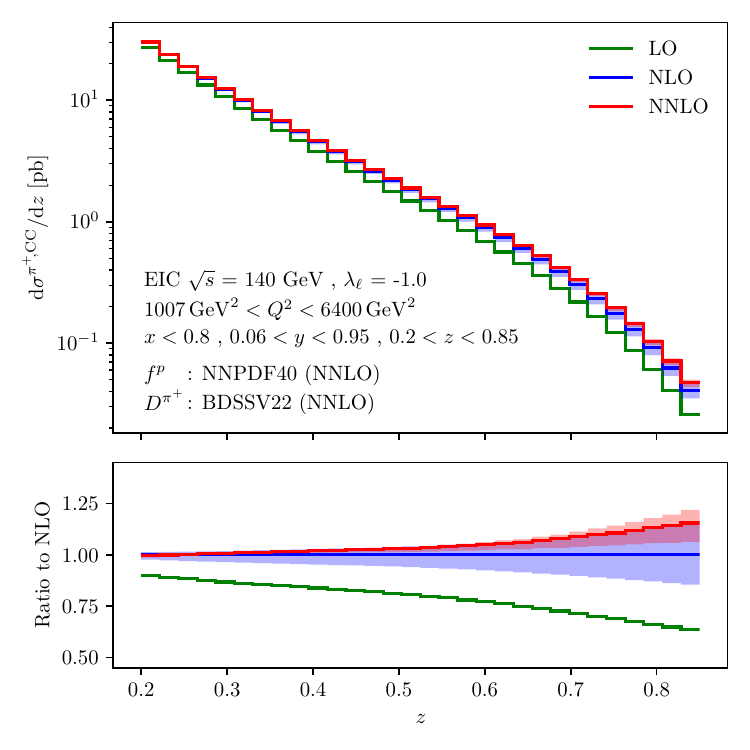}
\caption{High-$Q^2$}
\end{subfigure}
\begin{subfigure}{0.45\textwidth}
\centering
\includegraphics[width=1.0\linewidth]{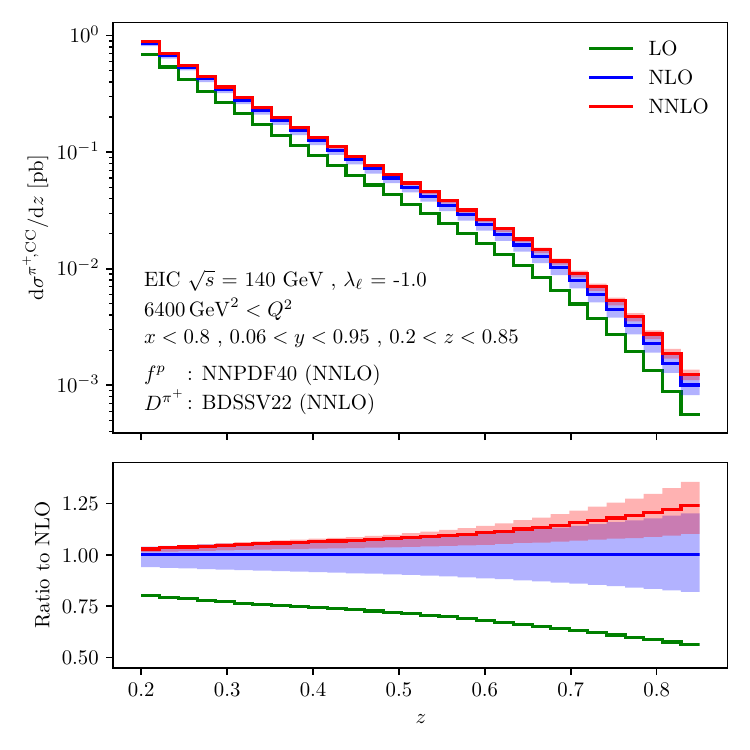}
\caption{Extreme-$Q^2$}
\end{subfigure}

\caption{
Electron-initiated CC $z$-distributions for $\pi^+$ production (unfavoured).
Top panels: total cross section.
Bottom panels: Ratio to NLO.
}
\label{fig:Z_CH_pip}

\end{figure}

\begin{figure}[tb]
\centering

\begin{subfigure}{0.45\textwidth}
\centering
\includegraphics[width=1.0\linewidth]{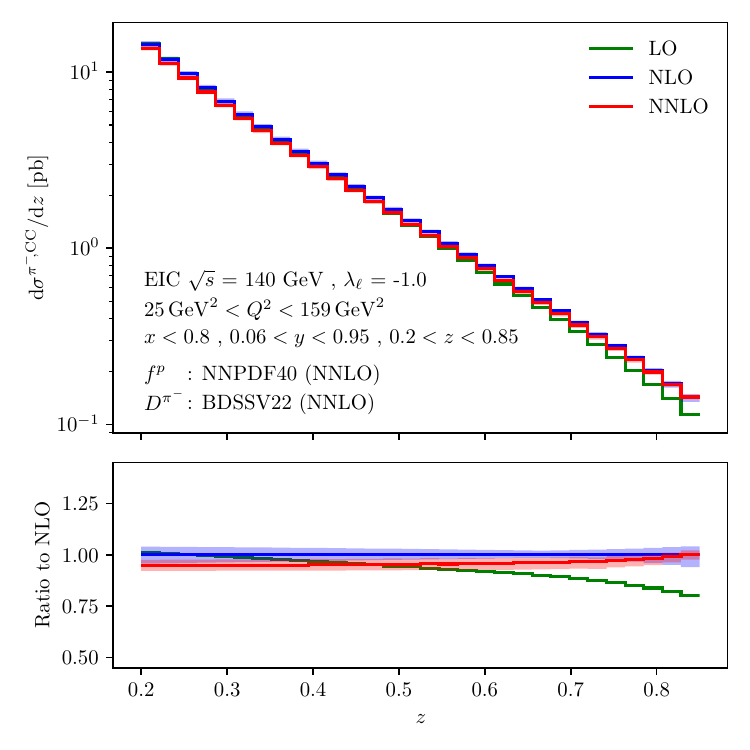}
\caption{Low-$Q^2$}
\end{subfigure}
\begin{subfigure}{0.45\textwidth}
\centering
\includegraphics[width=1.0\linewidth]{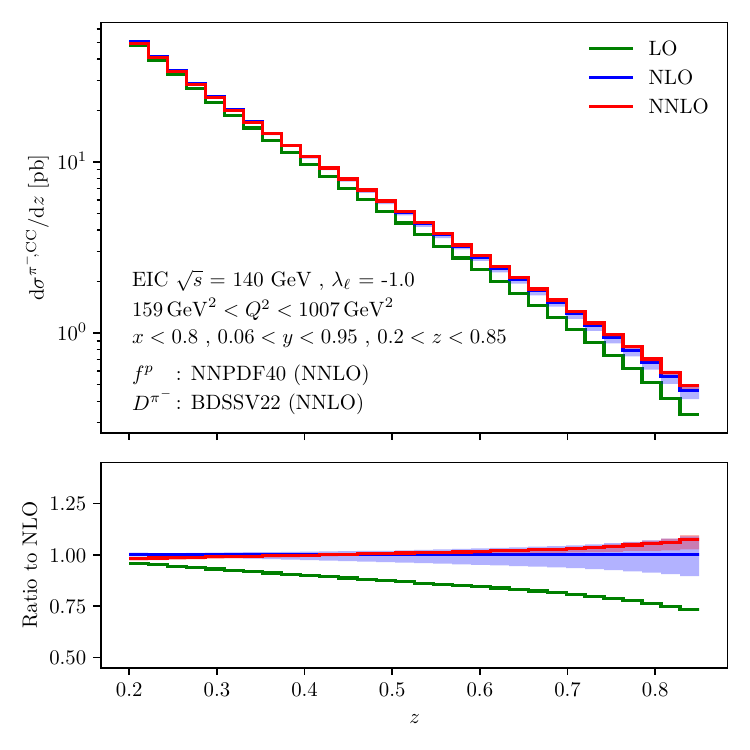}
\caption{Mid-$Q^2$}
\end{subfigure}
\begin{subfigure}{0.45\textwidth}
\centering
\includegraphics[width=1.0\linewidth]{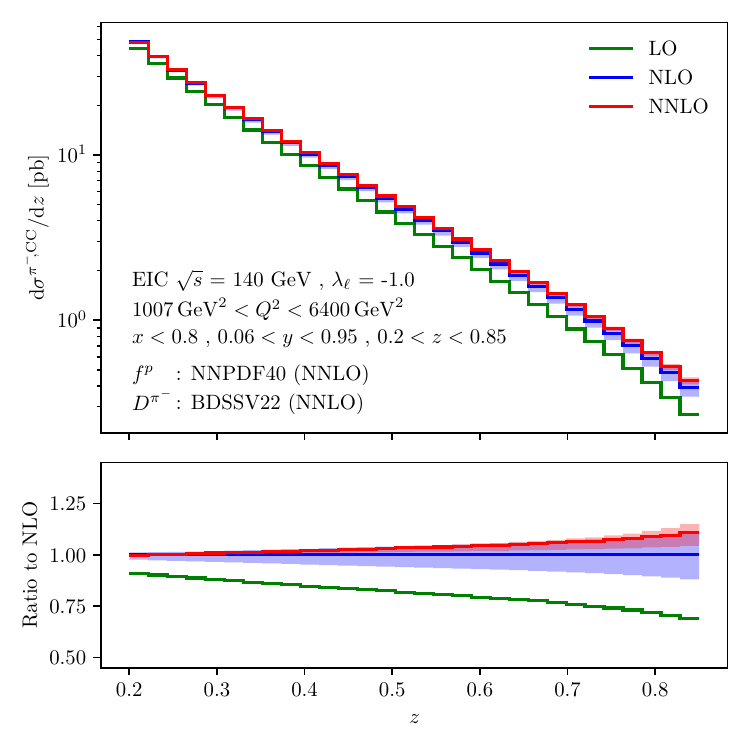}
\caption{High-$Q^2$}
\end{subfigure}
\begin{subfigure}{0.45\textwidth}
\centering
\includegraphics[width=1.0\linewidth]{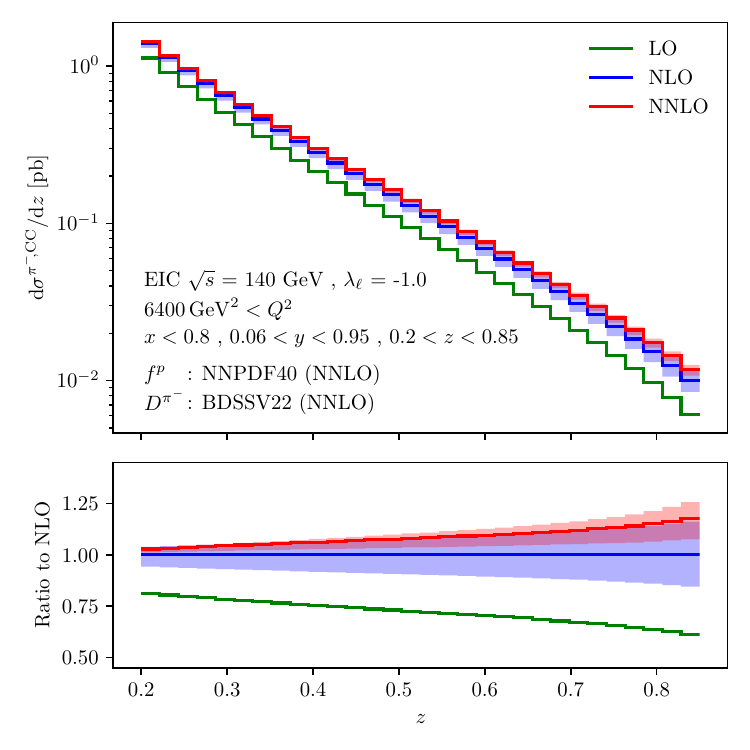}
\caption{Extreme-$Q^2$}
\end{subfigure}

\caption{
Electron-initiated CC $z$-distributions for $\pi^-$ production (favoured).
Top panels: total cross section.
Bottom panels: Ratio to NLO.
}
\label{fig:Z_CH_pim}

\end{figure}

In Fig.~\ref{fig:Z_CH_pip} and~\ref{fig:Z_CH_pim} and the $z$-distributions for the electron-initiated CC production cross section for the four different ranges in $Q^2$, cf.\ eq.~\eqref{eq:cutsQ2}, are shown, for $\pi^+$ and $\pi^-$ respectively.
The $z$-differential cross sections remain stable under the increase of
virtuality, peaking in the intermediate to high range, and subsequently decreasing.
They share a common functional shape, hinting at a low influence of the other kinematic variables on the distributions, which are dominated by the shape of the FFs.
The perturbative corrections increase with the energy, from a few percent in the Low-$Q^2$ bin over $\sim 15\,\%$ in the Mid-$Q^2$ bin to $\sim 25\,\%$ in the High-$Q^2$ bin for $z\sim 0.4$.
The NNLO corrections largely follow the functional shape of the NLO corrections with an offset.
The NNLO corrections become particularly large at high $z$ in the High-$Q^2$ regime, where they start deviating more
strongly from the NLO corrections.
Overlapping error bands throughout hint at a reliable estimate of uncertainties at NLO and
NNLO, and shrinking error bands from the intermediate energy range promise a good convergence of the perturbative series.
Size and
behaviour of the NNLO corrections are
 compatible with previous observations for (anti-)neutrino induced SIDIS in~\cite{Bonino:2025tnf}.

The cross sections are approximately $50\,\%$ larger for $\pi^-$ production than for $\pi^+$ production but have a similar overall shape.
A significant difference between $\pi^+$ and $\pi^-$ production cross sections is only visible in the high-$z$ tails above $\sim 0.5$.
In this region the cross section for flavour-unfavoured
$\pi^+$ production declines more rapidly, despite the perturbative corrections becoming around $15\,\%$ larger than for
flavour-favoured $\pi^-$ production.

\begin{figure}[p]
\centering
\begin{subfigure}{\textwidth}
\centering
\includegraphics[width=\textwidth]{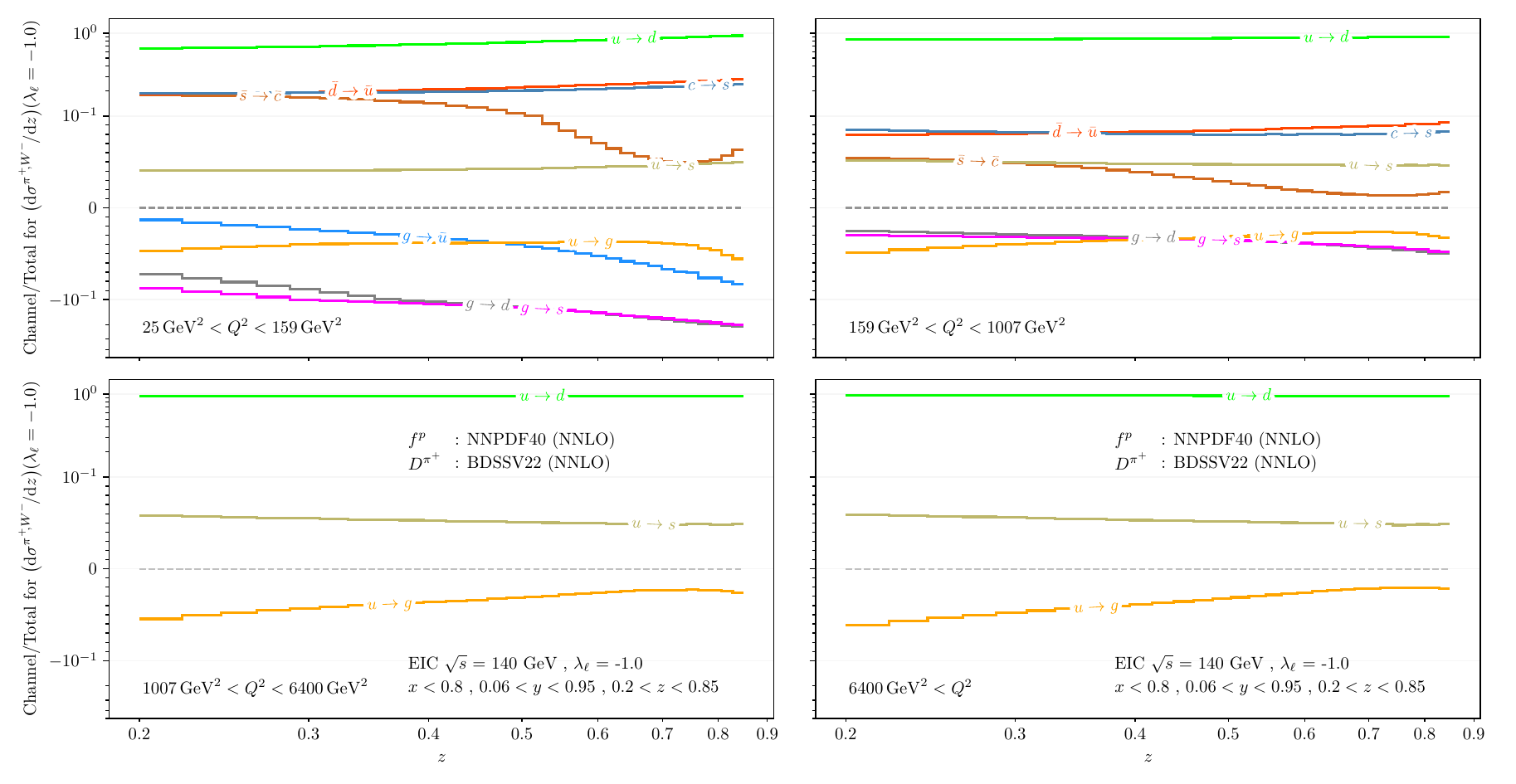}
\caption{Flavour-unfavoured transition $\dd \sigma^{\pi^+, W^-}/\dd z$}
\label{fig:z_CH_channels_pip}
\end{subfigure}
\begin{subfigure}{\textwidth}
\centering
\includegraphics[width=\textwidth]{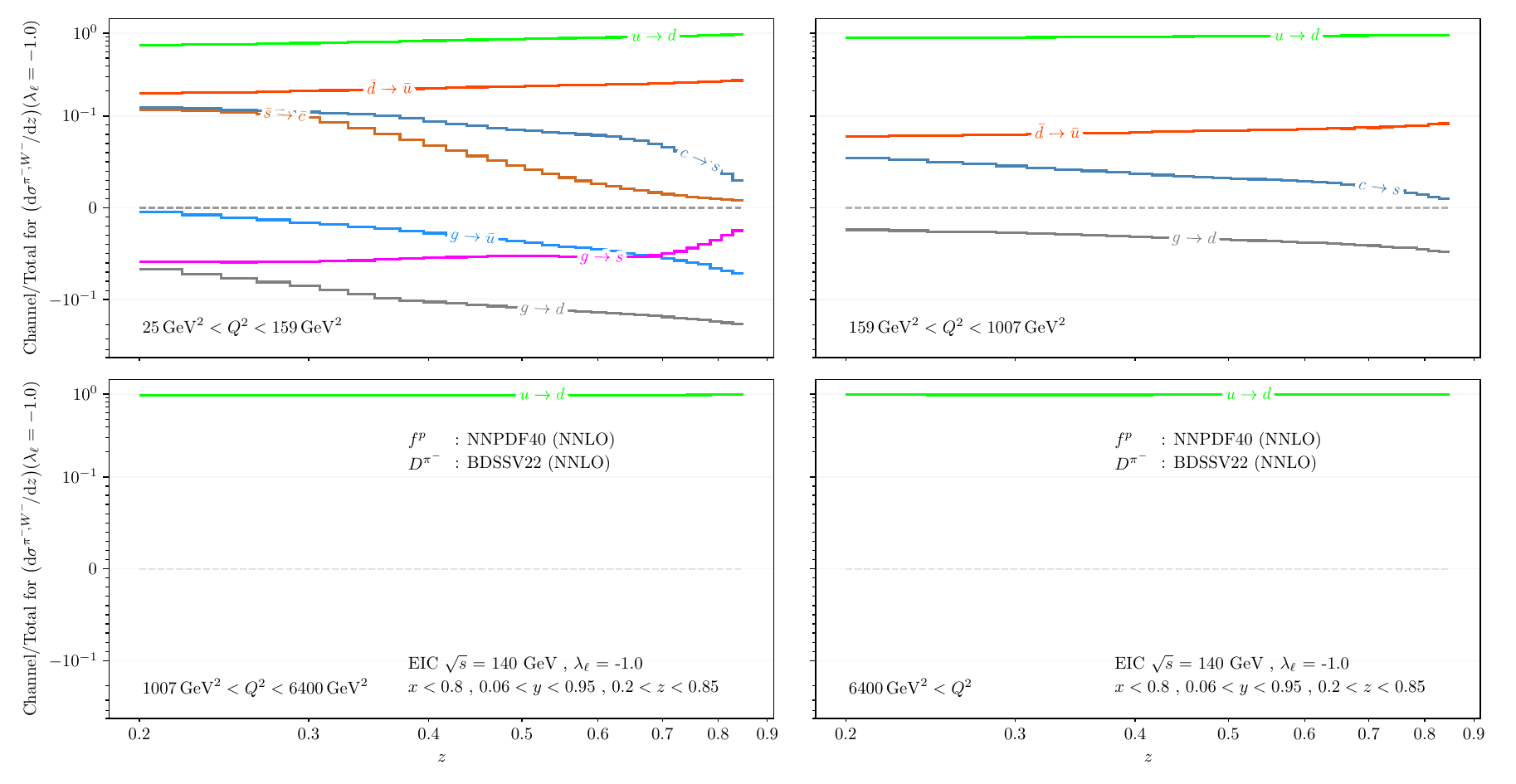}
\caption{Flavour-favoured transition $\dd \sigma^{\pi^-, W^-}/\dd z$}
\label{fig:z_CH_channels_pim}
\end{subfigure}
\caption{
Channel decomposition for electron-initiated $\CC$ $\pi^\pm$ production for $\dd \sigma^{\pi^\pm,W^-}$ as a function of $z$.
Only channels contributing more than $4 \,\%$ in any bin are displayed.
The range $(-0.1, 0.1)$ on the vertical axis is linear, the ranges above and below are plotted logarithmically.
}
\label{fig:z_CH_channels}
\end{figure}

In Fig.~\ref{fig:z_CH_channels} the channel decomposition for $\pi^\pm$ production in $\CC$ SIDIS is shown.
The flavour-favoured $\CC$ $\pi^-$ production may be compared to the $\NC$ $\pi^+$ production, cf.\ Fig.~\ref{fig:z_NC_channel_pip}.
We observe similarities with respect to the $\NC$ channels with initial state up-type quarks, down-type antiquarks and gluons.
The corresponding final state quarks are the weak isospin conjugates of the $\NC$ final state quarks.
The channels corresponding to up-to-up-type and down-to-down-type quark transitions are not present in $\CC$ up to NLO. At NNLO none of these channels contributes more than $4\,\%$ to the total cross section.
The same correspondence is also observed between the flavour-unfavoured $\pi^+$ production mode in $\CC$ SIDIS and $\pi^-$ production in $\NC$ SIDIS.

\begin{figure}[tb]
\centering

\begin{subfigure}{0.45\textwidth}
\centering
\includegraphics[width=1.0\linewidth]{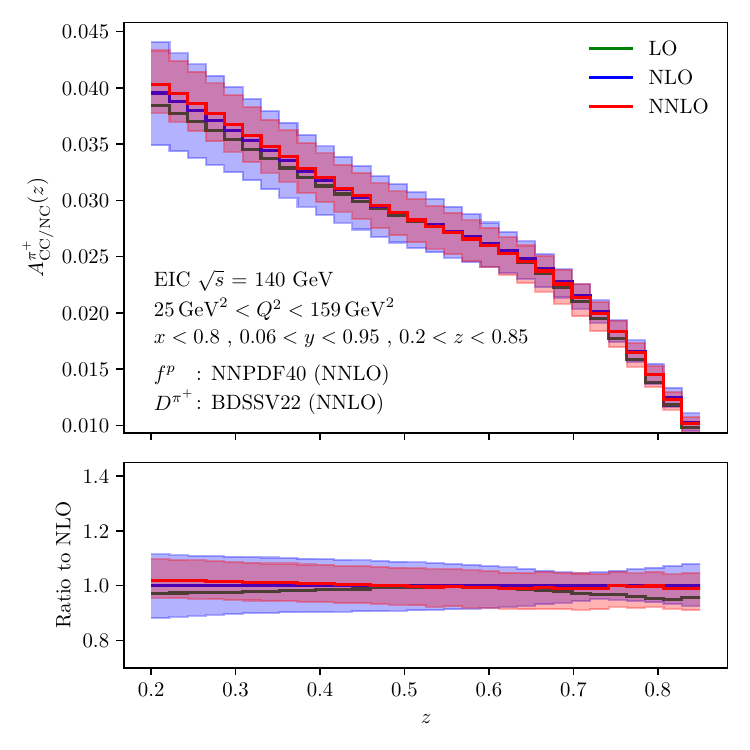}
\caption{Low-$Q^2$}
\end{subfigure}
\begin{subfigure}{0.45\textwidth}
\centering
\includegraphics[width=1.0\linewidth]{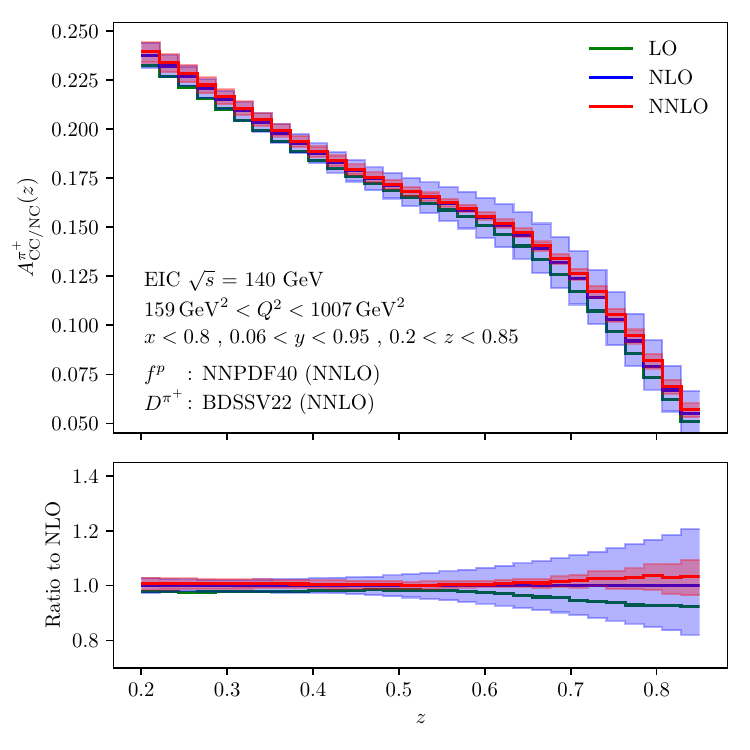}
\caption{Mid-$Q^2$}
\end{subfigure}
\begin{subfigure}{0.45\textwidth}
\centering
\includegraphics[width=.993\linewidth]{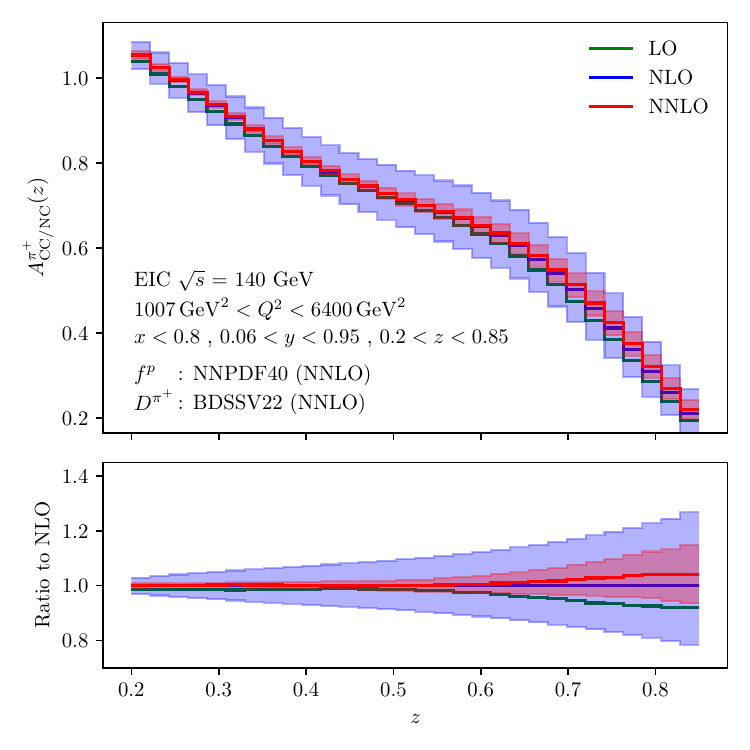}
\caption{High-$Q^2$}
\end{subfigure}
\begin{subfigure}{0.45\textwidth}
\centering
\includegraphics[width=1.0\linewidth]{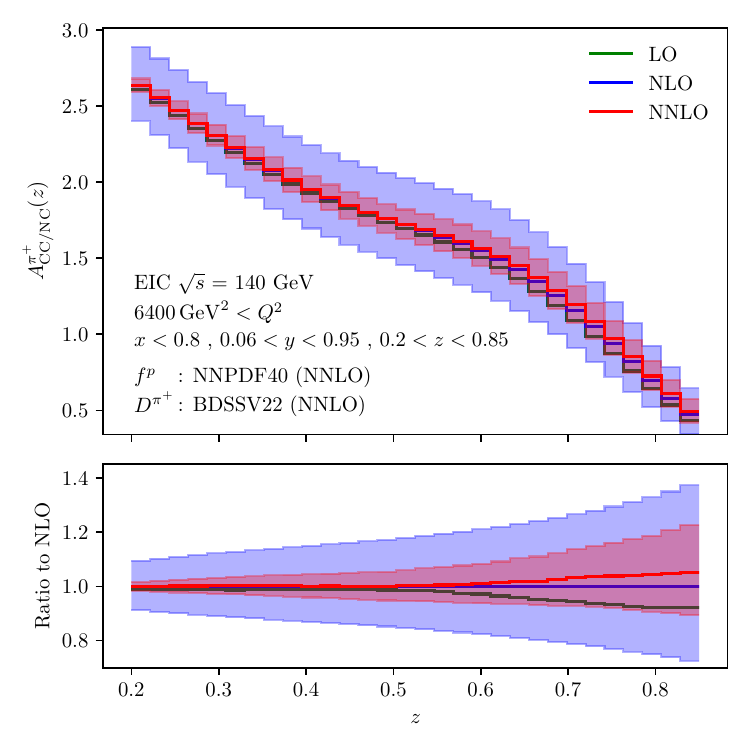}
\caption{Extreme-$Q^2$}
\end{subfigure}

\caption{
Electron-initiated asymmetry $A^{\pi^+}_\mathrm{CC/NC}$  in $\pi^+$ production as a function of $z$.
Top panels: total asymmetry.
Bottom panels: Ratio to NLO.
}
\label{fig:Z_A_CCNC_pip}

\end{figure}

\begin{figure}[tb]
\centering

\begin{subfigure}{0.45\textwidth}
\centering
\includegraphics[width=1.0\linewidth]{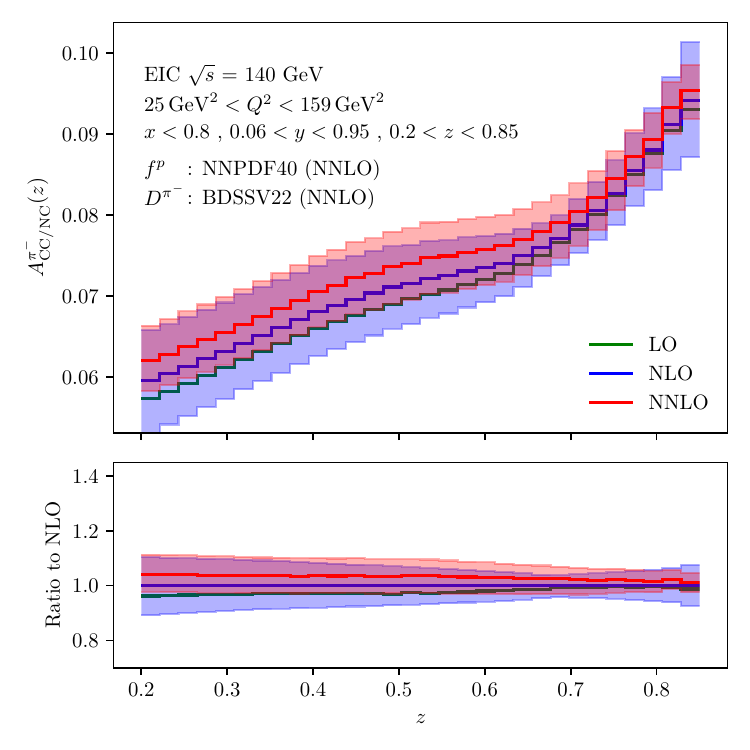}
\caption{Low-$Q^2$}
\end{subfigure}
\begin{subfigure}{0.45\textwidth}
\centering
\includegraphics[width=1.0\linewidth]{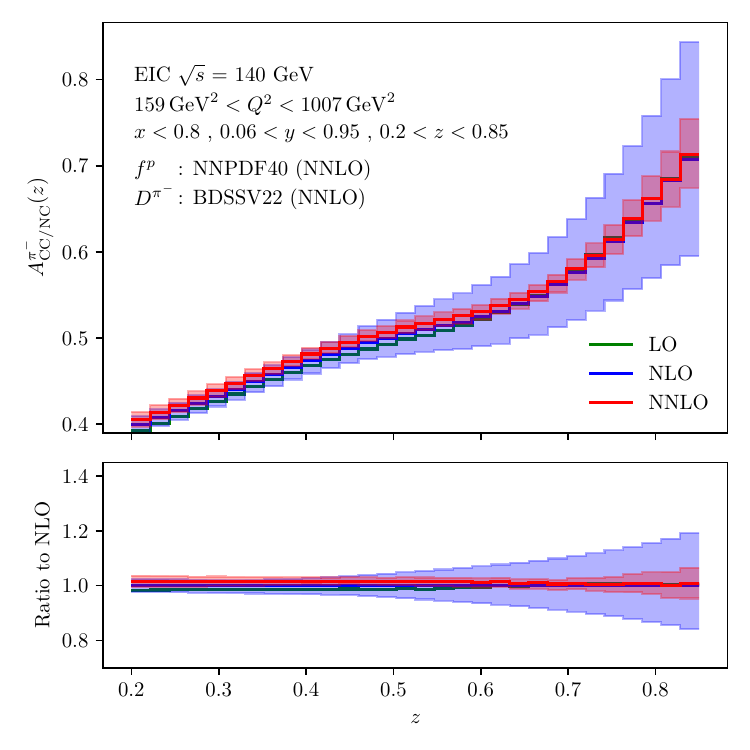}
\caption{Mid-$Q^2$}
\end{subfigure}
\begin{subfigure}{0.45\textwidth}
\centering
\includegraphics[width=.995\linewidth]{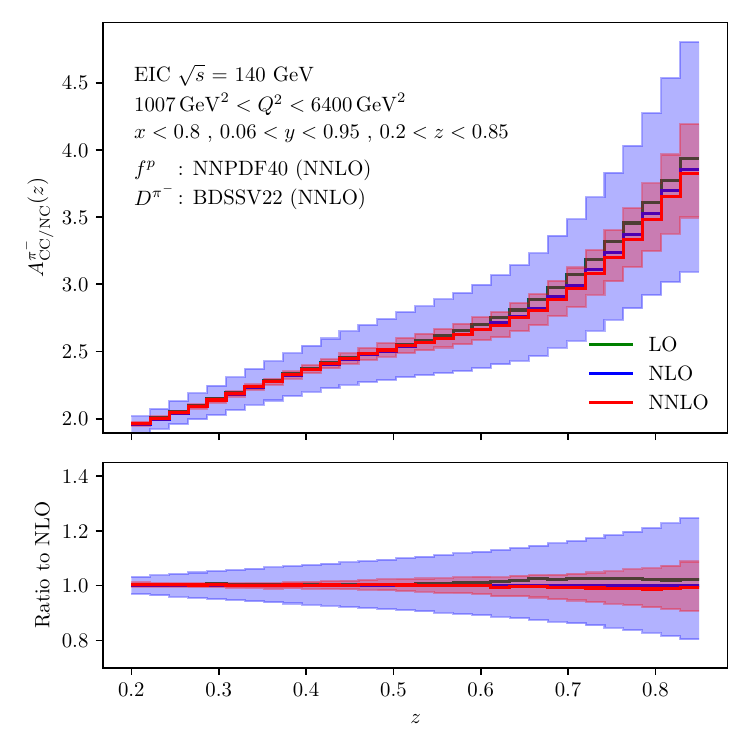}
\caption{High-$Q^2$}
\end{subfigure}
\begin{subfigure}{0.45\textwidth}
\centering
\includegraphics[width=1.0\linewidth]{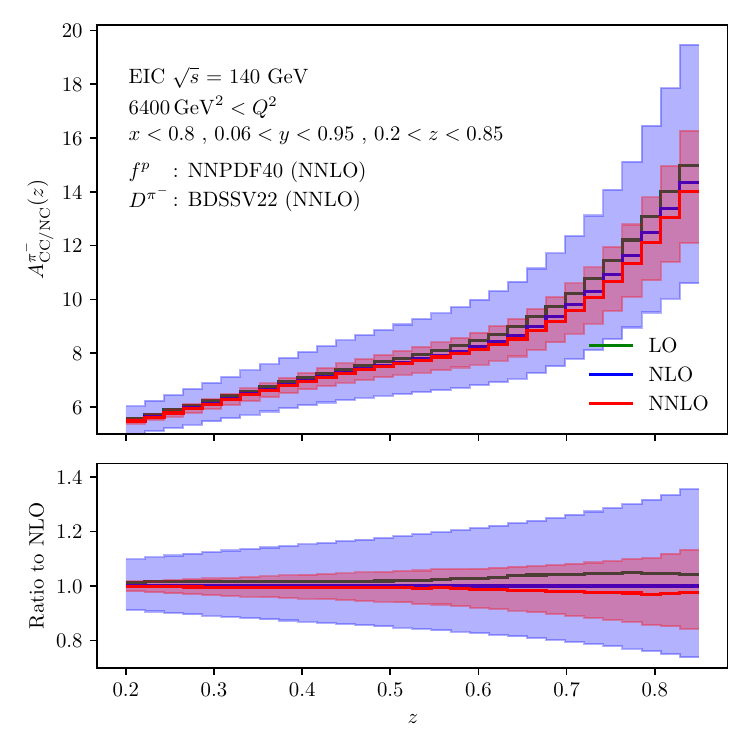}
\caption{Extreme-$Q^2$}
\end{subfigure}

\caption{
Electron-initiated asymmetry $A^{\pi^-}_\mathrm{CC/NC}$ in $\pi^-$ production as a function of $z$.
Top panels: total asymmetry.
Bottom panels: Ratio to NLO.
}
\label{fig:Z_A_CCNC_pim}

\end{figure}

In Fig.~\ref{fig:Z_A_CCNC_pip} and~\ref{fig:Z_A_CCNC_pim} the asymmetries $A^{\pi^{\pm}}_\mathrm{CC/NC}$
are displayed
as function of $z$ for four ranges in $Q^2$.
We observe a mild dependency on the perturbative order with up to $10\, \%$ corrections at NNLO in the high-$z$ region.
Below $z<0.6$ the perturbative corrections are quite moderate.
In the intermediate and high $Q^2$ regions the large-$z$ region is enhanced due to the emission of soft gluons, potentially requiring resummation.

Especially in the intermediate and high $Q^2$ region the $A^{\pi^{\pm}}_\mathrm{CC/NC}(z)$ asymmetry is of comparable size to the $\lambda_\ell$-odd cross section, which will enable constraints on flavour-off-diagonal combinations between PDFs and FFs in this region.

The $A_{\CC/\NC}^{\pi^-}(z)$ asymmetry for the favoured production mode is around $50\,\%$ larger for $z\simeq 0.2$ than $A_{\CC/\NC}^{\pi^+}(z)$.
Up to $z\sim 0.55$ we witness a moderate rise of $A_{\CC/\NC}^{\pi^-}(z)$, before it starts growing quickly towards $z\sim 0.85$.
In contrast to the unfavoured production mode $A_{\CC/\NC}^{\pi^+}(z)$ at first decreases moderately up to $z\sim 0.5$ before entering a rapid decline.
These trends are visible in each of the $Q^2$ regions, but become more pronounced with increasing $Q^2$.
The $K$-factors are mostly flat, with deviations commonly only appearing in the regions of strong rise or slumps for $z\gtrsim 0.5$.

\subsubsection[\texorpdfstring{$x$-distributions}{x-distributions}]{\boldmath $x$-distributions}

\begin{figure}[tbp]
\centering

\begin{subfigure}{0.45\textwidth}
\centering
\includegraphics[width=1.0\linewidth]{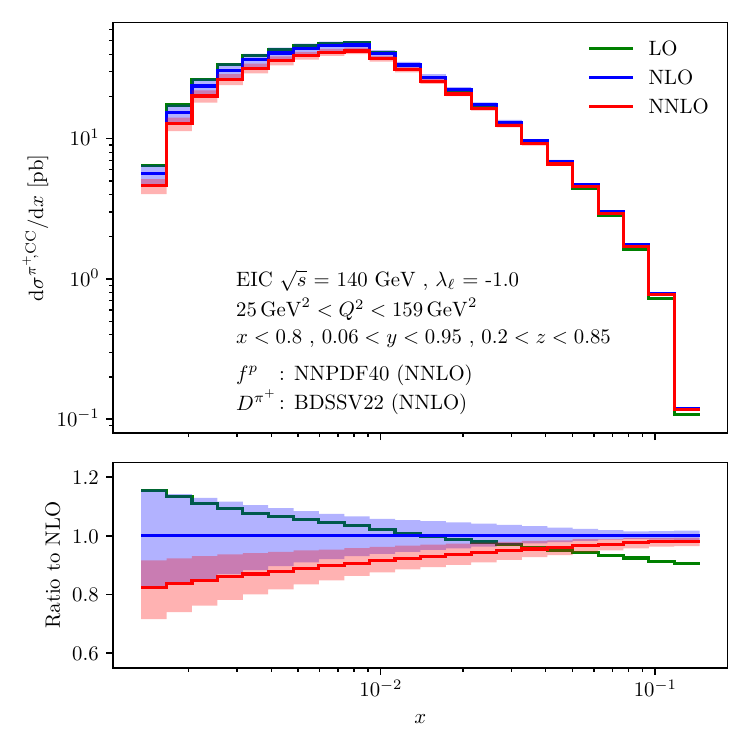}
\caption{Low-$Q^2$}
\end{subfigure}
\begin{subfigure}{0.45\textwidth}
\centering
\includegraphics[width=1.0\linewidth]{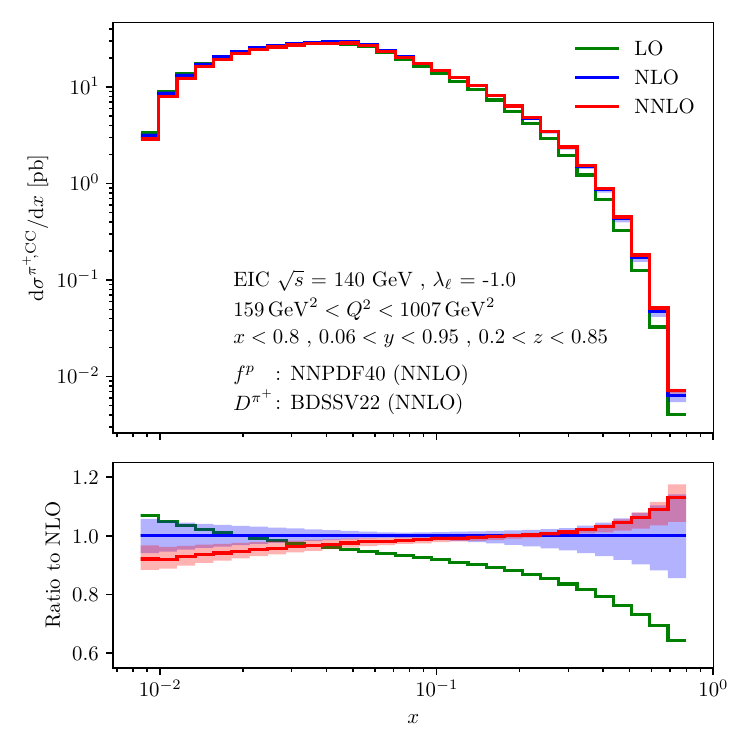}
\caption{Mid-$Q^2$}
\end{subfigure}
\begin{subfigure}{0.45\textwidth}
\centering
\includegraphics[width=1.0\linewidth]{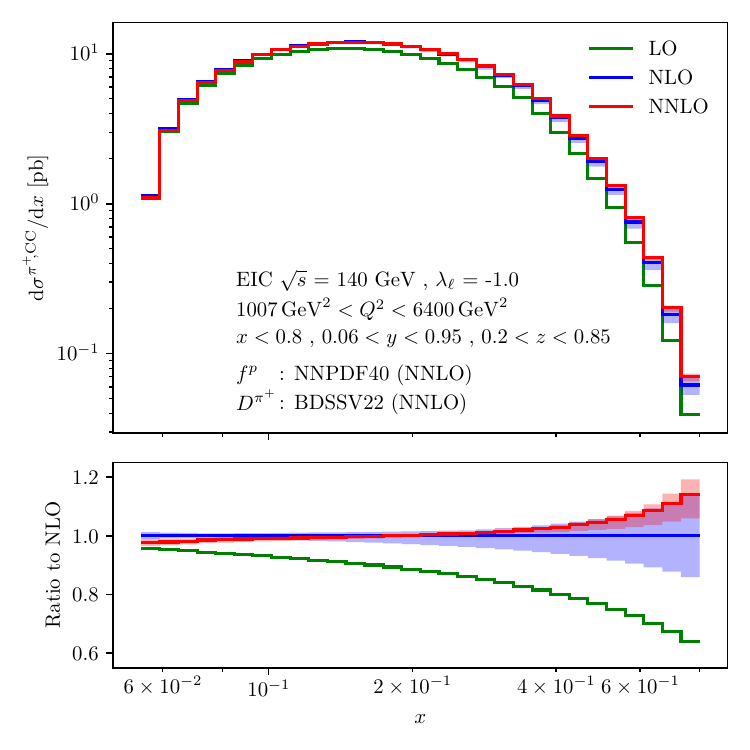}
\caption{High-$Q^2$}
\end{subfigure}
\begin{subfigure}{0.45\textwidth}
\centering
\includegraphics[width=1.0\linewidth]{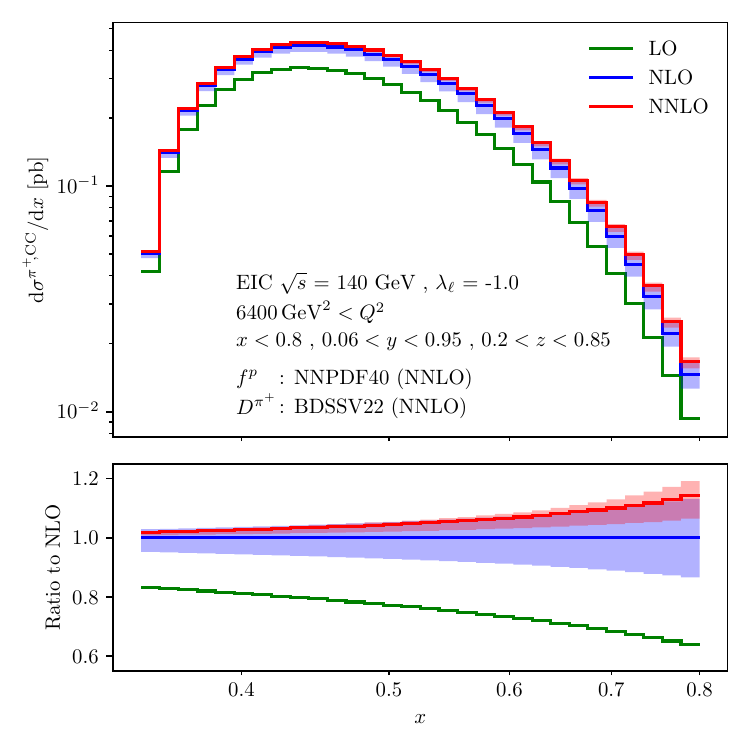}
\caption{Extreme-$Q^2$}
\end{subfigure}

\caption{
Electron-initiated CC $x$-distributions in $\pi^+$ production.
Top panels: total cross section.
Bottom panels: Ratio to NLO.
}
\label{fig:X_CH_pip}

\end{figure}

\begin{figure}[tbp]
\centering

\begin{subfigure}{0.45\textwidth}
\centering
\includegraphics[width=1.0\linewidth]{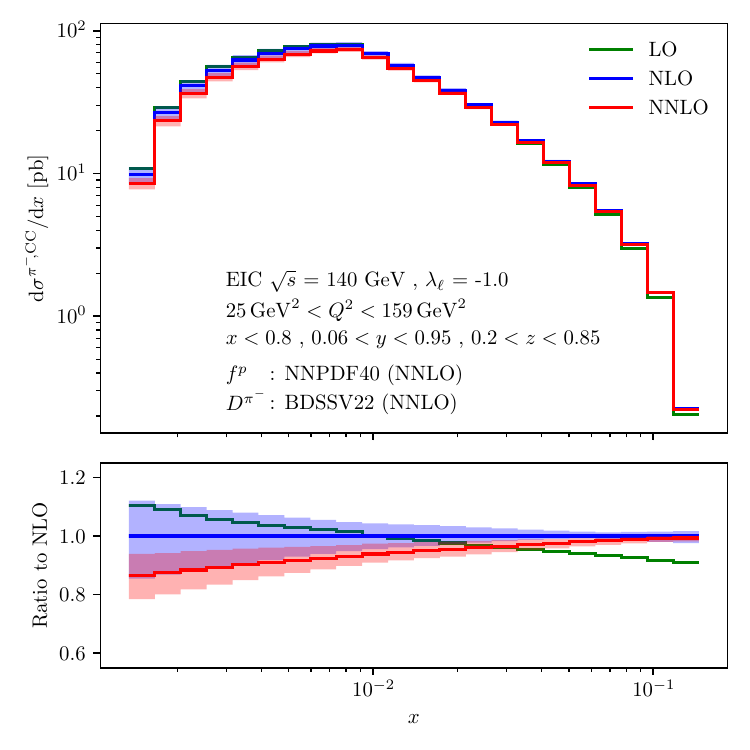}
\caption{Low-$Q^2$}
\end{subfigure}
\begin{subfigure}{0.45\textwidth}
\centering
\includegraphics[width=1.0\linewidth]{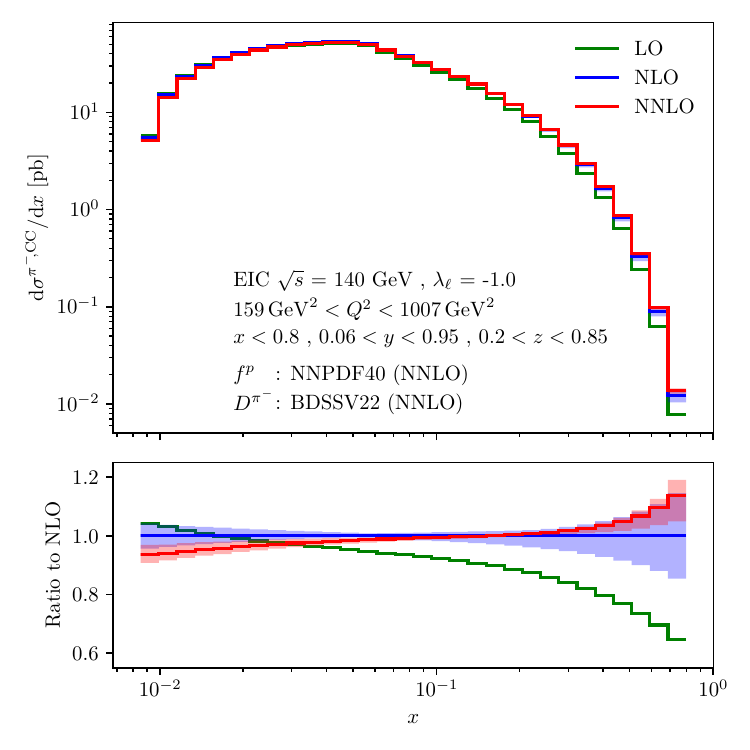}
\caption{Mid-$Q^2$}
\end{subfigure}
\begin{subfigure}{0.45\textwidth}
\centering
\includegraphics[width=.993\linewidth]{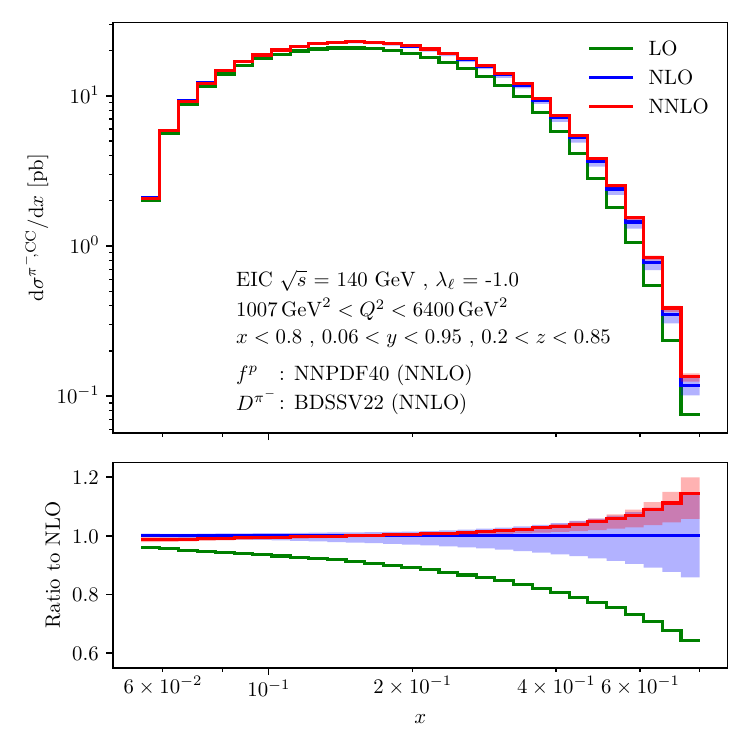}
\caption{High-$Q^2$}
\end{subfigure}
\begin{subfigure}{0.45\textwidth}
\centering
\includegraphics[width=1.0\linewidth]{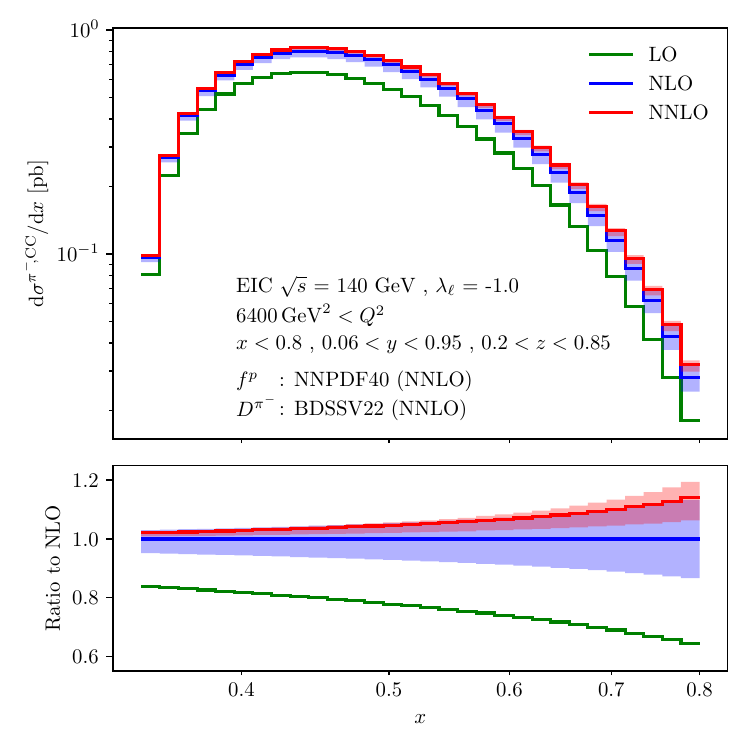}
\caption{Extreme-$Q^2$}
\end{subfigure}

\caption{
Electron-initiated CC $x$-distributions in $\pi^-$ production.
Top panels: total cross section.
Bottom panels: Ratio to NLO.
}
\label{fig:X_CH_pim}

\end{figure}

In Fig.~\ref{fig:X_CH_pip} and~\ref{fig:X_CH_pim} we show the $x$-distribution for the electron-initiated CC cross section in $\pi^+$ and $\pi^-$ production for the four different ranges in $Q^2$.
The differential cross sections at the considered perturbative orders show a similar behaviour in the flavour-favoured and flavour-unfavoured production mode.
In terms of size the favoured $\pi^-$ production mode is around $60\,\%$ larger in most energy ranges.
In the Extreme-$Q^2$ $\pi^-$ production is slightly more enhanced.
The shape of the CC cross sections exhibits the same kinematical dependence as the fiducial NC cross sections,
related to the accessible ranges in $y$.
Below the kink visible in the centre region for Low-$Q^2$ and Mid-$Q^2$, the fiducial cross section increases due to the loosening of the upper bound in $y$.
For the first few bins after the kink a moderate downward trend is visible before the lower bound in $y$ restricts the kinematic range and rapidly depletes the cross section.
The LO and NLO distributions differ significantly starting from Mid-$Q^2$ energies.
In the same virtuality ranges the NNLO corrections are small and substantially reduce the scale variation bands.
Also at Low-$Q^2$ we observe overlapping scale variation bands, but the corrections and associated uncertainties become particularly large for $x$-values favouring smaller virtualities.

\begin{figure}[tbp]
\centering

\begin{subfigure}{0.45\textwidth}
\centering
\includegraphics[width=.993\linewidth]{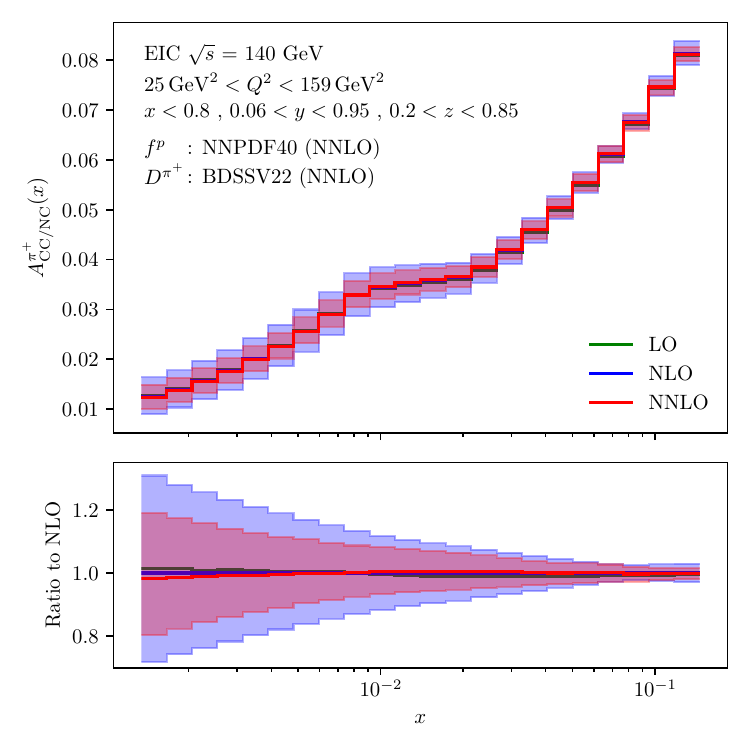}
\caption{Low-$Q^2$}
\end{subfigure}
\begin{subfigure}{0.45\textwidth}
\centering
\includegraphics[width=1.0\linewidth]{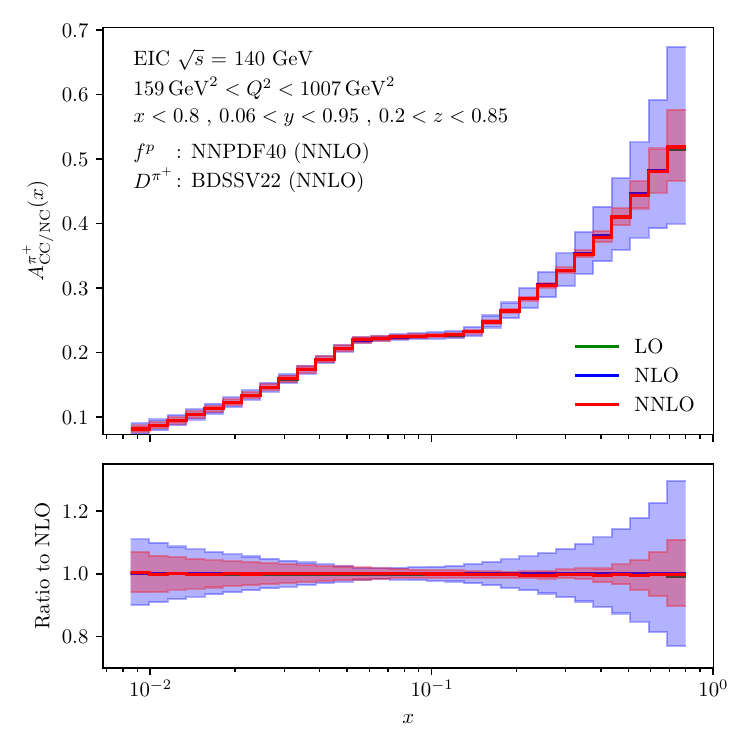}
\caption{Mid-$Q^2$}
\end{subfigure}
\begin{subfigure}{0.45\textwidth}
\centering
\includegraphics[width=1.0\linewidth]{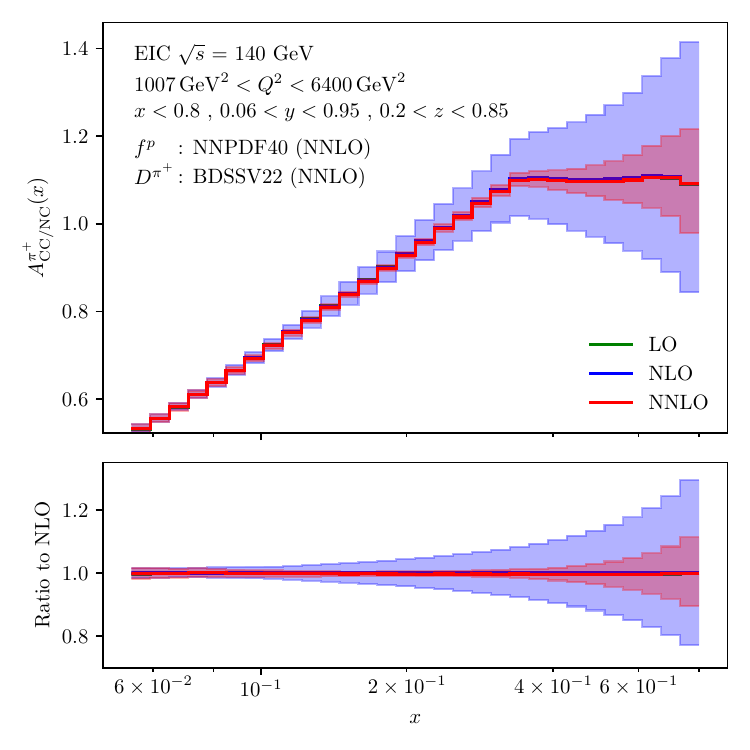}
\caption{High-$Q^2$}
\end{subfigure}
\begin{subfigure}{0.45\textwidth}
\centering
\includegraphics[width=1.0\linewidth]{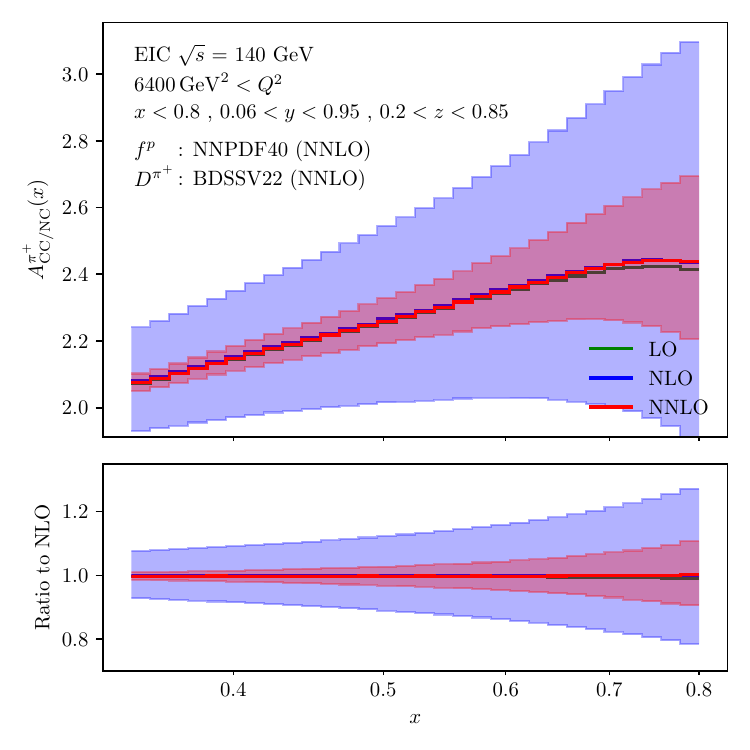}
\caption{Extreme-$Q^2$}
\end{subfigure}

\caption{
Electron-initiated asymmetry $A^{\pi^+}_\mathrm{CC/NC}$ as a function of $x$ in $\pi^+$ production.
Top panels: total asymmetry.
Bottom panels: Ratio to NLO.
}
\label{fig:X_A_CCNC_pip}

\end{figure}

\begin{figure}[tbp]
\centering

\begin{subfigure}{0.45\textwidth}
\centering
\includegraphics[width=1.0\linewidth]{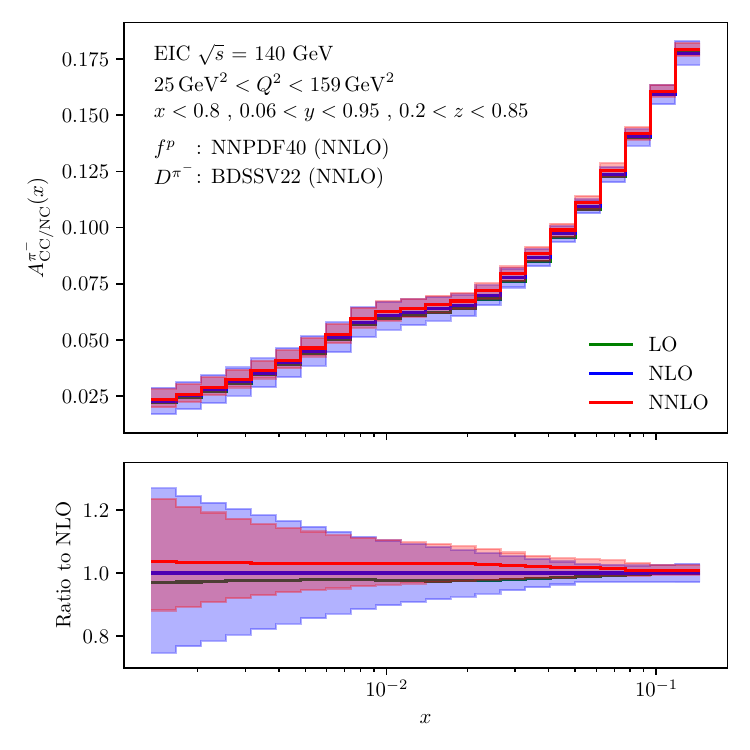}
\caption{Low-$Q^2$}
\end{subfigure}
\begin{subfigure}{0.45\textwidth}
\centering
\includegraphics[width=1.0\linewidth]{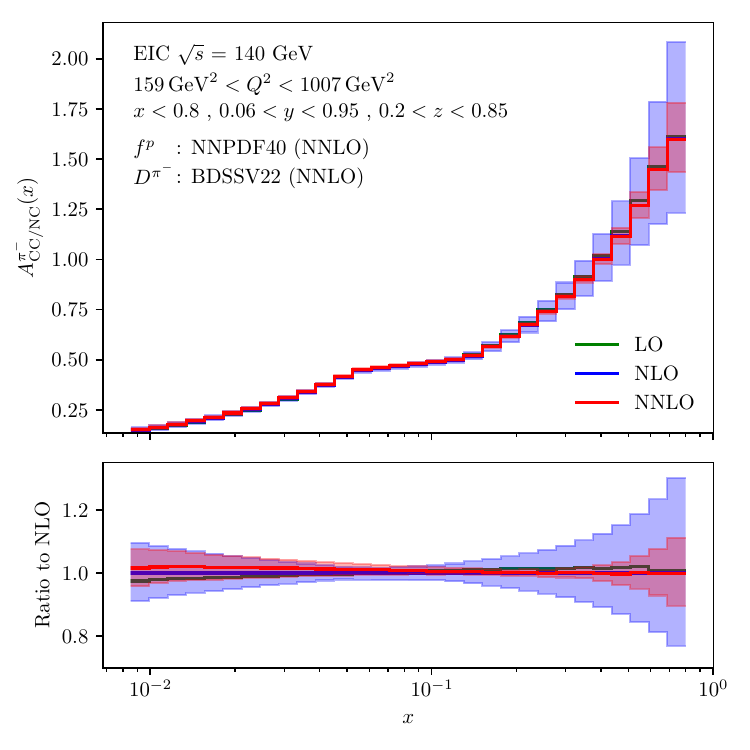}
\caption{Mid-$Q^2$}
\end{subfigure}
\begin{subfigure}{0.45\textwidth}
\centering
\includegraphics[width=1.0\linewidth]{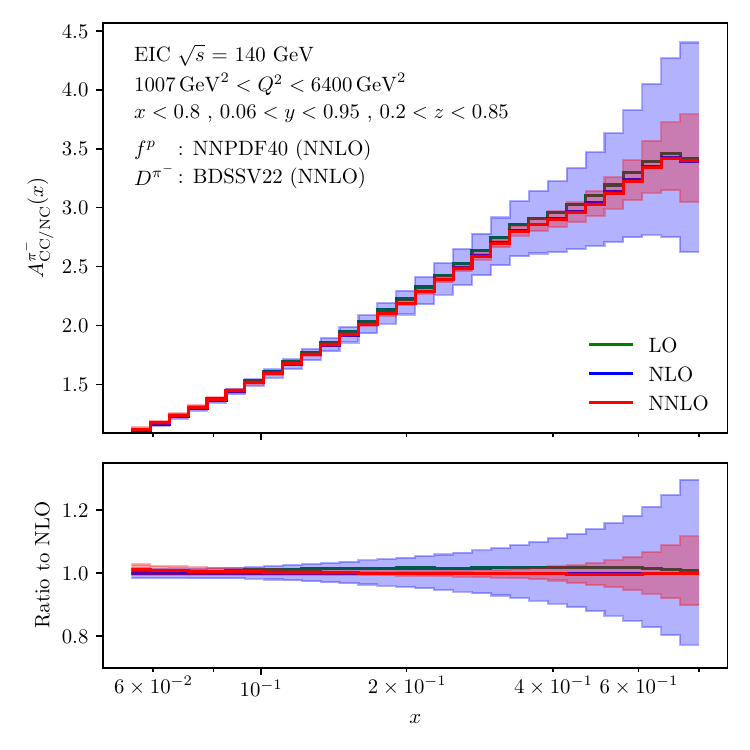}
\caption{High-$Q^2$}
\end{subfigure}
\begin{subfigure}{0.45\textwidth}
\centering
\includegraphics[width=1.0\linewidth]{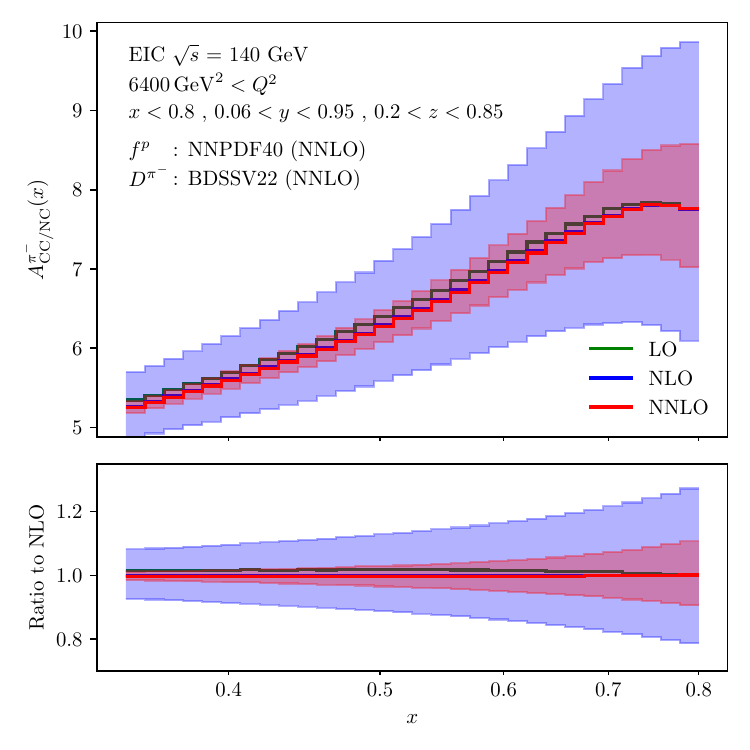}
\caption{Extreme-$Q^2$}
\end{subfigure}

\caption{
Electron-initiated asymmetry $A^{\pi^-}_\mathrm{CC/NC}$ as a function of $x$ in $\pi^-$ production.
Top panels: total asymmetry.
Bottom panels: Ratio to NLO.
}
\label{fig:X_A_CCNC_pim}
\end{figure}

In Fig.~\ref{fig:X_A_CCNC_pip} and~\ref{fig:X_A_CCNC_pim} the asymmetries $A^{\pi^\pm}_\mathrm{CC/NC}$ are presented as functions of $x$ in the four ranges in $Q^2$.
They are highly dependent on $Q^2$, and consequently also on $x$.

At Low-$Q^2$ and Mid-$Q^2$ they increase particularly strongly with $x$, while at increased $Q^2$ the rate of ascent in $x$ is dampened.
The charged-neutral current asymmetries for $\pi^-$ and $\pi^+$ production display a similar functional form at Low-$Q^2$ and Mid-$Q^2$, with a strong impact from the allowed values of $y$ in the intermediate-$x$ range.
Especially prominent are the plateaus, which are found in the same ranges of $x$ as for the neutral current asymmetry $A_\NC^{\pi^\pm}(x)$ and result from the kinematically allowed $y$-range.
This dependence may be explained with the increase of the accessible minimum virtuality due to the cut in $y$, which boosts the asymmetry in these regions.
Conversely, in the plateau region no strict cut on $Q^2$ is imposed and the allowed $Q^2$ range varies only moderately.

At $x>0.3$ for High-$Q^2$ and Extreme-$Q^2$ the growth of the asymmetries is moderated for flavour-favoured $\pi^-$ production mode and nearly halted for $\pi^+$ production.
The charged-neutral current asymmetry is largely independent of the perturbative order, and shows good perturbative convergence with a sizeable reduction of theoretical uncertainties especially for higher $Q^2$.
The size of theory uncertainties is comparable to the uncertainties in $A^{\pi^\pm}_\NC(x)$.

\clearpage
\newpage

\section{Conclusion}
\label{sec:conc}

In this paper, we presented the derivation of the
NNLO QCD corrections to single-inclusive hadron production in deeply
inelastic scattering, accounting for electroweak NC and CC current exchanges and initial-state lepton polarisation.
We introduced a new partonic channel decomposition for the SIDIS coefficient functions, which allows a
unified treatment of the NNLO corrections to NC and CC processes.
The
analytical results for the coefficient functions up to NNLO
are distributed as an ancillary file with the \texttt{arXiv} submission of this paper.

We performed detailed phenomenological studies
of NC and CC SIDIS cross sections for EIC kinematics at its
maximal center-of-mass energy value. By introducing lepton polarisation asymmetries and
CC/NC asymmetries, we were able to single out electroweak effects in SIDIS processes, thereby
potentially enabling their future study at the EIC.
The various processes display substantially different
compositions in terms of partonic channels, highlighting their potential to probe specific flavour combinations of fragmentation functions.

For EIC kinematics, NNLO corrections are typically found to be of moderate size, and the resulting NNLO predictions are
within the uncertainties of the previously known NLO predictions. A reduction of uncertainties from NLO to NNLO is
observed throughout.
In the NC case, we observe that the perturbative corrections to the
full electroweak SIDIS cross sections are very well described by re-scaling the respective higher-order predictions
for pure photon exchange processes by the electroweak-to-electromagnetic ratio of the Born-level cross sections.
A similar rescaling is still feasible but less accurate for the NC polarisation asymmetries.

Our work will enable NNLO precision studies of parton distributions and fragmentation functions with
future SIDIS data from the EIC, which will cover a substantially larger kinematical range than previous
fixed-target experiments in NC and CC processes, combined with excellent particle identification
capabilities required for SIDIS studies.

\acknowledgments

We would like to thank Ignacio Borsa for insightful discussions and
Valerio Bertone for validating parts of our numerical code implementation.
TG thanks the Institute for Nuclear Theory at the University of Washington for its
hospitality and the US Department of Energy for partial support through grant No.\ DE-FG02- 00ER41132
during the completion of this work.
This work has received funding from the Swiss National Science Foundation (SNF) under contract 200020-204200 and from the European Research Council (ERC) under the European Union’s Horizon 2020 research and innovation program grant agreement 101019620 (ERC Advanced Grant TOPUP).
KS was supported by the UZH Postdoc Grant, grant no.~[FK-24-115].

\newpage
\appendix
\section{Content of the ancillary file}\label{app:ancillary}
The ancillary file is in \texttt{FORM} readable format and contains all coefficient functions up to NNLO as listed in Sections \ref{sec:NCSymmCF}, \ref{sec:NCAntisymmCF},  \ref{sec:CCSymmCF} and \ref{sec:CCAntisymmCF}. 
We use the same notation and channel decomposition adopted in the above mentioned sections, which is explained in detail in the header of the ancillary file.  
The coefficient functions contained in the ancillary file are stripped of their couplings.
The coefficient functions are written in terms of plus distributions and contain the full dependence on the renormalization scale and the initial-state and final-state factorization scale.
Special functions appearing in the coefficient functions are also specified in the header of the ancillary file.
The naming scheme for the contributions in the ancillary file is
\begin{align}
  \texttt{\textcolor{MidnightBlue}{[{Structure function}]}\textcolor{Green}{C}\textcolor{BrickRed}{[order]}\textcolor{Fuchsia}{[{PDF}]}2\textcolor{Red}{[{FF}]}\textcolor{BlueViolet}{[labels]}} \,,
\end{align}
e.g.
\begin{align}
  \texttt{\textcolor{MidnightBlue}{FT}\textcolor{Green}{C}\textcolor{BrickRed}{0}\textcolor{Fuchsia}{q}2\textcolor{Red}{q}\textcolor{BlueViolet}{M}}
  \leftrightarrow
  \textcolor{Green}{C}^{\textcolor{MidnightBlue}{T},\textcolor{BrickRed}{(0)}}_{\textcolor{Red}{q}\textcolor{Fuchsia}{q}}
\end{align}
with the label \textcolor{BlueViolet}{\texttt{M}} denoting that the coefficient function is in the $\overline{\mathrm{MS}}$ scheme.
Like throughout the paper, the label \texttt{NoW} denotes that the coefficient is only present if the flavour is conserved at the electroweak vertex.
The label \texttt{Fcon} denotes flavour-conserving contributions if the flavour is not conserved at the electroweak vertex, and the labels \texttt{A} and \texttt{AA} denote single and double anomaly insertions.
Table \ref{tab:ancillary} lists the names of the coefficient functions in this paper and their notation in the ancillary file.

\renewcommand{\arraystretch}{1.3}
\begin{table}[tb]
\centering
\resizebox{\textwidth}{!}{%
\begin{tabular}{c|c|c|c|c|c}
\toprule
  \multicolumn{2}{c|}{$\Fcal^h_T$ } &
  \multicolumn{2}{c|}{$\Fcal^h_L$} &
  \multicolumn{2}{c}{$\Fcal^h_3$} \\
\midrule
 paper & ancillary & paper & ancillary & paper & ancillary \\
 \midrule
 $C^{T,(0)}_{qq}$  & \texttt{FTC0q2qM} &  &  & $C^{3,(0)}_{qq}$ & \texttt{F3C0q2qM} \\
 \midrule 
 $C^{T,(1)}_{qq}$ & \texttt{FTC1q2qM} & $C^{L,(1)}_{qq}$ & \texttt{FLC1q2qM}  &  $C^{3,(1)}_{qq}$ & \texttt{F3C1q2qM} \\
 $C^{T,(1)}_{gq}$ & \texttt{FTC1q2gM} & $C^{L,(1)}_{gq}$ & \texttt{FLC1q2gM}  &  $C^{3,(1)}_{gq}$ & \texttt{F3C1q2gM} \\
 $C^{T,(1)}_{qg}$ & \texttt{FTC1g2qM} & $C^{L,(1)}_{qg}$ & \texttt{FLC1g2qM}  &  $C^{3,(1)}_{qg}$ & \texttt{F3C1g2qM} \\
  \midrule
 $C^{T,(2)}_{qq}$                                   & \texttt{FTC2q2qM}              &  $C^{L,(2)}_{qq}$                                   & \texttt{FLC2q2qM}             &  $C^{3,(2)}_{qq}$                                    &  \texttt{F3C2q2qM} \\
 $C^{T,(2)}_{qq,\Fcon,1}$                      & \texttt{FTC2q2qMFcon1}    & $C^{L,(2)}_{qq,\Fcon,1}$                       & \texttt{FLC2q2qMFcon1}   & $C^{3,(2)}_{qq,\Fcon,1}$                        & \texttt{F3C2q2qMFcon1} \\
 $C^{T,(2)}_{qq,\Fcon,2}$                     & \texttt{FTC2q2qMFcon2}    & $C^{L,(2)}_{qq,\Fcon,2}$                     & \texttt{FLC2q2qMFcon2}   &                                                                    & \\
 $C^{T,(2)}_{qq,\mathrm{A},\NoW}$   & \texttt{FTC2q2qMANoW}   & $C^{L,(2)}_{qq,\mathrm{A},\NoW}$   & \texttt{FLC2q2qMANoW}   & $C^{3,(2)}_{qq,\mathrm{A},\NoW}$     & \texttt{F3C2q2qMANoW} \\
 $C^{T,(2)}_{gq}$                                   & \texttt{FTC2q2gM}              & $C^{L,(2)}_{gq}$                                   & \texttt{FLC2q2gM}              & $C^{3,(2)}_{gq}$                                     & \texttt{F3C2q2gM} \\
 $C^{T,(2)}_{gq,\mathrm{A},\NoW}$   & \texttt{FTC2q2gMANoW}   & $C^{L,(2)}_{gq,\mathrm{A},\NoW}$   & \texttt{FLC2q2gMANoW}   & $C^{3,(2)}_{gq,\mathrm{A},\NoW}$      & \texttt{F3C2q2gMANoW} \\
 $C^{T,(2)}_{qg}$                                   & \texttt{FTC2g2qM}              & $C^{L,(2)}_{qg}$                                   & \texttt{FLC2g2qM}              & $C^{3,(2)}_{qg}$                                       & \texttt{F3C2g2qM} \\
 $C^{T,(2)}_{qg,\mathrm{A},\NoW}$   & \texttt{FTC2g2qMANoW}   & $C^{L,(2)}_{qg,\mathrm{A},\NoW}$   & \texttt{FLC2g2qMANoW}   & $C^{3,(2)}_{qg,\mathrm{A},\NoW}$        & \texttt{F3C2g2qMANoW} \\
 $C^{T,(2)}_{gg}$                                   & \texttt{FTC2g2gM}              & $C^{L,(2)}_{gg}$                                   & \texttt{FLC2g2gM}              &                                                                      & \\
 $C^{T,(2)}_{gg,\mathrm{AA},\NoW}$ & \texttt{FTC2g2gMAANoW} & $C^{L,(2)}_{gg,\mathrm{AA},\NoW}$ & \texttt{FLC2g2gMAANoW} &                                                                      & \\
 $C^{T,(2)}_{\qp q,1}$                            & \texttt{FTC2q2qpM1}          & $C^{L,(2)}_{\qp q,1}$                             & \texttt{FLC2q2qpM1}           & $C^{3,(2)}_{\qp q,1}$                                & \texttt{F3C2q2qpM1} \\
 $C^{T,(2)}_{\qp q,2}$                            & \texttt{FTC2q2qpM2}         & $C^{L,(2)}_{\qp q,2}$                            & \texttt{FLC2q2qpM2}           & $C^{3,(2)}_{\qp q,2}$                               & \texttt{F3C2q2qpM2} \\
 $C^{T,(2)}_{\qp q,\NoW,3}$                 & \texttt{FTC2q2qpMNoW3} & $C^{L,(2)}_{\qp q,\NoW,3}$                 & \texttt{FLC2q2qpMNoW3}  & $C^{3,(2)}_{\qp q,\NoW,3}$                    & \texttt{F3C2q2qpMNoW3} \\
 $C^{T,(2)}_{\qp q,\NoW,4}$                 & \texttt{FTC2q2qpMNoW4} & $C^{L,(2)}_{\qp q,\NoW,4}$                 & \texttt{FLC2q2qpMNoW4}  & $C^{3,(2)}_{\qp q,\NoW,4}$                     & \texttt{F3C2q2qpMNoW4} \\
 $C^{T,(2)}_{\qb q}$                               & \texttt{FTC2q2qbM}            & $C^{L,(2)}_{\qb q}$                               & \texttt{FLC2q2qbM}            & $C^{3,(2)}_{\qb q}$                              & \texttt{F3C2q2qbM} \\
 $C^{T,(2)}_{\qb q,\Fcon}$                     & \texttt{FTC2q2qbMFcon}    & $C^{L,(2)}_{\qb q,\Fcon}$                    & \texttt{FLC2q2qbMFcon}    & $C^{3,(2)}_{\qb q,\Fcon}$                     & \texttt{F3C2q2qbMFcon} \\
  \bottomrule
\end{tabular}
}
\caption{List of coefficient functions and ancillary file notation grouped by structure function and perturbative order.}
\label{tab:ancillary}
\end{table}

\clearpage
\newpage
\bibliographystyle{JHEP}
\bibliography{EW_SIDIS}

\providecommand{\href}[2]{#2}\begingroup\raggedright\begin{thebibliography}{10}

\bibitem{Accardi:2012qut}
A.~Accardi et~al., \emph{{Electron Ion Collider: The Next QCD Frontier}:
  {Understanding the glue that binds us all}},
  \href{https://doi.org/10.1140/epja/i2016-16268-9}{\emph{Eur. Phys. J. A}
  {\bfseries 52} (2016) 268} [\href{https://arxiv.org/abs/1212.1701}{{\ttfamily
  1212.1701}}].

\bibitem{AbdulKhalek:2021gbh}
R.~Abdul~Khalek et~al., \emph{{Science Requirements and Detector Concepts for
  the Electron-Ion Collider}: {EIC Yellow Report}},
  \href{https://doi.org/10.1016/j.nuclphysa.2022.122447}{\emph{Nucl. Phys. A}
  {\bfseries 1026} (2022) 122447}
  [\href{https://arxiv.org/abs/2103.05419}{{\ttfamily 2103.05419}}].

\bibitem{Bjorken:1968dy}
J.D.~Bjorken, \emph{{Asymptotic Sum Rules at Infinite Momentum}},
  \href{https://doi.org/10.1103/PhysRev.179.1547}{\emph{Phys. Rev.} {\bfseries
  179} (1969) 1547}.

\bibitem{Feynman:1969ej}
R.P.~Feynman, \emph{{Very high-energy collisions of hadrons}},
  \href{https://doi.org/10.1103/PhysRevLett.23.1415}{\emph{Phys. Rev. Lett.}
  {\bfseries 23} (1969) 1415}.

\bibitem{Moch:2004xu}
S.~Moch, J.A.M.~Vermaseren and A.~Vogt, \emph{{The Longitudinal structure
  function at the third order}},
  \href{https://doi.org/10.1016/j.physletb.2004.11.063}{\emph{Phys. Lett. B}
  {\bfseries 606} (2005) 123}
  [\href{https://arxiv.org/abs/hep-ph/0411112}{{\ttfamily hep-ph/0411112}}].

\bibitem{Vermaseren:2005qc}
J.A.M.~Vermaseren, A.~Vogt and S.~Moch, \emph{{The Third-order QCD corrections
  to deep-inelastic scattering by photon exchange}},
  \href{https://doi.org/10.1016/j.nuclphysb.2005.06.020}{\emph{Nucl. Phys. B}
  {\bfseries 724} (2005) 3}
  [\href{https://arxiv.org/abs/hep-ph/0504242}{{\ttfamily hep-ph/0504242}}].

\bibitem{Moch:2008fj}
S.~Moch, J.A.M.~Vermaseren and A.~Vogt, \emph{{Third-order QCD corrections to
  the charged-current structure function F(3)}},
  \href{https://doi.org/10.1016/j.nuclphysb.2009.01.001}{\emph{Nucl. Phys. B}
  {\bfseries 813} (2009) 220}
  [\href{https://arxiv.org/abs/0812.4168}{{\ttfamily 0812.4168}}].

\bibitem{Blumlein:2022gpp}
J.~{Bl\"umlein}, P.~Marquard, C.~Schneider and K.~{Sch\"onwald}, \emph{{The
  massless three-loop Wilson coefficients for the deep-inelastic structure
  functions F$_{2}$, F$_{L}$, xF$_{3}$ and g$_{1}$}},
  \href{https://doi.org/10.1007/JHEP11(2022)156}{\emph{JHEP} {\bfseries 11}
  (2022) 156} [\href{https://arxiv.org/abs/2208.14325}{{\ttfamily
  2208.14325}}].

\bibitem{Gehrmann:2018odt}
T.~Gehrmann, A.~Huss, J.~Niehues, A.~Vogt and D.M.~Walker, \emph{{Jet
  production in charged-current deep-inelastic scattering to third order in
  QCD}}, \href{https://doi.org/10.1016/j.physletb.2019.03.003}{\emph{Phys.
  Lett. B} {\bfseries 792} (2019) 182}
  [\href{https://arxiv.org/abs/1812.06104}{{\ttfamily 1812.06104}}].

\bibitem{Carli:2010cg}
T.~Carli, T.~Gehrmann and S.~Hoeche, \emph{{Hadronic final states in
  deep-inelastic scattering with Sherpa}},
  \href{https://doi.org/10.1140/epjc/s10052-010-1261-2}{\emph{Eur. Phys. J. C}
  {\bfseries 67} (2010) 73} [\href{https://arxiv.org/abs/0912.3715}{{\ttfamily
  0912.3715}}].

\bibitem{Hoche:2018gti}
S.~H\"oche, S.~Kuttimalai and Y.~Li, \emph{{Hadronic Final States in DIS at
  NNLO QCD with Parton Showers}},
  \href{https://doi.org/10.1103/PhysRevD.98.114013}{\emph{Phys. Rev. D}
  {\bfseries 98} (2018) 114013}
  [\href{https://arxiv.org/abs/1809.04192}{{\ttfamily 1809.04192}}].

\bibitem{Banfi:2023mhz}
A.~Banfi, S.~Ferrario~Ravasio, B.~J\"ager, A.~Karlberg, F.~Reichenbach and
  G.~Zanderighi, \emph{{A POWHEG generator for deep inelastic scattering}},
  \href{https://doi.org/10.1007/JHEP02(2024)023}{\emph{JHEP} {\bfseries 02}
  (2024) 023} [\href{https://arxiv.org/abs/2309.02127}{{\ttfamily
  2309.02127}}].

\bibitem{Borsa:2024rmh}
I.~Borsa and B.~J\"ager, \emph{{Parton-shower effects in polarized deep
  inelastic scattering}},
  \href{https://doi.org/10.1007/JHEP07(2024)177}{\emph{JHEP} {\bfseries 07}
  (2024) 177} [\href{https://arxiv.org/abs/2404.07702}{{\ttfamily
  2404.07702}}].

\bibitem{Buonocore:2024pdv}
L.~Buonocore, G.~Limatola, P.~Nason and F.~Tramontano, \emph{{An event
  generator for Lepton-Hadron deep inelastic scattering at NLO+PS with POWHEG
  including mass effects}},
  \href{https://doi.org/10.1007/JHEP08(2024)083}{\emph{JHEP} {\bfseries 08}
  (2024) 083} [\href{https://arxiv.org/abs/2406.05115}{{\ttfamily
  2406.05115}}].

\bibitem{FerrarioRavasio:2024kem}
S.~Ferrario~Ravasio, R.~Gauld, B.~J\"ager, A.~Karlberg and G.~Zanderighi,
  \emph{{An event generator for neutrino-induced Deep Inelastic Scattering and
  applications to neutrino astronomy}},
  \href{https://arxiv.org/abs/2407.03894}{{\ttfamily 2407.03894}}.

\bibitem{Goyal:2023zdi}
S.~Goyal, S.-O.~Moch, V.~Pathak, N.~Rana and V.~Ravindran,
  \emph{{Next-to-Next-to-Leading Order QCD Corrections to Semi-Inclusive
  Deep-Inelastic Scattering}},
  \href{https://doi.org/10.1103/PhysRevLett.132.251902}{\emph{Phys. Rev. Lett.}
  {\bfseries 132} (2024) 251902}
  [\href{https://arxiv.org/abs/2312.17711}{{\ttfamily 2312.17711}}].

\bibitem{Bonino:2024qbh}
L.~Bonino, T.~Gehrmann and G.~Stagnitto, \emph{{Semi-Inclusive Deep-Inelastic
  Scattering at Next-to-Next-to-Leading Order in QCD}},
  \href{https://doi.org/10.1103/PhysRevLett.132.251901}{\emph{Phys. Rev. Lett.}
  {\bfseries 132} (2024) 251901}
  [\href{https://arxiv.org/abs/2401.16281}{{\ttfamily 2401.16281}}].

\bibitem{Goyal:2024emo}
S.~Goyal, R.N.~Lee, S.-O.~Moch, V.~Pathak, N.~Rana and V.~Ravindran,
  \emph{{NNLO QCD corrections to unpolarized and polarized SIDIS}},
  \href{https://doi.org/10.1103/PhysRevD.111.094007}{\emph{Phys. Rev. D}
  {\bfseries 111} (2025) 094007}
  [\href{https://arxiv.org/abs/2412.19309}{{\ttfamily 2412.19309}}].

\bibitem{Bonino:2024wgg}
L.~Bonino, T.~Gehrmann, M.~L\"ochner, K.~Sch\"onwald and G.~Stagnitto,
  \emph{{Polarized Semi-Inclusive Deep-Inelastic Scattering at
  Next-to-Next-to-Leading Order in QCD}},
  \href{https://doi.org/10.1103/PhysRevLett.133.211904}{\emph{Phys. Rev. Lett.}
  {\bfseries 133} (2024) 211904}
  [\href{https://arxiv.org/abs/2404.08597}{{\ttfamily 2404.08597}}].

\bibitem{Goyal:2024tmo}
S.~Goyal, R.N.~Lee, S.-O.~Moch, V.~Pathak, N.~Rana and V.~Ravindran,
  \emph{{Next-to-Next-to-Leading Order QCD Corrections to Polarized
  Semi-Inclusive Deep-Inelastic Scattering}},
  \href{https://doi.org/10.1103/PhysRevLett.133.211905}{\emph{Phys. Rev. Lett.}
  {\bfseries 133} (2024) 211905}
  [\href{https://arxiv.org/abs/2404.09959}{{\ttfamily 2404.09959}}].

\bibitem{Amaldi:1979yh}
U.~Amaldi et~al., \emph{{REPORT FROM THE STUDY GROUP ON DETECTORS FOR CHARGED
  CURRENT EVENTS}},  in \emph{{ECFA Study of an ep Facility for Europe}},
  pp.~377--414, 1979.

\bibitem{Aachen-Bonn-CERN-Munich-Oxford:1982jrr}
{\scshape Aachen-Bonn-CERN-Munich-Oxford} collaboration, \emph{{An
  Investigation of Quark and Diquark Fragmentation in Neutrino $p$ and
  Anti-neutrino $p$ Charged Current Interactions in {BEBC}}},
  \href{https://doi.org/10.1016/0550-3213(83)90239-0}{\emph{Nucl. Phys. B}
  {\bfseries 214} (1983) 369}.

\bibitem{Helenius:2024fow}
I.~Helenius, H.~Paukkunen and S.~Yrj\"anheikki, \emph{{Dimuons from
  neutrino-nucleus collisions in the semi-inclusive DIS approach}},
  \href{https://doi.org/10.1007/JHEP09(2024)043}{\emph{JHEP} {\bfseries 09}
  (2024) 043} [\href{https://arxiv.org/abs/2405.12677}{{\ttfamily
  2405.12677}}].

\bibitem{Paukkunen:2025kjb}
H.~Paukkunen, I.~Helenius and S.~Yrj\"anheikki, \emph{{Improving the
  description of dimuon production in neutrino-nucleus collisions using the
  SACOT-$\chi$ scheme}},  \href{https://arxiv.org/abs/2506.09492}{{\ttfamily
  2506.09492}}.

\bibitem{Bonino:2025tnf}
L.~Bonino, T.~Gehrmann, M.~L\"ochner, K.~Sch\"onwald and G.~Stagnitto,
  \emph{{Identified Hadron Production in Deeply Inelastic Neutrino-Nucleon
  Scattering}},  \href{https://arxiv.org/abs/2504.05376}{{\ttfamily
  2504.05376}}.

\bibitem{ParticleDataGroup:2024cfk}
{\scshape Particle Data Group} collaboration, \emph{{Review of particle
  physics}}, \href{https://doi.org/10.1103/PhysRevD.110.030001}{\emph{Phys.
  Rev. D} {\bfseries 110} (2024) 030001}.

\bibitem{deFlorian:2012wk}
D.~de~Florian and Y.~Rotstein~Habarnau, \emph{{Polarized semi-inclusive
  electroweak structure functions at next-to-leading-order}},
  \href{https://doi.org/10.1140/epjc/s10052-013-2356-3}{\emph{Eur. Phys. J. C}
  {\bfseries 73} (2013) 2356}
  [\href{https://arxiv.org/abs/1210.7203}{{\ttfamily 1210.7203}}].

\bibitem{Anderle:2016kwa}
D.~Anderle, D.~de~Florian and Y.~Rotstein~Habarnau, \emph{{Towards
  semi-inclusive deep inelastic scattering at next-to-next-to-leading order}},
  \href{https://doi.org/10.1103/PhysRevD.95.034027}{\emph{Phys. Rev. D}
  {\bfseries 95} (2017) 034027}
  [\href{https://arxiv.org/abs/1612.01293}{{\ttfamily 1612.01293}}].

\bibitem{Altarelli:1979kv}
G.~Altarelli, R.K.~Ellis, G.~Martinelli and S.-Y.~Pi, \emph{{Processes
  Involving Fragmentation Functions Beyond the Leading Order in QCD}},
  \href{https://doi.org/10.1016/0550-3213(79)90062-2}{\emph{Nucl. Phys. B}
  {\bfseries 160} (1979) 301}.

\bibitem{Baier:1979sp}
R.~Baier and K.~Fey, \emph{{Finite corrections to quark fragmentation functions
  in perturbative QCD}}, \href{https://doi.org/10.1007/BF01545897}{\emph{Z.
  Phys. C} {\bfseries 2} (1979) 339}.

\bibitem{Furmanski:1981cw}
W.~Furmanski and R.~Petronzio, \emph{{Lepton - Hadron Processes Beyond Leading
  Order in Quantum Chromodynamics}},
  \href{https://doi.org/10.1007/BF01578280}{\emph{Z. Phys. C} {\bfseries 11}
  (1982) 293}.

\bibitem{Nogueira:1991ex}
P.~Nogueira, \emph{{Automatic Feynman Graph Generation}},
  \href{https://doi.org/10.1006/jcph.1993.1074}{\emph{J. Comput. Phys.}
  {\bfseries 105} (1993) 279}.

\bibitem{Vermaseren:2000nd}
J.A.M.~Vermaseren, \emph{{New features of FORM}},
  \href{https://arxiv.org/abs/math-ph/0010025}{{\ttfamily math-ph/0010025}}.

\bibitem{Larin:1993tq}
S.A.~Larin, \emph{{The Renormalization of the axial anomaly in dimensional
  regularization}},
  \href{https://doi.org/10.1016/0370-2693(93)90053-K}{\emph{Phys. Lett. B}
  {\bfseries 303} (1993) 113}
  [\href{https://arxiv.org/abs/hep-ph/9302240}{{\ttfamily hep-ph/9302240}}].

\bibitem{vonManteuffel:2012np}
A.~von Manteuffel and C.~Studerus, \emph{{Reduze 2 - Distributed Feynman
  Integral Reduction}},  \href{https://arxiv.org/abs/1201.4330}{{\ttfamily
  1201.4330}}.

\bibitem{Chetyrkin:1981qh}
K.G.~Chetyrkin and F.V.~Tkachov, \emph{{Integration by parts: The algorithm to
  calculate $\beta$-functions in 4 loops}},
  \href{https://doi.org/10.1016/0550-3213(81)90199-1}{\emph{Nucl. Phys. B}
  {\bfseries 192} (1981) 159}.

\bibitem{Laporta:2000dsw}
S.~Laporta, \emph{{High-precision calculation of multiloop Feynman integrals by
  difference equations}},
  \href{https://doi.org/10.1142/S0217751X00002159}{\emph{Int. J. Mod. Phys. A}
  {\bfseries 15} (2000) 5087}
  [\href{https://arxiv.org/abs/hep-ph/0102033}{{\ttfamily hep-ph/0102033}}].

\bibitem{Gehrmann:1999as}
T.~Gehrmann and E.~Remiddi, \emph{{Differential equations for two loop four
  point functions}},
  \href{https://doi.org/10.1016/S0550-3213(00)00223-6}{\emph{Nucl. Phys. B}
  {\bfseries 580} (2000) 485}
  [\href{https://arxiv.org/abs/hep-ph/9912329}{{\ttfamily hep-ph/9912329}}].

\bibitem{Gehrmann:2005pd}
T.~Gehrmann, T.~Huber and D.~Maitre, \emph{{Two-loop quark and gluon
  form-factors in dimensional regularisation}},
  \href{https://doi.org/10.1016/j.physletb.2005.07.019}{\emph{Phys. Lett. B}
  {\bfseries 622} (2005) 295}
  [\href{https://arxiv.org/abs/hep-ph/0507061}{{\ttfamily hep-ph/0507061}}].

\bibitem{Anastasiou:2002yz}
C.~Anastasiou and K.~Melnikov, \emph{{Higgs boson production at hadron
  colliders in NNLO QCD}},
  \href{https://doi.org/10.1016/S0550-3213(02)00837-4}{\emph{Nucl. Phys. B}
  {\bfseries 646} (2002) 220}
  [\href{https://arxiv.org/abs/hep-ph/0207004}{{\ttfamily hep-ph/0207004}}].

\bibitem{Bonino:2024adk}
L.~Bonino, T.~Gehrmann, M.~Marcoli, R.~Sch\"urmann and G.~Stagnitto,
  \emph{{Antenna subtraction for processes with identified particles at hadron
  colliders}}, \href{https://doi.org/10.1007/JHEP08(2024)073}{\emph{JHEP}
  {\bfseries 08} (2024) 073}
  [\href{https://arxiv.org/abs/2406.09925}{{\ttfamily 2406.09925}}].

\bibitem{Ahmed:2024owh}
T.~Ahmed, S.~Goyal, S.M.~Hasan, R.N.~Lee, S.-O.~Moch, V.~Pathak et~al.,
  \emph{{NNLO phase-space integrals for semi-inclusive deep-inelastic
  scattering}}, \href{https://doi.org/10.1103/22x4-dmmp}{\emph{Phys. Rev. D}
  {\bfseries 112} (2025) 014020}
  [\href{https://arxiv.org/abs/2412.16509}{{\ttfamily 2412.16509}}].

\bibitem{Gehrmann:2022cih}
T.~Gehrmann and R.~Sch\"urmann, \emph{{Photon fragmentation in the antenna
  subtraction formalism}},
  \href{https://doi.org/10.1007/JHEP04(2022)031}{\emph{JHEP} {\bfseries 04}
  (2022) 031} [\href{https://arxiv.org/abs/2201.06982}{{\ttfamily
  2201.06982}}].

\bibitem{Haug:2022hkr}
J.~Haug and F.~Wunder, \emph{{The massless single off-shell scalar box integral
  \textemdash{} branch cut structure and all-order epsilon expansion}},
  \href{https://doi.org/10.1007/JHEP02(2023)177}{\emph{JHEP} {\bfseries 02}
  (2023) 177} [\href{https://arxiv.org/abs/2211.14110}{{\ttfamily
  2211.14110}}].

\bibitem{Remiddi:1999ew}
E.~Remiddi and J.A.M.~Vermaseren, \emph{{Harmonic polylogarithms}},
  \href{https://doi.org/10.1142/S0217751X00000367}{\emph{Int. J. Mod. Phys. A}
  {\bfseries 15} (2000) 725}
  [\href{https://arxiv.org/abs/hep-ph/9905237}{{\ttfamily hep-ph/9905237}}].

\bibitem{Gehrmann:2000zt}
T.~Gehrmann and E.~Remiddi, \emph{{Two loop master integrals for gamma*
  ---\ensuremath{>} 3 jets: The Planar topologies}},
  \href{https://doi.org/10.1016/S0550-3213(01)00057-8}{\emph{Nucl. Phys. B}
  {\bfseries 601} (2001) 248}
  [\href{https://arxiv.org/abs/hep-ph/0008287}{{\ttfamily hep-ph/0008287}}].

\bibitem{Maitre:2005uu}
D.~Maitre, \emph{{HPL, a mathematica implementation of the harmonic
  polylogarithms}},
  \href{https://doi.org/10.1016/j.cpc.2005.10.008}{\emph{Comput. Phys. Commun.}
  {\bfseries 174} (2006) 222}
  [\href{https://arxiv.org/abs/hep-ph/0507152}{{\ttfamily hep-ph/0507152}}].

\bibitem{Duhr:2019tlz}
C.~Duhr and F.~Dulat, \emph{{PolyLogTools \textemdash{} polylogs for the
  masses}}, \href{https://doi.org/10.1007/JHEP08(2019)135}{\emph{JHEP}
  {\bfseries 08} (2019) 135}
  [\href{https://arxiv.org/abs/1904.07279}{{\ttfamily 1904.07279}}].

\bibitem{Borsa:2022vvp}
I.~Borsa, R.~Sassot, D.~de~Florian, M.~Stratmann and W.~Vogelsang,
  \emph{{Towards a Global QCD Analysis of Fragmentation Functions at
  Next-to-Next-to-Leading Order Accuracy}},
  \href{https://doi.org/10.1103/PhysRevLett.129.012002}{\emph{Phys. Rev. Lett.}
  {\bfseries 129} (2022) 012002}
  [\href{https://arxiv.org/abs/2202.05060}{{\ttfamily 2202.05060}}].

\bibitem{NNPDF:2021njg}
{\scshape NNPDF} collaboration, \emph{{The path to proton structure at 1\%
  accuracy}}, \href{https://doi.org/10.1140/epjc/s10052-022-10328-7}{\emph{Eur.
  Phys. J. C} {\bfseries 82} (2022) 428}
  [\href{https://arxiv.org/abs/2109.02653}{{\ttfamily 2109.02653}}].

\bibitem{Buckley:2014ana}
A.~Buckley, J.~Ferrando, S.~Lloyd, K.~Nordstr\"om, B.~Page, M.~R\"ufenacht
  et~al., \emph{{LHAPDF6: parton density access in the LHC precision era}},
  \href{https://doi.org/10.1140/epjc/s10052-015-3318-8}{\emph{Eur. Phys. J. C}
  {\bfseries 75} (2015) 132} [\href{https://arxiv.org/abs/1412.7420}{{\ttfamily
  1412.7420}}].

\bibitem{Borsa:2022cap}
I.~Borsa, D.~de~Florian and I.~Pedron, \emph{{NNLO jet production in neutral
  and charged current polarized deep inelastic scattering}},
  \href{https://doi.org/10.1103/PhysRevD.107.054027}{\emph{Phys. Rev. D}
  {\bfseries 107} (2023) 054027}
  [\href{https://arxiv.org/abs/2212.06625}{{\ttfamily 2212.06625}}].

\end{thebibliography}\endgroup

\end{document}